\documentclass[11pt,a4paper]{article}
\usepackage[a4paper, left=1in, right=1in, top=1in, bottom=1in, includehead, includefoot, foot = 0pt, head=0pt]{geometry} 
\parskip=\medskipamount 

\usepackage{booktabs} 
\usepackage{enumitem} 

\usepackage[round,colon,authoryear]{natbib} 
\usepackage[hyphens,spaces,obeyspaces]{url}
\usepackage[pdftex,colorlinks,citecolor=blue,urlcolor=blue,linkcolor=red]{hyperref} 
\DeclareUrlCommand\doi{\def\UrlLeft##1\UrlRight{doi:\hspace{0.1cm} \href{http://dx.doi.org/##1}{##1}}\urlstyle{rm}}

\usepackage{graphicx} 
\usepackage[small,bf,hang]{caption}	

\usepackage{tabularx}
\usepackage[export]{adjustbox}

\usepackage{amssymb} 
\usepackage{amsmath} 
\usepackage{accents}
\usepackage{bbm} 
\usepackage{amsbsy} 
\usepackage{multirow} 
\usepackage[dvipsnames]{xcolor} 
\usepackage{mathtools}
\usepackage{eurosym}

\usepackage[defaultcolor=black]{changes}
\usepackage{authblk} 

\usepackage{fancyhdr}
\newcommand{\sbt}{\,\begin{picture}(-1,1)(-1,-3)\circle*{3}\end{picture}\ }
\DeclareMathOperator*{\argmax}{arg\,max}
\DeclareMathOperator*{\argmin}{arg\,min}
\bibpunct[, ]{(}{)}{;}{a}{,}{,}

\usepackage{tikz,pgfplots}
\usetikzlibrary{fit, positioning,chains,shapes.arrows,fit}
\usetikzlibrary{shapes,arrows, arrows.meta}
\usetikzlibrary{decorations.pathreplacing}
\usetikzlibrary{calc} 
\usetikzlibrary{matrix} 
\usepackage{pgf}
\usetikzlibrary{decorations.text}
\tikzset{cross/.style={cross out, draw=black, minimum size=2*(#1-\pgflinewidth), inner sep=0pt, outer sep=0pt},
cross/.default={1pt}}
\usetikzlibrary{shapes.misc}
\usepackage{environ}
\makeatletter
\newsavebox{\measure@tikzpicture}
\NewEnviron{scaletikzpicturetowidth}[1]{%
  \def\tikz@width{#1}%
  \begin{lrbox}{\measure@tikzpicture}%
  \BODY
  \end{lrbox}%
  \pgfmathparse{#1/\wd\measure@tikzpicture}%
  \BODY
}
\makeatother
\tikzdeclarecoordinatesystem{page}{
    \parsecomma#1\endparsecomma
    \pgfpointanchor{current page}{north east}
    \pgf@xc=\pgf@x%
    \pgf@yc=\pgf@y%
    \pgfpointanchor{current page}{south west}
    \pgf@xb=\pgf@x%
    \pgf@yb=\pgf@y%
    \pgfmathparse{(\pgf@xc-\pgf@xb)/2.*\page@x+(\pgf@xc+\pgf@xb)/2.}
    \expandafter\pgf@x\expandafter=\pgfmathresult pt
    \pgfmathparse{(\pgf@yc-\pgf@yb)/2.*\page@y+(\pgf@yc+\pgf@yb)/2.}
    \expandafter\pgf@y\expandafter=\pgfmathresult pt
}
\makeatother
\usetikzlibrary{arrows,automata}
\usepackage{ragged2e}
\definecolor{arrowcolor}{RGB}{201,216,232}
\definecolor{circlecolor}{RGB}{79,129,189}
\colorlet{textcolor}{white}
\colorlet{bordercolor}{white}

\pgfdeclarelayer{background}
\pgfsetlayers{background,main}

\definecolor{otherblue}{rgb}{0.337, 0.706, 0.914}
\definecolor{airforceblue}{rgb}{0.36, 0.54, 0.66}
\definecolor{forestgreen}{rgb}{0.13, 0.55, 0.13}\definecolor{fulvous}{rgb}{0.86, 0.52, 0.0}
\definecolor{gray}{rgb}{0.5, 0.5, 0.5}
\definecolor{bistre}{rgb}{0.24, 0.17, 0.12}\definecolor{bostonuniversityred}{rgb}{0.8, 0.0, 0.0}
\definecolor{purpleheart}{rgb}{0.41, 0.21, 0.61}
\definecolor{lightsalmonpink}{rgb}{1.0, 0.6, 0.6}\definecolor{arrowcolor}{rgb}{0.92, 0.92, 0.92}
\tikzset{
inner/.style={
  on chain,
  circle,
  inner sep=4pt,
  fill=circlecolor,
  line width=1.5pt,
  draw=bordercolor,
  text width=1.2em,
  align=center,
  text height=1.25ex,
  text depth=0ex
},
on grid
}

\newcommand\drawarrow{

\node[on chain] (f) {};
\begin{pgfonlayer}{background}
\node[
  inner sep=10pt,
  single arrow,
  single arrow head extend=0.6cm,
  draw=none,
  fill=arrowcolor,
  fit= (c1) (f)
] (arrow) {};
\fill[white] 
  (arrow.before tail) -- (c1|-arrow.west) -- (arrow.after tail) -- cycle;
\end{pgfonlayer}
}

\newenvironment{timeline}[1][]
  {\par\noindent\begin{tikzpicture}[start chain,#1]}
  {\drawarrow\end{tikzpicture}\par}
\numberwithin{equation}{section}

\begin{document}
\title{Catastrophe risk in a stochastic multi-population mortality model}

\author[1,3,*]{Jens Robben}
\author[1,2,3,4,5]{Katrien Antonio}
\affil[1]{Faculty of Economics and Business, KU Leuven, Belgium.}
\affil[2]{Faculty of Economics and Business, University of Amsterdam, The Netherlands.}
\affil[3]{LRisk, Leuven Research Center on Insurance and Financial Risk Analysis, KU Leuven, Belgium.}
\affil[4]{LStat, Leuven Statistics Research Center, KU Leuven, Belgium.}
\affil[5]{RCLR, Research Centre for Longevity Risk, University of Amsterdam, The Netherlands.}
\affil[*]{Corresponding author: \href{mailto:jens.robben@kuleuven.be}{jens.robben@kuleuven.be}}
\date{\vspace{-1.25cm}} 
\maketitle
\thispagestyle{empty}

\begin{abstract}
\noindent 
\added{This paper incorporates mortality shocks in the scenarios for future mortality rates produced by a stochastic multi-population mortality model. Hereto, the proposed model combines a decreasing stochastic mortality trend with a mechanism that switches between regimes of low and high volatility. During the high volatility regimes, mortality shocks occur that last from one to several years and temporarily impact the mortality rates before returning to the overall mortality trend. Furthermore, we account for the age-specific impact of these mortality shocks on mortality rates. Actuaries and risk managers can tailor this scenario generator to their specific needs, risk management objectives, or supervisory requirements. The generated scenarios allow (re)insurers, policymakers, or actuaries to evaluate the effects of different catastrophe risk scenarios on, e.g., the calculation of solvency capital requirements.}
\end{abstract}

\vspace{-0.25cm}
\vspace{-0.25cm}
\paragraph{Data and code availability statement:} The data sets used in this paper are publicly available from the Human Mortality Database (\url{https://www.mortality.org/}) and Eurostat (\url{https://ec.europa.eu/eurostat}). The code used for the implementation and analysis of the case study presented in this paper is available on the GitHub repository \url{https://github.com/RobbenJ/catastrophe-risk}. The code is written in \texttt{R} and can be accessed and downloaded for further reference and replication of the obtained results.
\vspace{-0.25cm}
\paragraph{Funding statement:} The authors acknowledge funding from the FWO and F.R.S.-FNRS under the Excellence of Science (EOS) programme, project ASTeRISK (40007517). \added{The FWO WOG network W001021N is acknowledged as well.} Katrien Antonio acknowledges support from the Chair of Excellence on Digital Insurance And Long‑term risk (DIALog) by CNP Assurances and from the RCLR, the Research Centre for Longevity Risk at the University of Amsterdam.
\vspace{-0.25cm}
\paragraph{Conflict of interest disclosure:} The authors declare no conflict of interest.

\section{Introduction} \label{section:introduction}
\begin{figure}[ht!]
\centering
\includegraphics[width=0.75\textwidth]{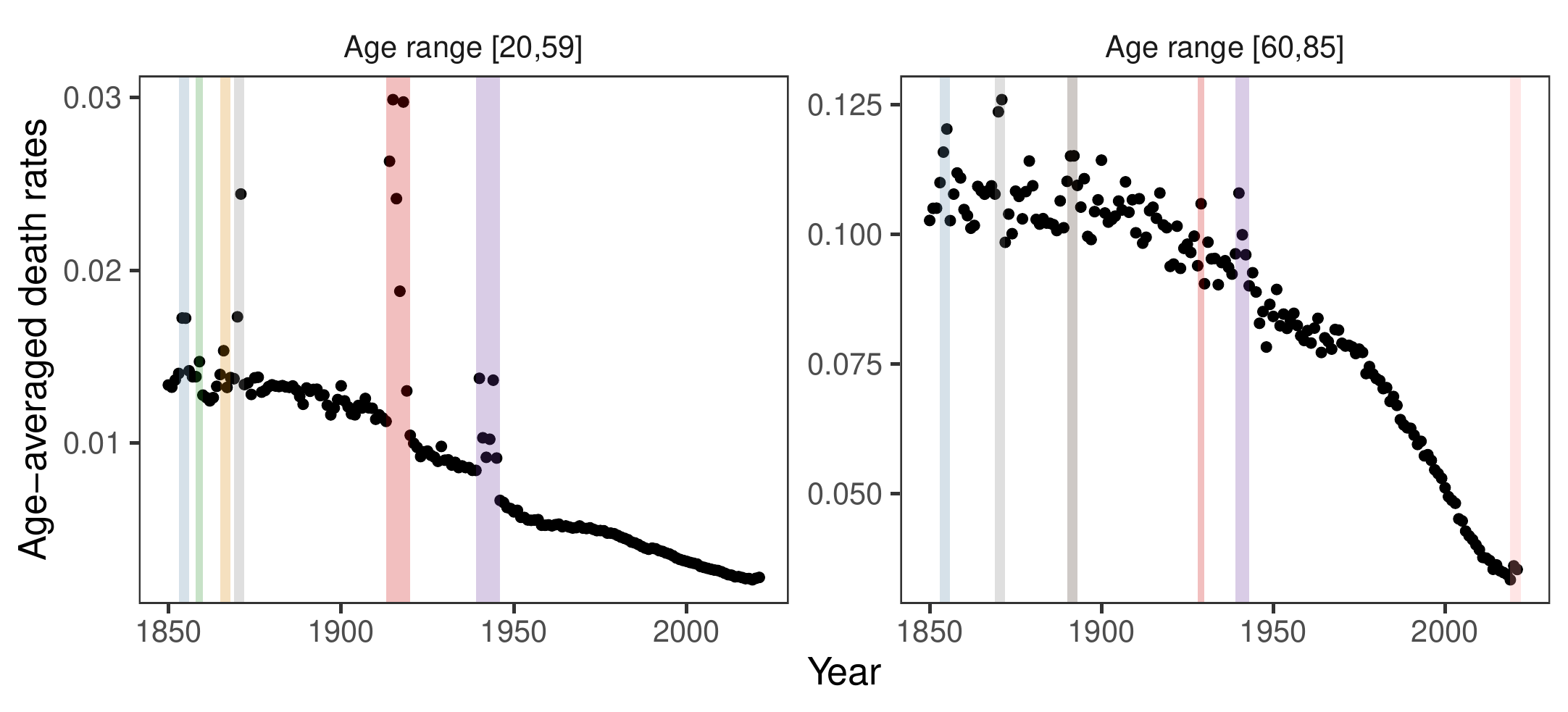}
\caption{The age-averaged death rates, defined as $1/|\mathcal{X}| \sum_{x\in \mathcal{X}} d_{x,t}^A/E_{x,t}^A$, with $\mathcal{X}$ the age range under consideration, $d_{x,t}^A$ the total deaths, and $E_{x,t}^A$ the population exposure, aggregated over a set of West and North European countries, i.e.~Austria, Belgium, Denmark, Germany, Finland, France, Ireland, Iceland, Luxembourg, the Netherlands, Norway, Sweden, Switzerland, and the United Kingdom. We average the death rates over the age range 20-59 (left panel) and 60-85 (right panel). The color scheme used for highlighting the main mortality shocks is also used in Figure~\ref{fig:intro:overview} to visualize the occurrence of the shocks over time.\label{fig:shockrates}}
\end{figure}

\begin{figure}[t]
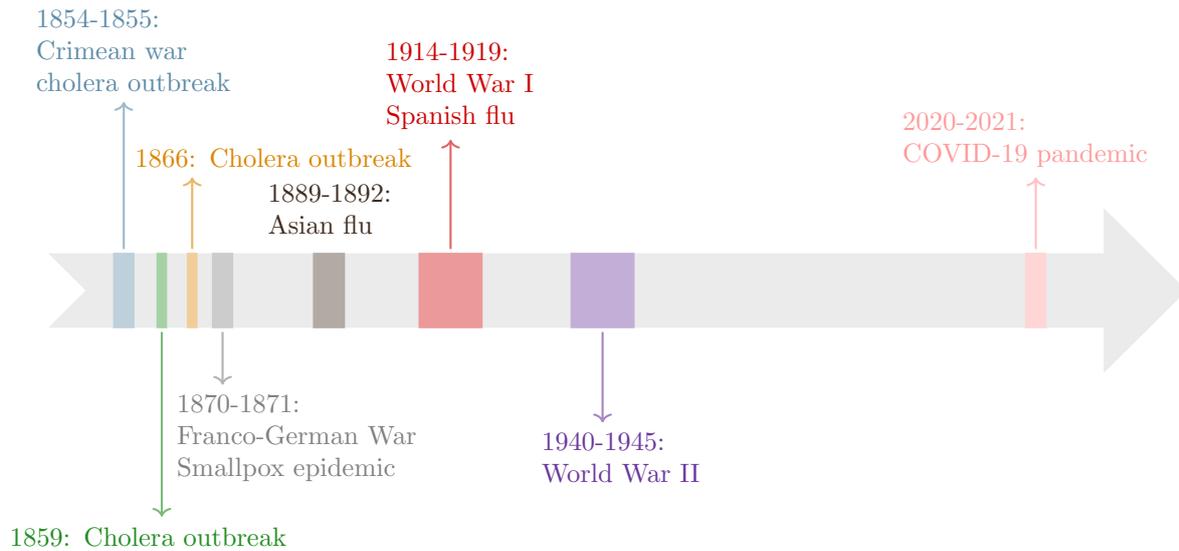
 
\begin{timeline}
  \node[on chain] (c1) {}; 
  \node[inner, fill = arrowcolor,color = arrowcolor] at (11.5,0) {};
  \draw[white!60!airforceblue, line width = 8pt] (0.5,-0.5) -- (0.5,0.5); 
  \draw[white!60!forestgreen, line width = 4pt] (1.0,-0.5) -- (1.0,0.5); 
  \draw[white!60!fulvous, line width = 4pt] (1.4,-0.5) -- (1.4,0.5); 
  \draw[white!60!gray, line width = 8pt] (1.8,-0.5) -- (1.8,0.5); 
  \draw[white!60!bistre, line width = 12pt] (3.2,-0.5) -- (3.2,0.5); 
  \draw[white!60!bostonuniversityred, line width = 24pt] (4.8,-0.5) -- (4.8,0.5); 
  \draw[white!60!purpleheart, line width = 24pt] (6.8,-0.5) -- (6.8,0.5); 
  \draw[white!60!lightsalmonpink, line width = 8pt] (12.5,-0.5) -- (12.5,0.5);
  
  \draw[->,thick,white!40!airforceblue] (0.5,0.55) -- (0.5,2.5) node[above, text width = 3cm, font = \small, xshift = 0.35cm] {\textcolor{airforceblue}{1854-1855: \mbox{Crimean war} \mbox{cholera outbreak}}};
 
  \draw[->,thick,white!40!forestgreen] (1,-0.55) -- (1,-3) node[below, text width = 4cm, font = \small] {\textcolor{forestgreen}{1859: \mbox{Cholera outbreak}}};
  
    \draw[->,thick,white!40!fulvous] (1.4,0.55) -- (1.4,1.5) node[above, text width = 4cm, xshift =1.25cm, font = \small] {\textcolor{fulvous}{1866: \mbox{Cholera outbreak}}}; 
  
  \draw[->,thick,white!40!gray] (1.8,-0.55) -- (1.8,-1.25) node[below, text width = 4cm, xshift = 1.4cm, font = \small] {\textcolor{gray}{1870-1871: \mbox{Franco-German War} \mbox{Smallpox epidemic}}}; 
  
    \draw[->,thick,white] (3.4,0.55) -- (3.4,1) node[below, text width = 2cm, xshift = 0cm, yshift = 0.55cm, font = \small] {\textcolor{bistre}{1889-1892: \mbox{Asian flu}}}; 
  
\draw[->,thick,white!40!bostonuniversityred] (4.8,0.55) -- (4.8,2) node[above, text width = 2cm, xshift =0.15cm, font = \small] {\textcolor{bostonuniversityred}{1914-1919: \mbox{World War I} \mbox{Spanish flu}}};   

\draw[->,thick,white!40!purpleheart] (6.8,-0.55) -- (6.8,-1.75) node[below, text width = 2cm, xshift = 0.2cm, font = \small] {\textcolor{purpleheart}{1940-1945: \mbox{World War II}}};   

\draw[->,thick,white!40!lightsalmonpink] (12.5,0.55) -- (12.5,1.5) node[above, text width = 2cm, xshift =-0.75cm, font = \small] {\textcolor{lightsalmonpink}{2020-2021: \mbox{COVID-19 pandemic}}}; 
\end{timeline}
\caption{Historical mortality shocks in Europe from 1850-2021.\label{fig:intro:overview}}
\end{figure}

The overall evolution of mortality rates over time is referred to as the mortality trend. Since the beginning of the 19th century, mortality rates have declined in Europe, but occasional and temporary mortality jumps or shocks have disrupted this trend \citep{lee2003demographic}. Figure~\ref{fig:shockrates} shows the age-averaged death rates by aggregating total deaths and population exposures over age ranges 20-59 and 60-85 for a set of West and North European countries. The figure highlights the main mortality shocks that happened throughout the historical period 1850-2021. The timeline in Figure~\ref{fig:intro:overview} reveals the nature of these shocks: some of the observed historical shocks are related to conflicts like World War I and World War II, while others are related to pandemics and epidemics such as a cholera outbreak, the Spanish flu, or, more recently, the COVID-19 pandemic. \added{As Figure~\ref{fig:shockrates} indicates, these shocks can be seen as sudden, upward spikes in the mortality rates, potentially spanning multiple years, followed by a subsequent return to the overall declining mortality trend.} 

Accurately modeling future mortality rates is crucial for the valuation of life-contingent risks. \added{To safeguard the solvability within the European (re)insurance industry, the solvency capital requirements (SCRs) in the Solvency II directive require the calculation of a separate SCR for mortality risk, longevity risk, and catastrophe risk.} Mortality risk refers to the risk of deterioration in mortality rates, longevity risk is the risk of people living longer than expected, and catastrophe risk captures sudden extreme events that lead to a sharp increase in mortality rates, such as a nuclear explosion or pandemic. The main objective of this paper is \added{the development of a toolbox that allows (re)insurers to generate scenarios for future mortality rates, equipped with shocks that can impact the mortality trend \added{over} several years while taking into account the age-specific impact of such a shock on mortality rates. This enables the valuation of risks in the presence of catastrophic shocks. Scenarios involving abrupt improvements, such as a sudden cure for diseases like cancer \citep{Gielen2014cancer}, are not considered in this paper.}

\added{Many insurers and reinsurers worldwide express concerns regarding the potential threats posed by catastrophic mortality events. \cite{toole2007potential} finds that under an extreme pandemic scenario, the US life insurance industry would face additional claims expenses up to $25\%$ of its risk-based capital, which could cause insolvency in the (re)insurance industry. Therefore, (re)insurers might consider, in an attempt to mitigate their exposure, investing in so-called mortality-linked securities, e.g., the Swiss Re CAT bond (2003). Pricing such mortality-linked securities requires insight into the modeling and projection of mortality rates. The inclusion of mortality shocks into mortality projection models is of significant importance. \cite{chencox} provide a case study on pricing mortality-linked securities with their proposed mortality model, including a shock component. \cite{zhou2013pricing} use an economic-pricing framework to investigate how mortality shocks can affect the supply and demand of mortality-linked securities. Moreover, \cite{liuli} provide an application for pricing hypothetical catastrophic mortality bonds.} 

The proposed mortality modeling framework consists of two building blocks. The first building block captures the overall mortality trend using a stochastic \added{multi-population} model. The second building block accounts for catastrophe risk via mortality shocks. We utilize the projection model equipped with mortality shocks to generate scenarios for future mortality rates \added{that are exposed to the occurrence of shocks}. \added{Our model introduces shocks that can last from one to several years, temporarily impacting the mortality rates before returning to the overall mortality trend. Furthermore, we account for the age-specific impact of a mortality shock on mortality rates.} \added{As such, we enable actuaries and risk managers to tailor the occurrence and impact of these shocks to their specific needs, risk appetite, or supervisory requirements.} \added{This flexibility builds on the calibration of the model, where users can make informed decisions about the inclusion or exclusion of historical mortality shocks. For instance, users may choose to disregard the mortality shocks caused by both world wars or assign more weight to recent observations. Such modifications directly influence the frequency and severity of future mortality shocks in the model projection step, empowering users to tailor the framework to their risk management objectives.}

The use of stochastic single-population mortality models originated from the seminal work of \cite{LeeCarter}, with extensions proposed by (among others) \cite{renshaw2003lee}, \cite{currie2006smoothing}, \cite{RENSHAW2006556}, \cite{cairns2009quantitative}, \cite{plat2009stochastic}, \cite{HABERMAN201135}, \cite{levantesi2019application} and \cite{dowd2020cbdx}. In addition to single population models,  incorporating mortality data from multiple populations with similar characteristics has been proposed as an effective way to enhance the stability of mortality projections. While it is generally expected that variations in mortality rates between such populations will stabilize over time, unforeseen factors or events can still lead to \added{temporary divergences.} The importance of multi-population models has been discussed in \cite{antonio2017producing} among others. A widely used multi-population model is the one proposed by \cite{LiandLee}. The Dutch Actuarial Association and the Institute of Actuaries in Belgium have both adopted a stochastic multi-population mortality model of type Li-Lee, as reported in \cite{KAG2022} and \cite{IABE2020} respectively. \cite{haberman2014longevity} discuss a variety of other multi-population mortality models as extensions of commonly used single-population models. Stochastic mortality models usually incorporate a series of age and period-specific parameters to model the logarithm of the age and period-specific force of mortality. This paper will refer to these parameters as age and period effects. Subsequently, time series models are used to extrapolate the calibrated period effects, which enables the projection of future mortality rates.

We use a mortality improvement model to examine the changes in mortality rates over time. This is in contrast to a classic mortality model that directly specifies the mortality rate or force of mortality. We motivate this set-up by previous work from \cite{mitchell2013modeling} and \cite{hunt2021mortality}. \cite{mitchell2013modeling} conduct an extensive empirical study comparing their proposed improvement model with traditional mortality models for the logarithm of the central death rate such as the Lee-Carter model \citep{LeeCarter}, (variants of) the model proposed by \cite{RENSHAW2006556} and the mortality model of \cite{plat2009stochastic}. \added{Their study covers} a comprehensive dataset spanning 100 years across 11 countries. The results of their investigation demonstrate the superiority of the improvement model over the aforementioned models using two in-sample goodness-of-fit measures, i.e.~the root sum of squared errors and the unexplained variance.\footnote{The unexplained variance at age (group) $x$ represents the proportion of variability in the observed mortality rates at age $x$ that is not accounted for by the calibrated model.} \cite{hunt2021mortality} argue that practitioners, such as life insurers and pension funds, are primarily interested in mortality improvement rates, especially when assessing longevity risk. A better-than-average mortality improvement rate that persists for several years indicates that life expectancy may exceed initial projections. As a result, reserves may have to increase to safeguard the solvability of the company. In the UK, the Continuous Mortality Investigation (CMI) introduced the concept of mortality improvement rates in 2002 \citep{continuous2002interim} and used it in their subsequent mortality projections. Furthermore, in the US, the Society of Actuaries' Retirement Plans Experience Committee \added{has published} since the year 2014 annual mortality improvement scales for projecting mortality rates.\footnote{The Mortality Improvement Scale MP-2014 can be consulted on \url{https://www.soa.org/resources/experience-studies/2014/research-2014-mp/}, where the mortality improvement model and corresponding assumptions are discussed.} This shows the interest in as well as practical relevance of mortality improvement models.

A limited body of literature explores the incorporation of mortality shocks into stochastic mortality models. \added{A first} stream of research primarily puts focus on detecting outliers in the time series for the calibrated period effects. These outliers, or mortality shocks, are considered non-repetitive, exogenous intervention effects that are not accounted for in future mortality projections. For instance, \cite{LeeCarter} utilize a random walk with drift in conjunction with an intervention term to capture the 1918 influenza epidemic. \cite{LiChan} extend this work and unravel the underlying mortality trend by identifying different types of outliers in the residuals of an ARIMA($p$,$d$,$q$) model fitted to the calibrated period effects. Their outlier detection strategy identifies different types of outliers. A second stream of literature argues that mortality shocks, including pandemic shocks, should not be seen as exogenous, one-time events unlikely to occur again in the future. Rather, they should be considered as recurrent events and explicitly incorporated into mortality projection models. For this purpose, \cite{coxlinwang} and \cite{chencox} augment the Lee-Carter framework with jump effects. The magnitude of these jump effects is modeled with independent and normally distributed random variables, while their occurrence is determined by a Bernoulli-distributed random variable. In \cite{coxlinwang}, the jump effect is considered to be permanent as the magnitude of a particular jump impacts all future mortality rates. However, \cite{chencox} argue that most mortality jumps or shocks are transitory in nature and implement the jump effect accordingly, affecting the mortality rates only during a single year. \cite{zhou2013pricing} further extend the work of \cite{chencox} to a two-population framework. Since these jump effects are included in the time series model for the period effect(s), \cite{coxlinwang}, \cite{chencox}, and \cite{zhou2013pricing} implicitly assume in their work that the age effect of a mortality shock is identical to the age effect observed in the general mortality trend of the Lee-Carter model. \cite{liuli} point out that this is a too restrictive assumption that does not align with empirical observations. Their analysis reveals that mortality shocks in the United States and England and Wales exhibit varying age patterns, deviating from the age effect calibrated with a Lee-Carter model. Furthermore, distinct mortality shocks also tend to impact different age groups, e.g., a war mainly affects younger ages, whereas the recent COVID-19 pandemic particularly affected older ages. Hence, \cite{liuli} extend the Lee-Carter modeling framework with a jump effect and dedicated age-specific parameters. \added{\cite{goes2023bayesian} introduce a Bayesian generalization of the model proposed by \cite{liuli} and integrate a vanishing jump effect on the mortality rates.} Given the scarcity of severe mortality improvements and shocks, a third stream of literature builds on the principles of extreme value theory. \cite{chencummins} and \cite{gungah} use a generalized Pareto distribution to characterize the first-order differences of the Lee-Carter's calibrated period effect above a specific threshold, while a random walk with drift is used below the threshold. Hence, these models effectively capture both extreme longevity improvements as well as mortality shocks. Furthermore, by taking random draws from the fitted generalized Pareto distribution, the model \added{accounts for} extreme mortality events \added{in the future}. The research avenues discussed so far fail in situations involving a prolonged calibration period during which mortality rates may be impacted by shocks as well as shifts in the mortality trend. In this regard, a fourth stream of research focuses on the use of regime-switching models for the calibrated period effects. \cite{mortregimepricing} employ a regime-switching model to capture regimes in the mortality trend for the US population (1901-1999). Hereto, the calibrated period effect of a Lee-Carter model is governed by a regime-switching random walk model, where the volatility of the random walk's error term varies across regimes. Similarly, \cite{HAINAUT2012236} proposes a regime-switching random walk model for the calibrated period effects of a multi-factor Lee-Carter model for the French population (1946-2007). Here, the drift of the random walk and the volatility of the error term vary across regimes. 

Our paper contributes to this literature in four ways. First, we directly incorporate the regime switch in the mortality model specifications rather than in the calibrated period effects, as is done by \cite{mortregimepricing} and \cite{HAINAUT2012236}. To the best of our knowledge, this has not been tackled yet in the current literature. In this way, we aim to construct a mortality model that switches between a low volatility regime in which no mortality shocks occur and a high volatility regime in which mortality shocks take place. In contrast to \cite{mortregimepricing} and \cite{HAINAUT2012236}, we model at the same time a dedicated set of age-specific effects that describe the impact of the mortality shocks on the mortality rates. The latter is in line with the extension of \cite{liuli} to the work of \cite{coxlinwang} and \cite{chencox}. However, we now apply this idea to regime-switching models rather than jump models to easily switch between periods of high and low volatility. We further add geometrically decaying weights in the calibration process of our mortality model to put more weight on the most recent mortality improvement rates. This enables us to capture and extrapolate the dynamic nature of mortality trends more effectively. Second, we calibrate our proposed model in a multi-population setting \added{with} a substantially long calibration period, starting from the year 1850 onwards. We use a multi-factor stochastic multi-population mortality improvement model to capture the dynamics over this lengthy calibration period. Third, we hereby derive an invariant set of parameter transformations in a two-factor mortality improvement model such that the transformed parameters satisfy the imposed identifiability constraints, inspired by \cite{hunt_blake_2020}, and lead to the same model fit. Fourth, in a case study, we shed light on the calculation of the SCR \added{in the Solvency II directive} for longevity, mortality, and catastrophe risk with both the standard model as well as our proposed mortality model equipped with mortality shocks. 

This paper is organized as follows: Section~\ref{sec:notation} introduces the notation. Section~\ref{sec:modelspecification} presents the proposed mortality improvement model, which is composed of a baseline mortality improvement model and a regime-switching model with age-specific impact. We explain the methodology for calibrating the baseline model in Section~\ref{sec:constr.baseline}, while Section~\ref{sec:cal.regime} covers the calibration of the regime-switching model. Section~\ref{sec:timedyn} explains how projections can be made with the calibrated mortality improvement model. In the case study discussed in Section~\ref{sec:case}, we model and project Dutch mortality rates based on the proposed multi-population mortality model equipped with mortality shocks. Furthermore, we compare the SCR calculated using our proposed mortality improvement model with the SCR obtained with the standard model.

\section{Notations} \label{sec:notation}
Let $d_{x,t}^{(c)}$ denote the observed number of deaths in a country $c$ at age $x$ in year $t$.\footnote{The data on deaths can be categorized as male, female, or total (unisex) deaths. Although our case study in Section~\ref{sec:case} concentrates solely on male data, it is important to note that the methodology is also applicable to both female and unisex data. Therefore, we will omit the dependence on gender in all notations used throughout the paper.} We consider a set of possible ages $x \in \mathcal{X}$ and a range of years $t \in \mathcal{T}$. The observed exposure in country $c$, denoted by $E_{x,t}^{(c)}$, reflects the total amount of person-years lived by individuals aged $[x,x+1)$ through year $[t,t+1)$. We denote the country-specific force of mortality by $\mu_{x,t}^{(c)}$ and the central death rate by $m_{x,t}^{(c)}$. The empirical estimate of the central death rate, also called the crude central death rate, at age $x$ in year $t$ is defined as:
$$ \hat{m}_{x,t}^{(c)} = \dfrac{d_{x,t}^{(c)}}{E_{x,t}^{(c)}}.$$
Furthermore, we define $q_{x,t}^{(c)}$ as the mortality rate at exact age $x$ in year $t$, i.e.~the probability that an individual from country $c$, born on January 1 of year $t-x$, who is alive at January 1 of year $t$ will die within the next year. In this paper, we assume that the number of deaths random variable $D_{x,t}^{(c)}$ follows a Poisson distribution \citep{BrouhnsDenuit}:
\begin{align} \label{eq:PoisDist}
D_{x,t}^{(c)} \sim \text{Pois}\left(E_{x,t}^{(c)} \cdot \mu_{x,t}^{(c)}\right).
\end{align} 
The force of mortality is assumed \added{to be} constant over each age $x$ and year $t$, i.e.~$\mu_{x+s,t+s}^{(c)} = \mu_{x,s}^{(c)}$ for $s \in [0,1)$. Under these assumptions, the maximum likelihood estimate of the force of mortality coincides with the crude central death rate, i.e.~$ \hat{\mu}_{x,t}^{(c)} = \hat{m}_{x,t}^{(c)}$ \citep{pitacco}. Additionally, the mortality rates can then be approximated by:
\begin{align} \label{eq:qxt.mort.rate}
\hat{q}_{x,t}^{(c)} = 1 - \exp\left(- \hat{\mu}_{x,t}^{(c)}\right).
\end{align} 

Single population stochastic mortality models typically structure the country-specific force of mortality $\mu_{x,t}^{(c)}$ using a combination of age-specific effects $\beta_x^{(j,c)}$, period-specific effects $\kappa_t^{(j,c)}$ and, occasionally, cohort-specific effects $\gamma_{t-x}^{(j,c)}$, for $j\in\{1,...,l\}$, as:
\begin{align} \label{eq:singlepopmort}
\log \mu_{x,t}^{(c)} = \beta_{x}^{(1,c)} \kappa_{t}^{(1,c)} \gamma_{t-x}^{(1,c)} + \ldots + \beta_{x}^{(l,c)} \kappa_{t}^{(l,c)} \gamma_{t-x}^{(l,c)},
\end{align}
where, potentially, some age, period, or cohort effects are fixed to zero or one. A common approach for estimating the involved parameters is to maximize the Poisson log-likelihood derived from Equation~\eqref{eq:PoisDist}. A stochastic mortality model then typically specifies and calibrates time series models to project the estimated period and cohort-specific effects. 

Multi-population stochastic mortality models are widely used to examine mortality patterns across populations that share similar characteristics, offering enhanced stability in mortality projections. These models structure the country-specific force of mortality by combining a set of country-specific age, period, and cohort effects, as in Equation~\eqref{eq:singlepopmort}, with a set of common age, period, and cohort effects. The latter structure a common, multi-population mortality trend that is shared by all the populations under consideration, whereas the country-specific parameters capture deviations from this common trend for the particular country of interest. 

Recently, more attention has been given to stochastic mortality improvement models. Such improvement models structure the (force of) mortality improvement rates, defined as:
\begin{align} \label{eq:mortimprvrates}
\log \left(\frac{\mu_{x,t}^{(c)}}{\mu_{x,t-1}^{(c)}} \right) = \log \mu_{x,t}^{(c)} - \log \mu_{x,t-1}^{(c)}.
\end{align}
A positive value for the mortality improvement rate at year $t$ indicates a mortality deterioration relative to the preceding year. Conversely, a negative value indicates an improvement in mortality compared to year $t-1$. Mortality improvement models therefore aim to capture the variability in the rate of change in mortality over time, as opposed to the level of mortality at a specific time point.

\section{A stochastic multi-population mortality improvement model with a shock regime} \label{sec:modelspecification}
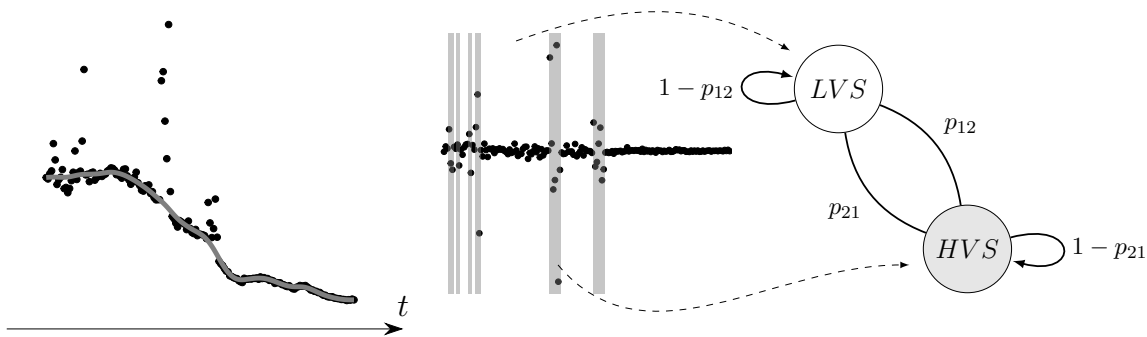
\begin{figure}[ht!]
\centering
\begin{adjustbox}{width=0.95\textwidth}
\begin{tikzpicture}[node distance=2cm,>=latex,auto,every edge/.append style={thick}]

\node[inner sep=0pt] (baseline) at (0,0)
    {\includegraphics[width=.35\textwidth]{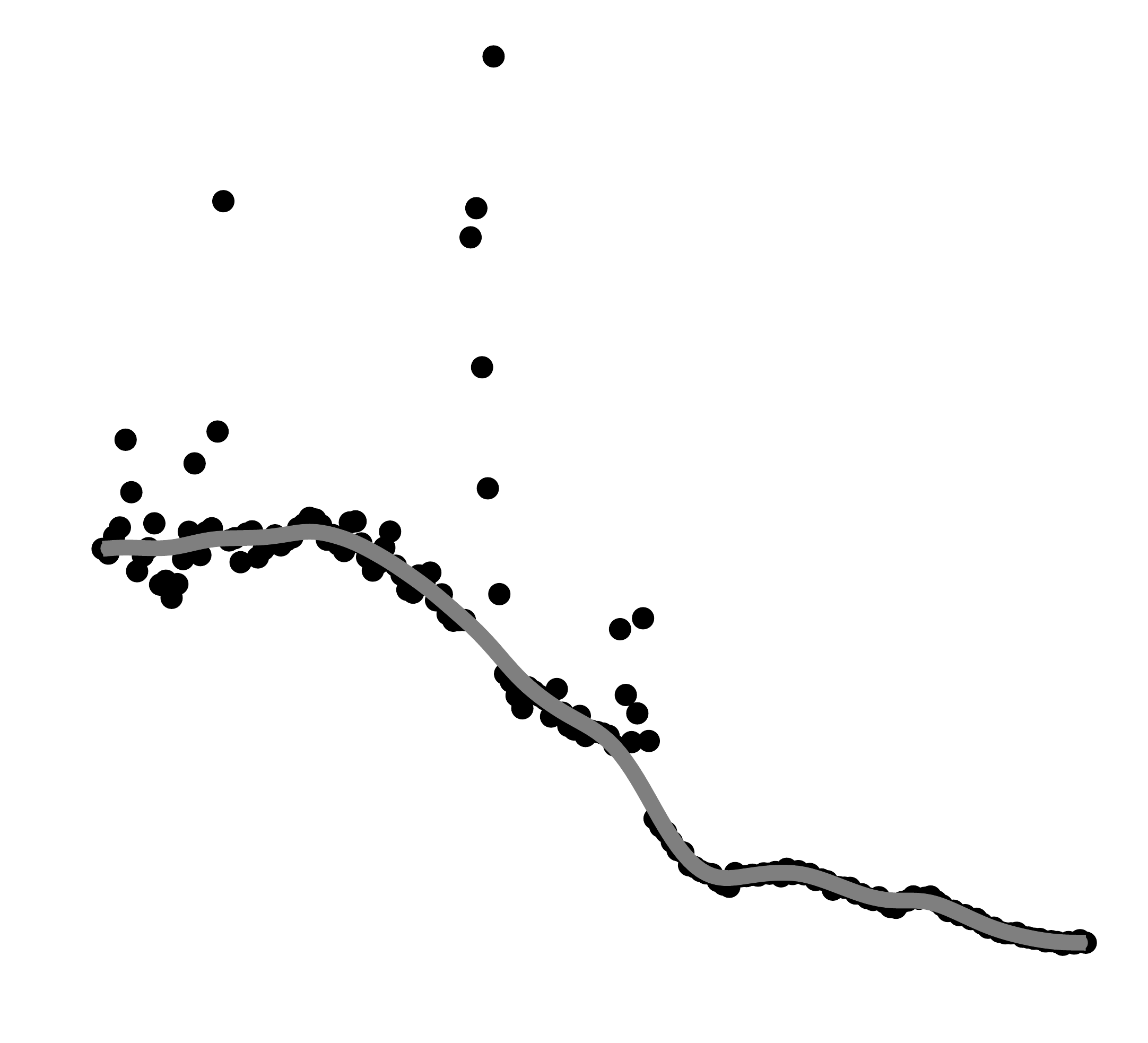}};

\node[inner sep=0pt] (markov) at (6,0)
    {\includegraphics[width=.30\textwidth]{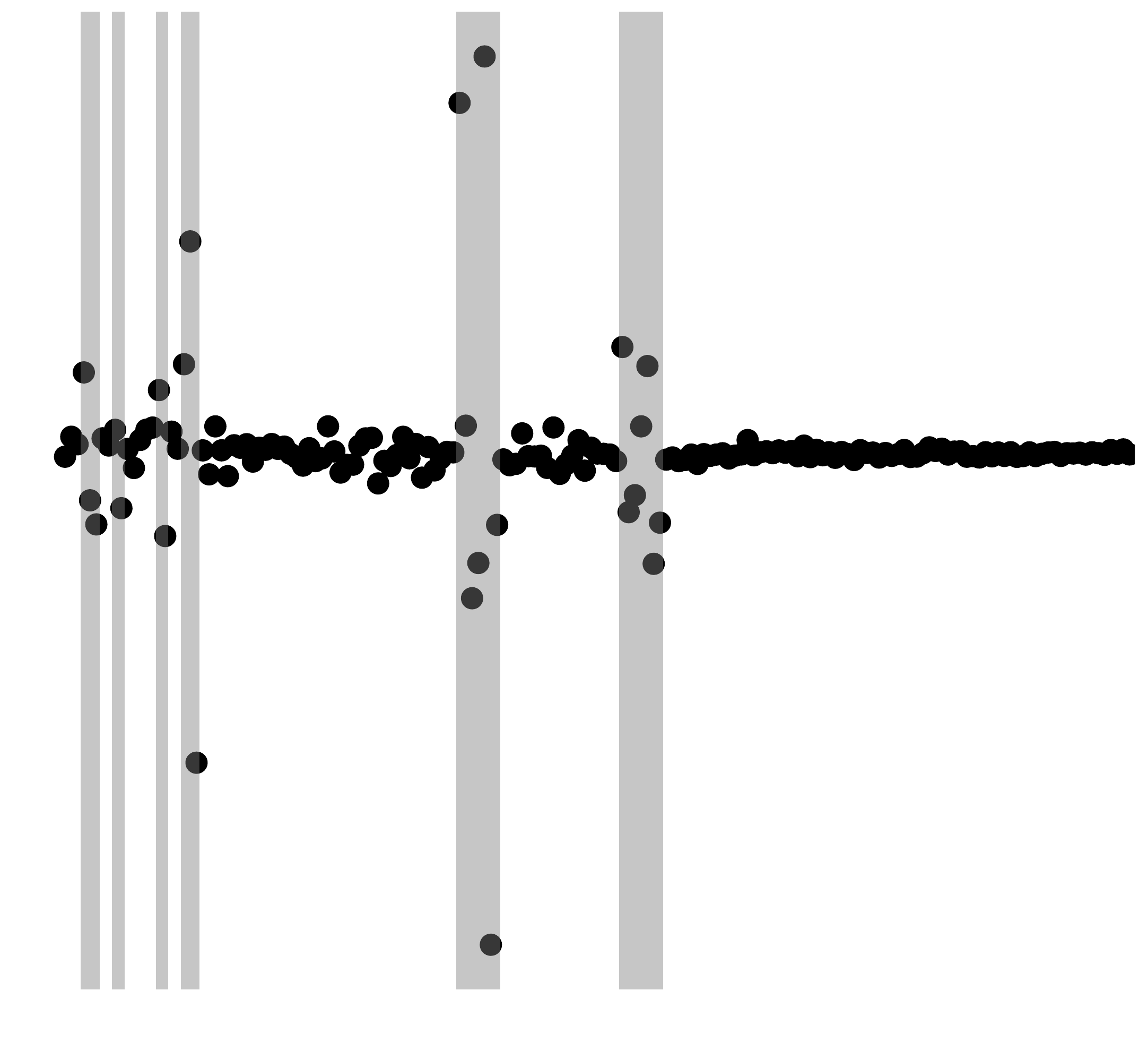}};
        


\node[state,text width = 1cm,align=center] (1) at (10,1.25) {\large $LVS$};
\node[state,text width = 1cm,align=center,fill = gray!20] (2) at (12,-1.25) {\large $HVS$};
\path (1) edge[loop left]  node {$1-p_{12}$} (1)
          edge[bend left]  node{$p_{12}$}   (2)
      (2) edge[loop right] node{$1-p_{21}$}  (2)
          edge[bend left] node{$p_{21}$}     (1);
          
\node[draw=white] at (3.25,-2.15) {{\Large $t$}};
\draw [->,black,dashed] (5,2) to[out=20,in=150] (9.25,1.85);
\draw [->,black,dashed] (5.65,-1.5) to[out=-50,in=180] (11.1,-1.5);
\draw [-{Stealth[length=3mm, width=2mm]},black] (-2.9,-2.5) to (3.25,-2.5); 
\end{tikzpicture}
\end{adjustbox}
\caption{Visual representation of the proposed mortality improvement model, consisting of a baseline mortality improvement model (left) and a regime-switching model (right). The baseline improvement model captures the country-specific mortality decline. The regime-switching model captures the country-specific mortality shocks and is constructed on the residuals of the estimated baseline model. In this way, the mortality improvement rates alternate between a low volatility state (LVS), \added{which} tracks the country-specific mortality decline, and a high volatility state (HVS), which captures the occurrence of mortality shocks. \label{tikz:model}}
\end{figure}

The proposed mortality model comprises a baseline mortality improvement model and a regime-switching model, as pictured in Figure~\ref{tikz:model}. In the first stage, we design and estimate the baseline mortality improvement model to capture the country-specific overall mortality decline, which is illustrated on the left-hand side of the figure. The residuals obtained from this estimated baseline improvement model are vulnerable to periods of high and low volatility, as evidenced on the right-hand side of Figure~\ref{tikz:model}. The high volatility periods arise due to country-specific mortality shocks caused by war or pandemic events. Therefore, in the second stage, we use a regime-switching model to capture such shocks present in the residuals. Hereto, an underlying Markov chain alternates between a low volatility state (LVS) and a high volatility state (HVS). The LVS mode tracks the country-specific mortality decline, while the HVS mode accounts for the shocks. 

\subsection{The baseline mortality improvement model} \label{sec:baselinemodel}
\paragraph{Model specification.} We introduce our mortality improvement model as an extension of a multi-factor multi-population model \citep{renshaw2003lee}:
\begin{align} \label{eq:modelspec:baseline}
\log \mu_{x,t}^{(c)} - \log \mu_{x,t-1}^{(c)} = 
A_x + \displaystyle \sum_{i=1}^m B_x^{(i)} K_t^{(i)} + 			\displaystyle \sum_{j=1}^l \beta_x^{(j,c)} \kappa_t^{(j,c)}.
\end{align}
We use a set of $m$ age/period effects to describe a common, multi-population mortality improvement trend. Subsequently, a set of $l$ age/period effects captures country-specific deviations from this common improvement trend. Depending on the area of interest, other types of mortality models can be chosen for the baseline mortality improvement trend. For instance, when the focus is only on modeling the mortality for the pensioner age range, a valid option would be to use a mortality model from the CBDX family \citep{dowd2020cbdx}. 
\paragraph{Identifiability constraints.} To ensure model identifiability, we need to impose parameter constraints in the model specifications of Equation~\eqref{eq:modelspec:baseline}. Specifically, we first impose straightforward extensions of the identifiability constraints applied in a classic Lee-Carter model:\footnote{Since we do not include a country-specific age-dependent intercept, we do not impose a location constraint on the $\kappa_t$-parameter.}
\begin{equation}
\begin{aligned} \label{eq:constr1}
\displaystyle \sum_{x\:\in\: \mathcal{X}} &\left(B_x^{(i)}\right)^2 = 1, \hspace{0.5cm} \sum_{t \:\in \:\mathcal{T}} K_t^{(i)} = 0, \hspace{1cm} &\text{for} \: i \in \{1,\ldots,m\} \\
&\hspace{1cm} \displaystyle \sum_{x\:\in \:\mathcal{X}} \left(\beta_x^{(j,c)}\right)^2 = 1,  &\text{for} \: j \in \{1,\ldots,l\}\:.
\end{aligned}
\end{equation}
However, as we show in Appendix~\ref{secA:constraintsLC2}, these constraints are not sufficient to make the mortality improvement model fully identifiable. To address this issue, we rely on additional orthogonality constraints proposed by \cite{hunt_blake_2020}. Specifically, for $m = 2$, we impose:
\begin{align} \label{eq:constr2}
\displaystyle \sum_{x\:\in\:\mathcal{X}} B_x^{(1)} B_x^{(2)} = 0, \hspace{0.5cm}\displaystyle \sum_{t\:\in\:\mathcal{T}} K_t^{(1)}  K_t^{(2)} = 0.
\end{align}
We also impose similar constraints on the parameters that model the country-specific deviation from the common mortality improvement rates. Appendix~\ref{secA:constraintsLC2} outlines the implementation details of the constraints in Equations~\eqref{eq:constr1} and~\eqref{eq:constr2} in the Newton-Raphson algorithm that is used to fit the baseline mortality improvement model.
\vspace{-0.4cm}
\paragraph{Equivalent mortality model specification.} We translate the baseline mortality improvement model into a baseline mortality model for the logarithm of the force of mortality $\mu_{x,t}^{(c)}$. Appendix~\ref{appendix:indirect.estimation} shows that the model specification in Equation~\eqref{eq:modelspec:baseline} is equivalent to: 
\begin{equation} \label{eq:modelspec:tf.baseline}
\log \mu_{x,t}^{(c)} = \log \mu_{x,t_{\min}}^{(c)}  + \left(t - t_{\min}\right) A_x + \displaystyle \sum_{i=1}^m B_x^{(i)} L_t^{(i)} + \displaystyle \sum_{j=1}^l \beta_x^{(j,c)} \lambda_t^{(j,c)},
\end{equation}
with $L_{t_{\min}}^{(i)} = 0$ for all $i=1,...,m$, $\lambda_{t_{\min}}^{(j,c)} = 0$ for all $j=1,...,l$, and
\begin{equation} \label{eq:definition.period}
L_t^{(i)} = \displaystyle \sum_{\tau = t_{\min}+1}^{t} K_{\tau}^{(i)}, \hspace{0.5cm} \lambda_t^{(j,c)} = \displaystyle \sum_{\tau = t_{\min}+1}^{t} \kappa_{\tau}^{(j,c)},
\end{equation}
for $t > t_{\min}$. In this expression, $\mu_{x,t_{\min}}^{(c)}$ is the force of mortality at age $x$ in the initial year of the calibration period, typically estimated by the crude central death rate $\hat{m}_{x,t_{\min}}^{(c)} = d_{x,t_{\min}}^{(c)}/E_{x,t_{\min}}^{(c)}$. Appendix~\ref{appendix:indirect.estimation} discusses the corresponding identifiability constraints of the model specified for the (log-transformed) force of mortality in Equation~\eqref{eq:modelspec:tf.baseline}.
\vspace{-0.4cm}
\paragraph{Distributional assumptions.} \cite{hunt2021mortality} present two methods for calibrating a mortality improvement model under a Poisson assumption for the number of deaths. As suggested by \cite{hunt2021mortality}, we follow the indirect estimation procedure for stability and reliability reasons. This indirect approach relies on the model specification in Equation~\eqref{eq:modelspec:tf.baseline}. We assume:
\begin{align}\label{eq:PoisAssumption}
D_{x,t}^{(c)} \sim \text{Pois}\left(E_{x,t}^{(c)} \cdot \mu_{x,t_{\min}}^{(c)} \cdot \exp\left(\left(t - t_{\min}\right) A_x + \displaystyle \sum_{i=1}^m B_x^{(i)} L_t^{(i)} + \displaystyle \sum_{j=1}^l \beta_x^{(j,c)} \lambda_t^{(j,c)}\right) \right).
\end{align}
We estimate the involved parameters using maximum likelihood (ML) estimation. In Section~\ref{sec:constr.baseline}, we further outline the calibration strategy using the distributional assumption in Equation~\eqref{eq:PoisAssumption}.

\subsection{The regime-switching model with age-specific effect} \label{sec:regimeswitch}
\paragraph{Model specification.} To construct the baseline model's residuals, we replace the mortality improvement rates in Equation~\eqref{eq:mortimprvrates} with their empirical counterpart and subtract the fitted multi-population model, as specified in Equation~\eqref{eq:modelspec:baseline}:
\begin{align*}
z_{x,t}^{(c)} := \log \hat{m}_{x,t}^{(c)} - \log \hat{m}_{x,t-1}^{(c)} - \left(\hat{A}_x + \displaystyle \sum_{i=1}^m \hat{B}_x^{(i)} \hat{K}_t^{(i)} + \displaystyle \sum_{j=1}^l \hat{\beta}_x^{(j,c)} \hat{\kappa}_t^{(j,c)}\right),
\end{align*}
with $t \in \mathcal{T}\backslash \left\{t_{\min}\right\}$, $x\in\mathcal{X}$. We assume that the $z_{x,t}^{(c)}$ are realizations of a random variable $Z_{x,t}^{(c)}$, defined as:
\begin{align} \label{eq:spec.ztx}
Z_{x,t}^{(c)} := \mathfrak{B}_x^{(c)} Y_t^{(c)} + \epsilon_{x,t}^{(c)},
\end{align}
where the value of $Y_t^{(c)}$ depends on the state of an underlying Markov chain ${\rho_t^{(c)}}$. At each time point $t$, $\rho_t^{(c)}$ can take one of two possible values, $\text{LVS}$ (low volatility state) or $\text{HVS}$ (high volatility state), which determines the value of $Y_t^{(c)}$. In the LVS, $Y_t^{(c)}$ is equal to zero, such that the mortality dynamics are described by the baseline mortality improvement model. However, in the HVS, $Y_t^{(c)}$ follows a normal distribution with mean $\mu_H$ and variance $\sigma_H^2$, allowing for normally distributed mortality shocks that affect the mortality improvement trend at the country-specific level. We define:
\begin{align} \label{eq:Ytc.spec}
Y_t^{(c)} \sim \begin{cases} 0 & \hspace{1cm} \text{if} \: \rho_t^{(c)} = \text{LVS} \\
\mathcal{N}(\mu_H,\sigma_H^2) & \hspace{1cm} \text{if} \: \rho_t^{(c)} = \text{HVS}.
\end{cases}
\end{align}
Empirical evidence suggests that the impact of mortality shocks on mortality (improvement) rates is not uniform across all age groups, as we witnessed e.g., throughout the COVID-19 pandemic. Therefore, we model the age-specific impact of the mortality shocks with a separate age effect $\mathfrak{B}_x^{(c)}$. The country-specific residuals $\epsilon_{x,t}^{(c)}$ represent the age- and time-specific variations in the mortality improvement rates that are not captured by the baseline and neither by the regime-switching model. We assume that these residuals $\epsilon_{x,t}^{(c)}$ are independent and normally distributed with mean $0$ and standard deviation $\sigma_e(x,t)$.\footnote{We allow $\sigma_e(x,t)$ to be dependent on age $x$ and time $t$. This assumption may be relaxed to a constant standard deviation.} In addition, we assume independence between $Y_t^{(c)}$ and $\epsilon_{x,t}^{(c)}$ for all $x \in \mathcal{X}$ and $t\in\mathcal{T}$.

The distribution of $Z_{x,t}^{(c)}$ is regime-switching and alternates as follows:
\begin{align} \label{eq:regimeswithcomp+}
Z_{x,t}^{(c)} \sim \begin{cases}
\mathcal{N}\left(0, \sigma_{e}^2(x,t)\right) & \hspace{1cm} \text{if} \: \rho_t^{(c)} = \text{LVS} \\
\mathcal{N}\left(\mathfrak{B}_x^{(c)} \mu_H , \left(\mathfrak{B}_x^{(c)}\right)^2 \sigma_H^2 + \sigma^2_e(x,t)\right) & \hspace{1cm} \text{if} \: \rho_t^{(c)} = \text{HVS}. \end{cases}
\end{align}

\paragraph{Identifiability constraints.} To make the model specifications of the regime-switching model identifiable, we additionally require that:\footnote{Multiplying $\mathfrak{B}_x^{(c)}$ by a constant factor $C$ and dividing both the mean and standard deviation of $Y_t^{(c)}$ by the same constant $C$ yields an equivalent model specification as described in Equation~\eqref{eq:spec.ztx}. To ensure model identifiability, we therefore use the same type of constraint as the one imposed on the age effects in Equation~\eqref{eq:constr1}.}
$$ \displaystyle \sum_{x\in \mathcal{X}} \left(\mathfrak{B}_x^{(c)}\right)^2 = 1.$$

\section{Calibration strategy}
\subsection{Calibrating the baseline mortality improvement model} \label{sec:constr.baseline}
In Section~\ref{sec:cal.trad.mort.model}, we first calibrate the baseline mortality improvement model on the complete calibration period $\mathcal{T}$. Then, Section~\ref{sec:detectoutlier} performs an outlier detection strategy on the calibrated common period effects. This allows us to identify the years in which mortality shocks took place. We remove the observations corresponding to these years and recalibrate the baseline mortality model. As such, we obtain the baseline improvement model that captures the overall country-specific mortality decline.

\subsubsection{Calibrating the baseline mortality model on the complete calibration period} \label{sec:cal.trad.mort.model} 
We calibrate the model in Equation~\eqref{eq:PoisAssumption} subject to the identifiability constraints in Equations~\eqref{eq:constr1.1} and~\eqref{eq:constr2.1} through a two-step procedure. The first step in this procedure estimates the parameters in the common trend, while the second step puts focus on the parameters in the country-specific deviation from that trend.

\begin{enumerate}
\item In the first step, we aggregate the number of deaths and exposures by age $x \in \mathcal{X}$ and year $t \in \mathcal{T}$ across the set $\mathcal{C}$ of countries that constitute the common, multi-population trend. We denote the aggregated age and period-specific deaths and exposures as:
\begin{align*}
D_{x,t}^{(A)} = \displaystyle \sum_{c \:\in \: \mathcal{C}} D_{x,t}^{(c)}, \hspace{0.5cm} E_{x,t}^{(A)} = \displaystyle \sum_{c \:\in\: \mathcal{C}} E_{x,t}^{(c)}.
\end{align*}
We assume that the aggregated deaths random variable $D_{x,t}^{(A)}$ follows a Poisson distribution with mean $E_{x,t}^{(A)} \cdot \exp(\eta_{x,t}^{(A)})$, with 
\begin{align*}
\eta_{x,t}^{(A)} := \log \mu_{x,t_{\min}}^{(A)} + \left(t-t_{\min}\right) A_x + \displaystyle \sum_{i=1}^m B_x^{(i)} L_t^{(i)},
\end{align*}
where $\mu_{x,t_{\min}}^{(A)}$ is replaced by its empirical counterpart $\hat{m}_{x,t_{\min}}^{(A)}:=d_{x,t_{\min}}^{(A)}/E_{x,t_{\min}}^{(A)}$, i.e.~the observed aggregated common death rate at age $x$ in the initial year of the calibration period $t_{\min}$. We estimate the common trend parameters by maximizing the following Poisson log-likelihood:
\begin{align*}
\displaystyle \max_{A_x, \{B_x^{(i)}\}_i, \{L_t^{(i)}\}_i} \: \displaystyle \sum_{t \in \mathcal{T}} \sum_{x\in \mathcal{X}} \left(d_{x,t}^{(A)} \eta_{x,t}^{(A)} - E_{x,t}^{(A)} \cdot \exp\left(\eta_{x,t}^{(A)} \right)\right),
\end{align*}
where we omit the terms in the log-likelihood that do not depend on the parameters of interest. We maximize this log-likelihood using the Newton-Raphson algorithm subject to the constraints on the common age and period effects, as given in Equations~\eqref{eq:constr1.1} and~\eqref{eq:constr2.1}. 

\item  In the second step, we assume that the number of deaths random variable for the country of interest $c$, i.e.~$D_{x,t}^{(c)}$, follows a Poisson distribution with mean $E_{x,t}^{(c)} \cdot \exp(\eta_{x,t}^{(c)})$. Since we already calibrated the parameters in the common part of the mortality model, we write:
\begin{align} \label{eq:abcdfghij}
\eta_{x,t}^{(c)} = \hat{\eta}_{x,t}^{(A)} + \eta_{x,t}^{(\text{dev})},
\end{align}
with 
\begin{align} \label{eq:abcdfghij}
\eta_{x,t}^{(\text{dev})} := \log \left(\frac{\mu_{x,t_{\min}}^{(c)}}{\mu_{x,t_{\min}}^{(A)}}\right) + \displaystyle \sum_{j=1}^l \beta_x^{(j,c)} \lambda_t^{(j,c)},
\end{align}
where $\mu_{x,t_{\min}}^{(c)}$ and $\mu_{x,t_{\min}}^{(A)}$ are replaced by their empirical counterparts. We then estimate the country-specific parameters by maximizing the following Poisson log-likelihood:
\begin{align*}
\displaystyle \max_{\{\beta_x^{(j,c)}\}_j, \{\lambda_t^{(j,c)}\}_j} \:  \displaystyle \sum_{t \in \mathcal{T}} \sum_{x\in \mathcal{X}} \left(d_{x,t}^{(c)} \eta_{x,t}^{(c)} - E_{x,t}^{(c)} \cdot \exp\left(\eta_{x,t}^{(c)}\right)\right),
\end{align*}
subject to the constraints on the country-specific age and period effects in Equations~\eqref{eq:constr1.1} and~\eqref{eq:constr2.1}. 
\end{enumerate}

\subsubsection{Towards an outlier-free baseline mortality improvement model} \label{sec:detectoutlier}
The quality of the fit of the baseline mortality improvement model may be influenced significantly by the presence of mortality shocks in the mortality data. Hereto, we investigate the presence of outliers in the calibrated common period effects $\hat{L}_t^{(i)}$, where $i \in \{1,\ldots,m\}$. Our focus on outlier detection in the common trend parameters only is motivated in the case study in Section~\ref{sec:case}.

Consider a calibrated period effect $\hat{L}_t^{(i)}$ and denote it as $L_t^{(i)}$ for notational convenience. We decompose the time series $L_t^{(i)}$ as \citep{timeseriesdecomp}:
$$ L_t^{(i)} = T_t^{(i)} + R_t^{(i)}, $$
for $i \in \{1,2,...,m\}$, where $T_t^{(i)}$ is a trend component and $R_t^{(i)}$ a so-called remainder component.\footnote{\cite{timeseriesdecomp} also consider a seasonal component. Since we work with annual data, we ignore the seasonal component.} After estimating the trend component, the remainder component equals:
$$ \hat{R}_t^{(i)} = L_t^{(i)} - \hat{T}^{(i)}_t.$$
Using an $m$-dimensional outlier detection technique, we determine the outliers in the set of remainder components: 
$$ \mathcal{R} = \left\{\hat{\boldsymbol{R}}_t := \left(\hat{R}_t^{(1)}, \hat{R}_t^{(2)}, \ldots, \hat{R}_t^{(m)}\right) \mid t \in \mathcal{T}\right\}.$$
Appendix~\ref{appendix:outlierdetection} explains our proposed strategy to detect the outlying years in $\mathcal{R}$ using the Mahalanobis distance. We denote the set of years that are identified as outlying with $\mathcal{T}^o$.

Next, we recalibrate the baseline mortality model in Equation~\eqref{eq:PoisAssumption} on the outlier-free calibration period $\widetilde{\mathcal{T}} = \mathcal{T} \backslash \left\{\mathcal{T}^o\right\}$. As a result, the recalibrated period effects $\hat{L}_t^{(i)}$ will be missing for the years detected as outlying. We account for this using missing value imputation.\footnote{We explain the details of the missing value imputation technique in Section~\ref{sec:casestudy:recal} of the case study.} We denote the final recalibrated parameters in our baseline mortality model as $\hat{A}_x$, $\hat{B}_x^{(1)}$, ..., $\hat{B}_x^{(m)}$, $\hat{L}_t^{(1)}$, ..., $\hat{L}_t^{(m)}$, $\hat{\beta}_x^{(1,c)}$, ..., $\hat{\beta}_x^{(l,c)}$, $\hat{\lambda}_t^{(1,c)}$, ..., $\hat{\lambda}_t^{(l,c)}$.  

After the recalibration of the age and period effects in the baseline mortality model, we recover the period effects of the original mortality improvement model via back-transformation starting from Equation~\eqref{eq:definition.period}:
\begin{align} \label{eq:recov.orig.mortmodel}
\hat{K}_t^{(i)} = \hat{L}_t^{(i)} - \hat{L}_{t-1}^{(i)}, \hspace{0.5cm} \hat{\kappa}_t^{(j,c)} = \hat{\lambda}_t^{(j,c)} - \lambda_{t-1}^{(j,c)},
\end{align}
for $t \in \mathcal{T}$, $i \in \{1,..,m\}$ and $j \in \{1,...,l\}$. The age effects in the original model specifications remain unchanged. Note that, by construction, the identifiability constraints in Equations~\eqref{eq:constr1.1} and~\eqref{eq:constr2.1} are fulfilled. In this way, a baseline mortality improvement model that captures the overall country-specific mortality decline results. We denote the final calibrated baseline improvement model as: 
\begin{align} \label{eq:cal.baseline.improvement}
\log \hat{\mu}_{x,t}^{(c)} - \log \hat{\mu}_{x,t-1}^{(c)} = \hat{A}_x + \displaystyle \sum_{i=1}^m \hat{B}_x^{(i)} \hat{K}_t^{(i)} + \displaystyle \sum_{j=1}^l \hat{\beta}_x^{(j,c)} \hat{\kappa}_t^{(j,c)}.
\end{align}
The trend line in Figure~\ref{fig:shockrates} visualizes the decrement in mortality rates attributed to the fitted baseline improvement model in Equation~\eqref{eq:cal.baseline.improvement}.

\subsection{Calibrating the regime-switching model for age-specific mortality shocks}  \label{sec:cal.regime}
We now explain the calibration strategy for the regime-switching model that drives the mortality shocks and their age-specific effects, see Section~\ref{sec:regimeswitch}. In the low volatility regime, the dynamics of mortality are modeled by the baseline mortality improvement model, which was calibrated in Section~\ref{sec:constr.baseline}. In the high volatility regime, the mortality trend experiences disruptions through normally distributed and age-specific mortality shocks.

Starting from the baseline model's residuals $Z_{x,t}$, as defined in Equation~\eqref{eq:spec.ztx}, we switch to vector notation: 
\begin{align*}
\boldsymbol{Z}_t^{(c)} = \boldsymbol{\mathfrak{B}}^{(c)} Y_t^{(c)} + \boldsymbol{E}_t^{(c)},
\end{align*}
where $\boldsymbol{Z}_t^{(c)} := (Z_{x,t}^{(c)})_{x\in\mathcal{X}}$. We denote $\boldsymbol{\mathfrak{B}}^{(c)} := (\mathfrak{B}_x^{(c)})_{x\in \mathcal{X}}$ for the vector of the mortality shocks' age-specific effects and $\boldsymbol{E}_t^{(c)} = (\epsilon_{x,t}^{(c)})_{x\in\mathcal{X}}$ for the vector of error terms. We obtain:\footnote{$|\mathcal{X}|$ denotes the number of ages in the age range $\mathcal{X}$.}
\begin{equation}\label{eq:cal:transdens}
\begin{aligned} 
\boldsymbol{Z}_t^{(c)} &\sim 
\begin{cases}
\mathcal{N}_{|\mathcal{X}|}\left(\boldsymbol{0}, \sigma_e^2(x,t) \boldsymbol{I}_{|\mathcal{X}|}\right) \hspace{1cm} &\text{if} \: \rho_t^{(c)} = \text{LVS} \\
\mathcal{N}_{|\mathcal{X}|}\left(\boldsymbol{\mathfrak{B}}^{(c)} \mu_H,  \boldsymbol{\mathfrak{B}}^{(c)} \left(\boldsymbol{\mathfrak{B}}^{(c)}\right)^T \sigma_H^2 + \sigma_e^2(x,t) \boldsymbol{I}_{|\mathcal{X}|}\right) \hspace{1cm} &\text{if} \: \rho_t^{(c)} = \text{HVS},
\end{cases}
\end{aligned}
\end{equation}
with $\boldsymbol{I}_{|\mathcal{X}|}$ the identity matrix of size $|\mathcal{X}|$ and $\mathcal{N}_{|\mathcal{X}|}(\cdot,\cdot)$ the $|\mathcal{X}|$-dimensional Gaussian distribution. We denote the time-invariant probabilities to transition between the states of the Markov chain $(\rho_t^{(c)})_{t\in\mathcal{T}}$ as:
\begin{align}\label{eq:cal:transprob}
p_{ij} = \mathbb{P}\left(\rho_t^{(c)} = j \mid \rho_{t-1}^{(c)} = i\right), \hspace{0.5cm} i,j \in \{1,2\} \: \: \text{and} \: \: t \in \mathcal{T}\backslash\{t_{\min}\},
\end{align}
where $p_{11} + p_{12} = 1$ and $p_{21} + p_{22} = 1$.\footnote{For notational convenience, state `1' refers to the LVS and state `2' refers to the HVS in the sequel of this paper.} If a mortality shock occurs in a given year $t$, we typically observe a positive outlier in the mortality improvement rates in year $t$, i.e.~for $\log \hat{\mu}_{x,t}^{(c)} - \log \hat{\mu}_{x,t-1}^{(c)}$, and a negative outlier in year $t+1$, i.e.~for $\log \hat{\mu}_{x,t+1}^{(c)} - \log \hat{\mu}_{x,t}^{(c)}$. Hence, at least two consecutive observations of the mortality improvement rates are impacted by the shock. To accommodate this observation, we introduce a memory state into the Markov chain, as detailed in Appendix~\ref{Appendix:calibrationRS}. As such, we guarantee that the Markov chain remains in the HVS for a minimum duration of two years once it transitions to that state. In this way, we better mimic situations that occur in real-world settings. 

The parameter vector in the regime-switching model is $|\mathcal{X}|+5$ dimensional and will be denoted as:\footnote{By default we assume a constant standard deviation $\sigma_e$ for the residuals $\epsilon_{x,t}$ in the mortality improvement model with shock regime. However, in Section~\ref{sec:case}, we relax this assumption and tailor it to the data at hand.}
\begin{align*}
\Theta = \left\{\mu_H, \sigma_H, \sigma_e, p_{12}, p_{21}, \boldsymbol{\mathfrak{B}}^{(c)}\right\} \in \mathbb{R}^{|\mathcal{X}|+5}.
\end{align*}

\paragraph{Constructing the log-likelihood.}  We obtain the observed sample $\{\boldsymbol{z}_{t}\}_{t\in\mathcal{T}}$ as the residuals from the calibrated baseline mortality improvement model in Equation~\eqref{eq:cal.baseline.improvement}. The log-likelihood of the sample equals:
\begin{equation} \label{eq:cal:loglikregimeswitch}
\begin{aligned} 
l(\Theta) &= \log f\left(\boldsymbol{z}_{t_{\min}}^{(c)}, \boldsymbol{z}_{t_{\min+1}}^{(c)}, ..., \boldsymbol{z}_{t_{\max}}^{(c)}\mid \Theta \right) \\
&= \displaystyle \sum_{t\in\mathcal{T}} \log f\left(\boldsymbol{z}_t^{(c)} \mid \boldsymbol{z}_{t -1}^{(c)}, ...,\boldsymbol{z}_{t_{\min}}^{(c)}, \Theta\right).
\end{aligned} 
\end{equation}
Appendix~\ref{Appendix:regime.loglik} demonstrates the calculation of the involved conditional probabilities in a recursive way. Additionally, we explain our strategy to determine starting values for the recursion.

\paragraph{Optimizing the log-likelihood.} Using the starting values and log-likelihood derived in Appendix~\ref{Appendix:regime.loglik}, we obtain the optimal set of parameters as:
$$ \hat{\Theta} = \argmax_{{\Theta}} \: l(\Theta).$$

\section{Mortality projection strategy} \label{sec:timedyn}
Next, we proceed with the mortality projection strategy. We first establish the time series models for the calibrated common period effects ($\hat{K}_t^{(1)}$, ..., $\hat{K}_t^{(m)}$) and the country-specific period effects ($\hat{\kappa}_t^{(1,c)}$, ..., $\hat{\kappa}_t^{(l,c)}$) for $t\in\mathcal{T}$. We use a multivariate time series model for the joint projection of both the common and country-specific period effects over time. Then, we generate the states that the Markov chain $\rho_t^{(c)}$ will occupy in the future as well as the severity $Y_t^{(c)}$ of future mortality shocks in the high volatility periods of this Markov chain. We combine the projected time series and mortality shocks with age-specific effects (see Equations~\eqref{eq:modelspec:baseline} and~\eqref{eq:spec.ztx}) and obtain as such the projections for the future mortality rates.

\subsection{Specifying the time series for the calibrated period effects} \label{sec:spectimedyn}
We represent the vector of period effects as:
\begin{align*}
\boldsymbol{K}_t = \begin{pmatrix}
K_t^{(1)} \\ \vdots \\ K_t^{(m)}
\end{pmatrix} \in \mathbb{R}^{m\times 1}, \hspace{0.5cm}
\boldsymbol{\kappa}_t^{(c)} = \begin{pmatrix}
\kappa_t^{(1,c)} \\ \vdots \\ \kappa_t^{(l,c)}
\end{pmatrix} \in \mathbb{R}^{l\times 1},
\end{align*}
and are interested in a multivariate time series model for these combined period effects. We denote:
\begin{align*}
\boldsymbol{\mathcal{K}}_t = \begin{pmatrix}
\boldsymbol{K}_t \\ \boldsymbol{\kappa}_t^{(c)}
\end{pmatrix} \in \mathbb{R}^{(m+l)\times 1}.
\end{align*}
Given our focus on a mortality improvement model, as motivated in Section~\ref{section:introduction}, we impose a VARMA($p$,$q$) model on $\boldsymbol{\mathcal{K}}_t$:\footnote{Due to our focus on mortality improvement rates, the period effects have already been implicitly differenced, as demonstrated in Equation~\eqref{eq:recov.orig.mortmodel}.}
\begin{equation}
\begin{aligned} \label{eq:TDF:timeseries}
\boldsymbol{\mathcal{K}}_t &= \boldsymbol{c} + \displaystyle \sum_{r=1}^{p} \boldsymbol{\Phi}_{r}  \boldsymbol{\mathcal{K}}_{t-r} + \displaystyle \sum_{s=1}^{q} \boldsymbol{\Psi}_{s} \boldsymbol{\delta}_{t-s} + \boldsymbol{\delta}_t, \\
\boldsymbol{\delta}_t &\sim \mathcal{N}_{m+l}\left(\boldsymbol{0},\boldsymbol{\Sigma}_{\delta}\right),
\end{aligned}
\end{equation}
for any $t\in \mathcal{T}$ and where $\boldsymbol{\Phi}_r$ and $\boldsymbol{\Psi}_s$ for all $r,s$ are $(m+l)$-dimensional matrices. Additionally, $\boldsymbol{\delta}_t$ denotes the joint $(m+l)$-dimensional vector of error terms at time $t$. We estimate the parameters in Equation~\eqref{eq:TDF:timeseries} using maximum likelihood estimation. The case study in Section~\ref{sec:case} adopts a weighting scheme in the maximum likelihood estimation of the VARMA model. As such, the modeler can put more weight on recent observations to better reflect the recent trends in the mortality decline.

\subsection{Mortality rate projections} \label{sec:projmortratesubsub}
We are now ready to generate a single trajectory for the mortality rates, $q_{x,t}^{(c)}$, for each age $x$ over the projection period $t \in \mathcal{T}^{\text{{\tiny pred}}} := \{t_{\max} + 1,\ldots, T\}$. We repeat this procedure $10\ 000$ times to assess the uncertainty in the projections. We denote each trajectory as $\hat{q}_{x,t,\iota}^{(c)}$, with $x\in \mathcal{X}$, $t \in \mathcal{T}^{\text{{\tiny pred}}}$ and $\iota$ the number of the generated path. Hereto, we first generate trajectories for both the estimated time series models from Section~\ref{sec:spectimedyn} as well as the Markov chain regime-switching model with estimated transition probabilities $\hat{p}_{ij}$ from Section~\ref{sec:cal.regime}. We proceed as follows:

\begin{enumerate}
\item  By generating new error terms $\boldsymbol{\delta}_{t,\iota}$ for $t \in \mathcal{T}^{\text{{\tiny pred}}}$ from the fitted multivariate normal distribution with mean $\boldsymbol 0$ and covariance matrix $\hat{\boldsymbol{\Sigma}}_{\delta}$, we obtain a trajectory for the future $\hat{\boldsymbol{\mathcal{K}}}_{t,\iota}$ using the calibrated time series model from Equation~\eqref{eq:TDF:timeseries}. This leads to generated trajectories for the common period effects $\hat{K}_{t,\iota}^{(i)}$ for $i=1,..,,m$ and the country-specific period effects $\hat{\kappa}_{t,\iota}^{(j,c)}$ for $j=1,...,l$.
\item To generate a trajectory for the regime-switching model over the projection period $\mathcal{T}^{\text{{\tiny pred}}}$, we first generate a trajectory for the underlying Markov chain $\rho_t^{(c)} \in \{1,2\}$ for $t\in \mathcal{T}^{\text{{\tiny pred}}}$. We initialize the trajectory of the Markov chain at time $t_{\max}$ with the most probable state as derived from the calibrated regime-switching model.\footnote{This is the state $r$ with the largest estimated probability $\mathbb{P}(\rho_{t_{\max}}^{(c)} = r \mid \boldsymbol{z}_{t_{\max}}^{(c)}, \ldots, \boldsymbol{z}_{t_{\min}}^{(c)}, \hat{\boldsymbol{\Theta}})$.} Based on this initial state and the calibrated transition probabilities, we can then generate the state the Markov chain will occupy at any future time point $t \in \mathcal{T}^{\text{{\tiny pred}}}$. This enables us to generate a trajectory for $\hat{Y}_{t,\iota}^{(c)}$ using the fitted normal distribution in the high volatility regime with mean $0$ and variance $\hat{\sigma}_H^2$, see Equation~\eqref{eq:Ytc.spec}. \added{Furthermore, we ensure that the generated mortality shocks lead to sudden increases in the mortality rates, possibly spanning multiple years, followed by a return to the overall baseline mortality decline, see Appendix~\ref{Appendix:C} for more details.}
\end{enumerate}

We generate 10 000 trajectories for the force of mortality of a person aged $x\in \mathcal{X}$ over projection period $t \in \mathcal{T}^{\text{{\tiny pred}}}$ in a recursive way: 
\begin{align} \label{eq:simmutsfs}
\log \hat{\mu}_{x,t,\iota}^{(c)} = \log \hat{\mu}_{x,t-1,\iota}^{(c)} + \displaystyle \sum_{i=1}^m \hat{B}_x^{(i)} \hat{K}_{t,\iota}^{(i)} + \displaystyle \sum_{j = 1}^l \hat{\beta}_x^{(j,c)} \hat{\kappa}_{t,\iota}^{(j,c)} +  \hat{\mathfrak{B}}_x^{(c)} \hat{Y}_{t,\iota}^{(c)},
\end{align}
where $\iota \in \{1,2,\dots, 10\ 000\}$, the generated paths. We start the recursion with the observed crude central death rates at time $t_{\max}$, i.e.~$\hat{\mu}_{x,t_{\max},\iota}^{(c)} = d_{x,t_{\max}}^{(c)}/E_{x,t_{\max}}^{(c)}$ for all $\iota$. The $\iota$-th generated mortality rate at age $x$ and time $t$ then equals:
\begin{align*}
\hat{q}_{x,t,\iota}^{(c)} &= 1 - \exp\left(-\hat{\mu}_{x,t,\iota}^{(c)}\right).
\end{align*}

\section{Case study} \label{sec:case}
\subsection{Multi-population mortality data set}
We use a multi-population mortality data set composed of a pre-selected set of countries. Hereto we rely on the set of countries selected by the Institute of Actuaries in Belgium \citep{IABE2020} and the Dutch Actuarial Association \citep{KAG2020,KAG2022}: Austria (AT), Belgium (BE), Denmark (DK), Germany (DE), Finland (FI), France (FR), Ireland (IE), Iceland (IS), Luxembourg (LU), the Netherlands (NL), Norway (NO), Sweden (SE), Switzerland (CH), and the United Kingdom (UK).\footnote{Ireland is excluded from our case study due to the unavailability of weekly mortality information, as elaborated later on in this section.} 

\begin{figure}[ht!]
\centering
\includegraphics[scale=0.45]{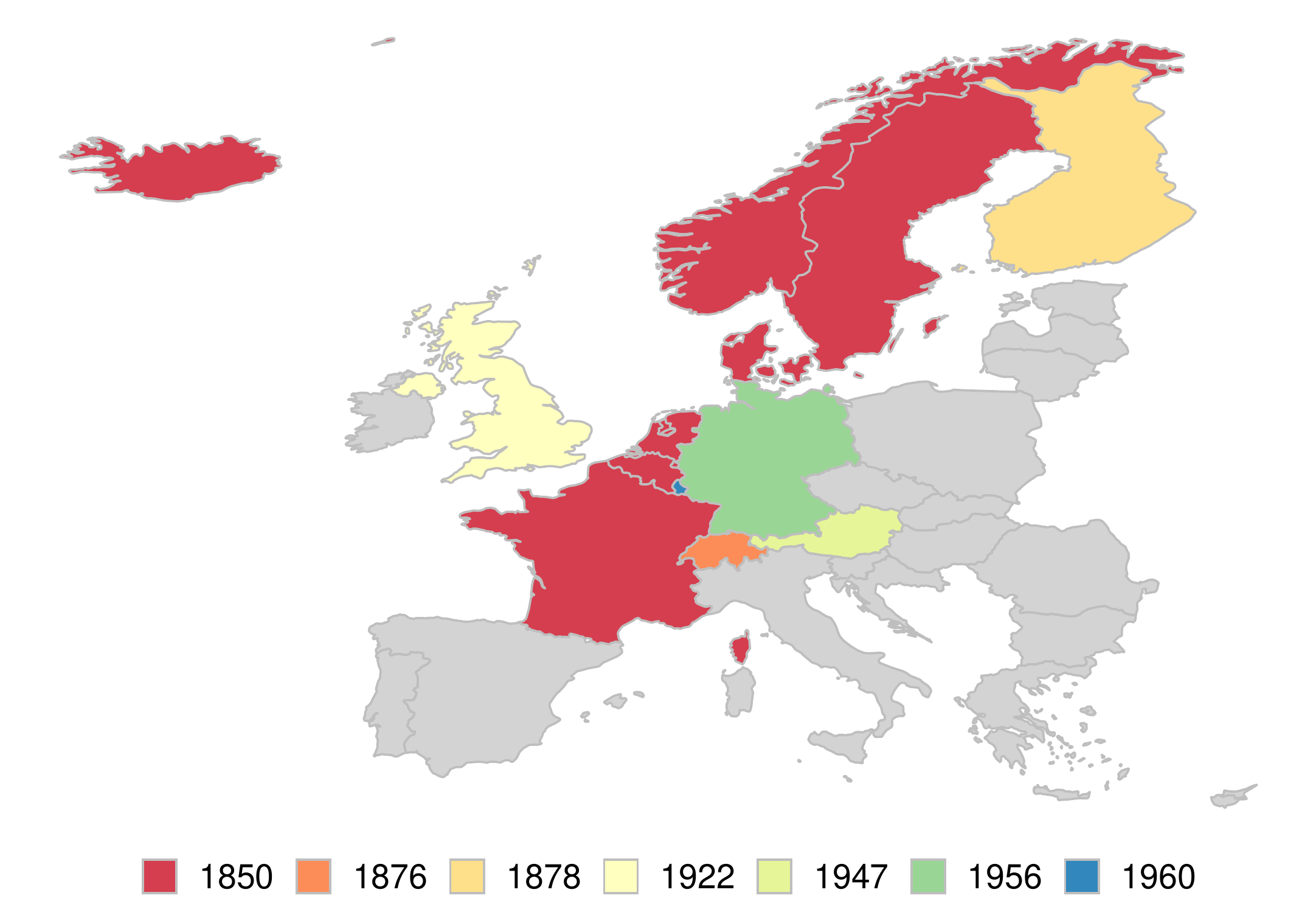}
\caption{Visualization of the thirteen European countries that are part of the common mortality improvement trend and indicate the year when mortality information becomes available.. \label{fig:mapeurope}}
\end{figure}

Whereas \cite{IABE2020} let the calibration period of their proposed Li-Lee model start in 1988 and \cite{KAG2020, KAG2022} in 1970, this study puts focus on a longer calibration period that starts in the year 1850. This enables access to more historical mortality shocks. However, it is not possible to retrieve mortality statistics for all thirteen considered European countries from the year 1850 onwards. To resolve this issue we work with a time-varying composition of the multi-population data set and sequentially add countries when more information becomes available. Figure~\ref{fig:mapeurope} colors the thirteen considered European countries and indicates the year from which each country is included in the multi-population mortality trend. We denote the set of countries that compose this mortality trend at year $t$ as $\mathcal{C}_t$. As an example, $\mathcal{C}_{1900}$ consists of nine countries: AT, BE, DK, FI, FR, IS, NL, NO and SE.

As such, our resulting calibration period is $\mathcal{T} = \{1850, 1851, ..., 2021\}$ and we focus on the age range $\mathcal{X} = \{20, 21, ...,85\}$. The exclusion of the oldest ages is a result of the limited exposure available for these ages \added{and is motivated by \cite{zhou2013pricing} and \cite{liuli}}. Furthermore, we exclude the youngest ages due to the high variability and limited number of deaths in this age range which can potentially result in unstable outcomes\added{, as discussed in \cite{zhou2013pricing} and \cite{HAINAUT2012236}}. In this case study, our focus is on modeling the Dutch male mortality rates.\footnote{\added{We analyze male mortality data due to the historical prevalence of more male victims in world wars, making the male mortality data set particularly relevant for this paper. Of course, a similar analysis can be performed on female or unisex data.}}

We retrieve the annual deaths, $d_{x,t}^{(c)}$, and exposures, $E_{x,t}^{(c)}$, at individual ages $x\in \mathcal{X}$ for each country $c$ from the Human Mortality Database (HMD).\footnote{The Human Mortality Database (HMD) can be accessed at \url{https://www.mortality.org/}.} We further complement the mortality data with annual mortality statistics from Eurostat when possible.\footnote{The Eurostat mortality data-base can be consulted on \url{https://ec.europa.eu/eurostat}.} Following the approach from \cite{robben2022assessing}, we complement the data sets, when necessary, with deaths and exposures up to the year 2021 using weekly mortality information from Eurostat and the Short Term Mortality Fluctuation (STMF) data series on the HMD.

\subsection{Calibrating the baseline mortality improvement model} \label{sec:casestudy:baseline}
Using the strategy outlined in Section~\ref{sec:constr.baseline}, we calibrate the baseline mortality improvement model over the complete calibration period $\mathcal{T}$. In this case study, we specify a two-factor age/period structure for both the common and the country-specific deviation trend, i.e.~we use $m=l=2$ in the baseline mortality improvement model in Equation~\eqref{eq:modelspec:baseline}:\footnote{We omit the superscript $c$ in the country-specific parameters for notational purposes.}
\begin{align} \label{eq:case:baseline.model}
\log \mu_{x,t}^{({\scriptscriptstyle \text{NL}})} - \log \mu_{x,t-1}^{({\scriptscriptstyle \text{NL}})} 
&= A_x + B_x^{(1)} K_t^{(1)} + B_x^{(2)} K_t^{(2)} + \beta_x^{(1)} \kappa_t^{(1)} + B_x^{(2)} \kappa_t^{(2)},
\end{align}
and in the equivalent baseline mortality model in Equation~\eqref{eq:modelspec:tf.baseline}:
\begin{align} \label{eq:case:tf.baseline.model}
\log \mu_{x,t}^{({\scriptscriptstyle \text{NL}})} &= \log \mu_{x,t_{\min}}^{({\scriptscriptstyle \text{NL}})}  + \left(t - t_{\min}\right) A_x + B_x^{(1)} L_t^{(1)} + B_x^{(2)} L_t^{(2)} + \beta_x^{(1)} \lambda_t^{(1)} + \beta_x^{(2)} \lambda_t^{(2)},
\end{align}
where $L_t^{(1)}$, $L_t^{(2)}$, $\lambda_t^{(1)}$ and $\lambda_t^{(2)}$ are defined as in Equation~\eqref{eq:definition.period}.
\subsubsection{Detecting outliers} \label{sec:casestudy:detectoutliers}
\begin{figure}[ht!]
\centering
\includegraphics[scale=0.65]{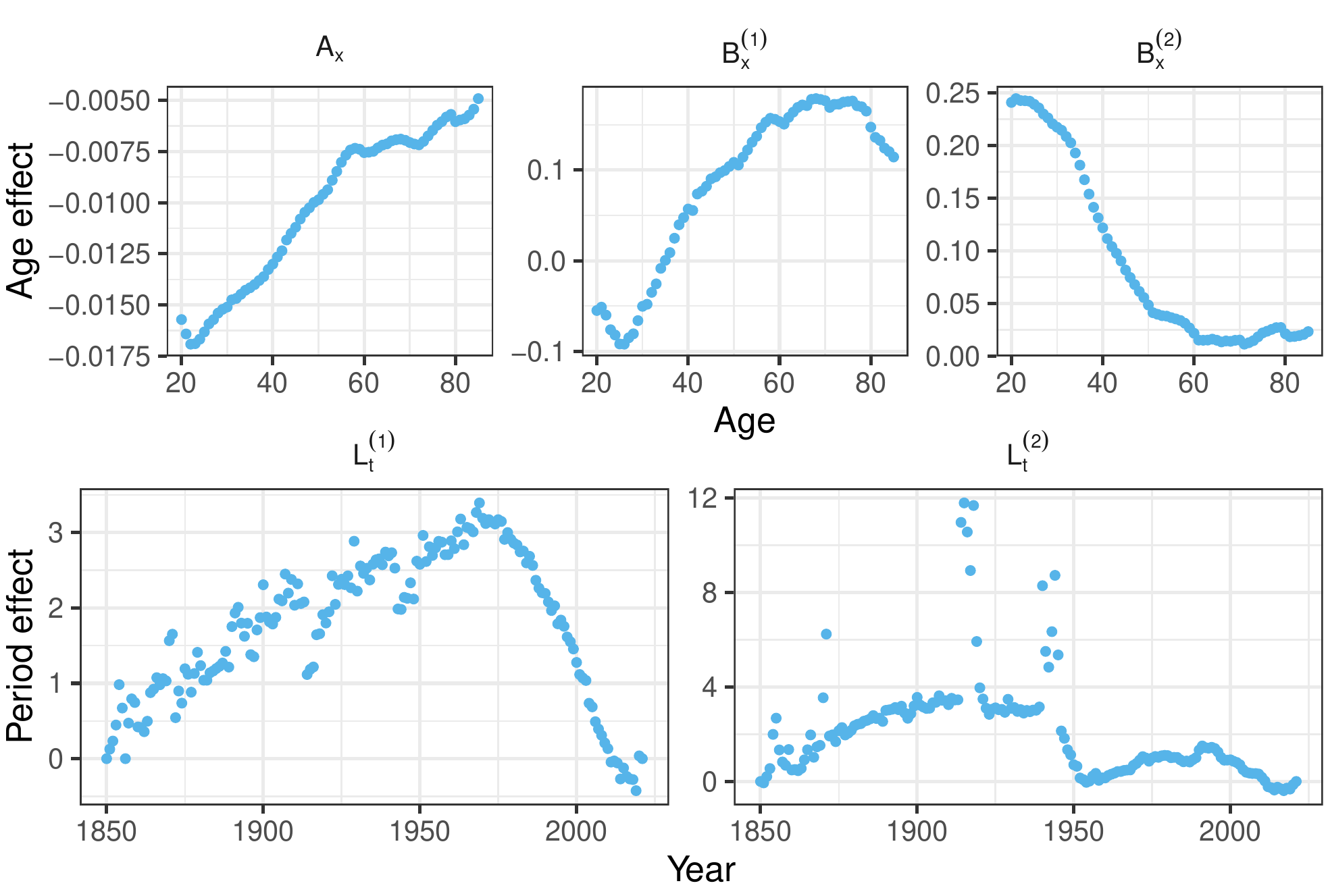}
\caption{The calibrated age (top) and period (bottom) effects in the common trend of the baseline mortality model in Equation~\eqref{eq:modelspec:tf.baseline}. Calibration period $1850$-$2021$, age range 20-85, male data.\label{fig:calibstep0}}
\end{figure}

Figure~\ref{fig:calibstep0} displays the calibrated parameters in the common part of the baseline mortality model. We observe clear outliers in the calibrated common period effects $\hat{L}_t^{(1)}$ and $\hat{L}_t^{(2)}$, representing mortality shocks.\footnote{For the remaining of this paper, we will use the terms `outliers' and `mortality shocks' interchangeably.} We observe the most pronounced outliers during the two world wars.

\begin{figure}[ht!]
\centering
\includegraphics[scale=0.6]{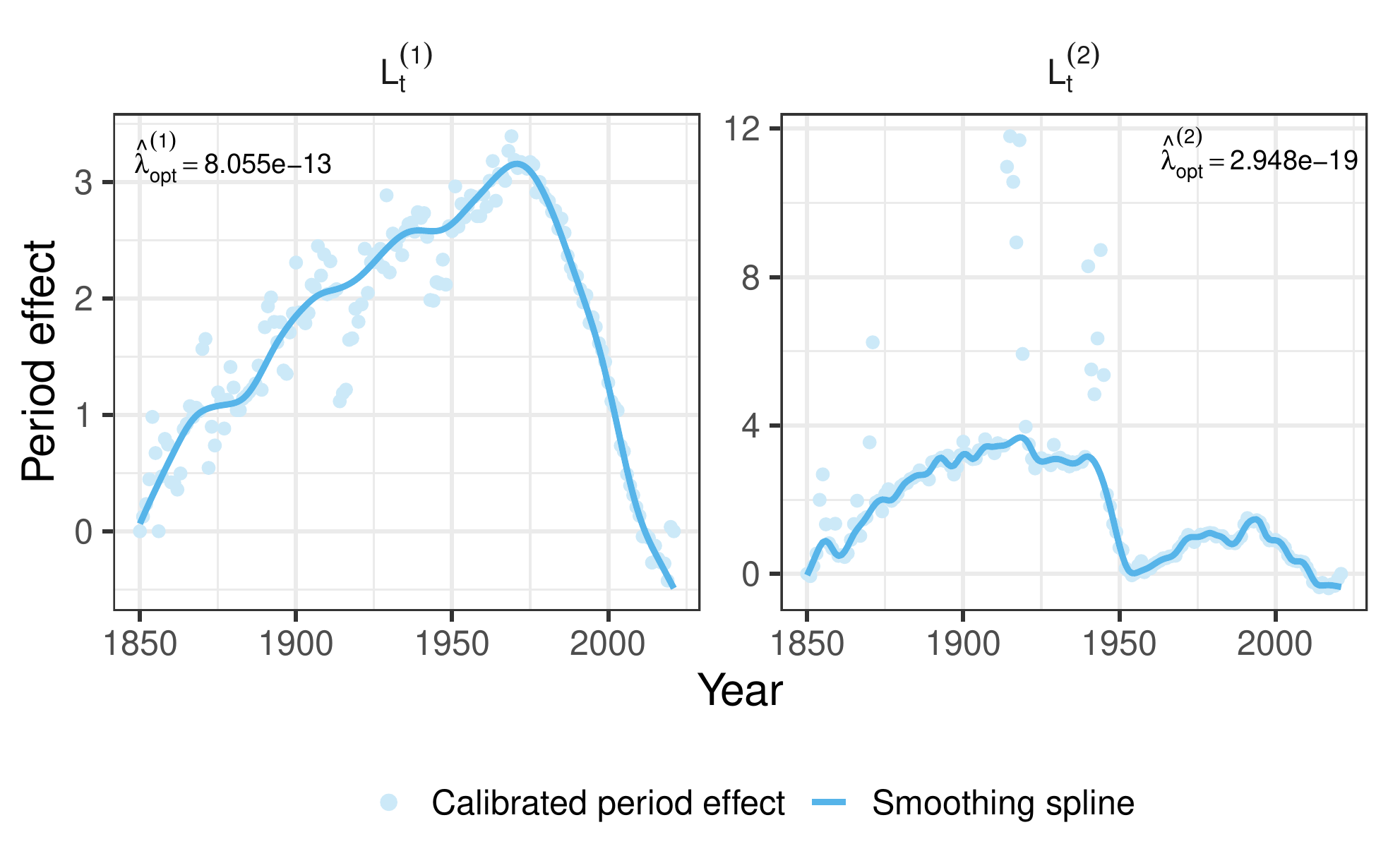}
\caption{The robust smoothing spline fitted to the calibrated common period effect $\hat{L}_t^{(1)}$ (left) and $\hat{L}_t^{(2)}$ (right) where we select the optimal smoothing parameter using the GCV criterion. Calibration period $1850$-$2021$, age range 20-85, male data.\label{fig:step0:ssplinesoptlam}}
\end{figure}

Following Section~\ref{sec:detectoutlier}, we first determine the trend component in each of the calibrated common period effects using robust cubic smoothing splines.\footnote{We use the function \texttt{robustSmoothPline} from the \texttt{aroma.light} package in \texttt{R}, which is a robust version of the commonly used \texttt{smooth.spline} function from the \texttt{stats} package.} Specifically, for $i=1,2$ we aim to find a function $f_i$ that minimizes the following objective function:
\begin{align} \label{eq:qsreg}
\displaystyle \hat{f}_i^{\lambda} = \argmin_{f_i} \displaystyle \sum_{t = 1850}^{2021} \left(L_t^{(i)} - f_i(t)\right)^2 + \lambda \displaystyle \int \left(f_i''(t)\right)^2dt. 
\end{align}
We approximate the function $f_i(\cdot)$ with a cubic smoothing spline with the years in the calibration period $\mathcal{T}$ as knots. To ensure the smoothness of the function $f_i$, we employ a penalty function based on the second-order derivative of $f_i$. The smoothing parameter $\lambda$ balances the goodness of the spline fit and the degree of smoothness of the cubic spline. We select $\lambda$ via the Generalized Cross Validation (GCV) criterion \citep{golub1979Generalized}. Figure~\ref{fig:step0:ssplinesoptlam} shows the resulting smoothing splines, as obtained with the optimally selected smoothing parameter $\hat{\lambda}_{\text{opt}}^{(i)}$.\footnote{We a priori remove the years of the major mortality shocks  (see Figure~\ref{fig:intro:overview}) to attain better robustness of our spline fits.} 

\begin{figure}[ht!]
\centering
\includegraphics[scale=0.45]{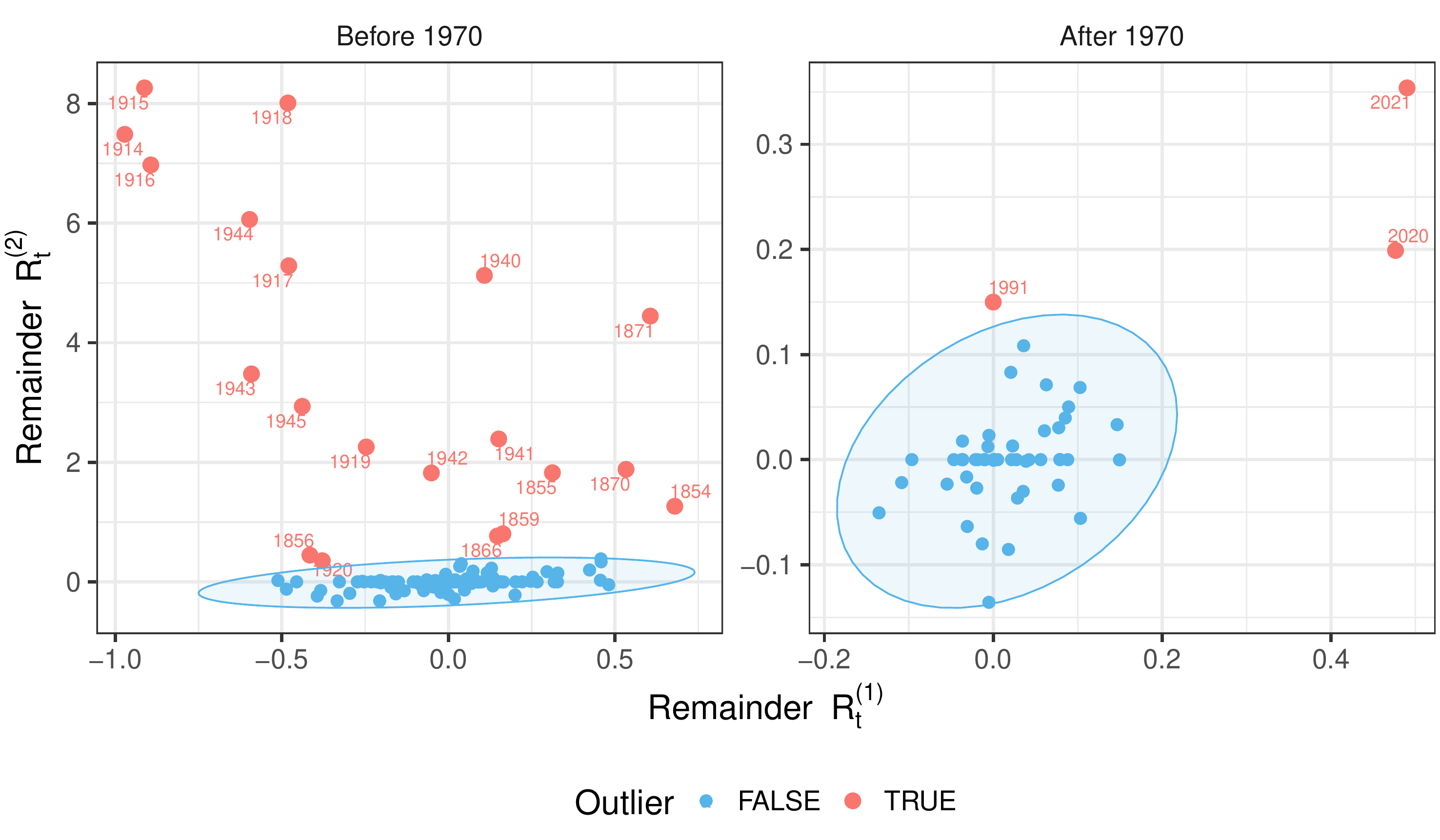}
\caption{Outlier detection using Mahalanobis distance on the remainder components before and after 1970. Calibration period $1850$-$2021$, age range 20-85, male data.\label{fig:mhdoutl}}
\end{figure}

Next, for each $i=1,2$, we define the remainder component of the calibrated period effect $\hat{L}_t^{(i)}$ after removing the trend captured by the smoothing spline:
$$ R_t^{(i)} = \hat{L}_t^{(i)} - \hat{f}_i^{\hat{\lambda}_{\text{opt}}^{(i)}}.$$
We consider the bivariate time series of these remainder components:
$$ \mathcal{R} =  \left\{ \boldsymbol{R}_t := \left(R_t^{(1)}, R_t^{(2)}\right) \mid t = 1850, ..., 2021\right\}. $$
On these remainder components, we apply the robust bivariate outlier detection procedure explained in Appendix~\ref{appendix:outlierdetection}. Since the mortality rates before the year 1970 are more volatile compared to the mortality rates after 1970, we perform separate outlier detection on the set of remainder components before and after 1970. Figure~\ref{fig:mhdoutl} shows the results. The set of detected outlying years equals: 
$$\mathcal{T}^o = \left\{1854{\text -}1856, \: 1859,\: 1866, \:1870{\text -}1871,\: 1914{\text -}1920,\: 1929,\: 1940{\text -} 1945,1991,\: 2020{\text -}2021 \right\}.$$
The detected outliers demonstrate a strong correspondence with the historical mortality shocks observed in Europe, as shown in Figures~\ref{fig:shockrates} and~\ref{fig:intro:overview}.

\subsubsection{Constructing the baseline mortality improvement model} \label{sec:casestudy:recal}
We now recalibrate the baseline mortality model from Equation~\eqref{eq:PoisAssumption} on the outlier-free calibration period $\widetilde{\mathcal{T}} := \mathcal{T} \backslash \left\{\mathcal{T}^o\right\}$. This ensures that the calibrated baseline mortality model is not influenced by the mortality shocks that have been observed and detected in Section~\ref{sec:casestudy:detectoutliers}. However, this procedure results in missing observations in the calibrated period effects $\hat{L}_t^{(j)}$ due to the exclusion of certain years from the calibration period. To address this issue, we intervene in the Newton-Raphson (NR) algorithm, used to calibrate the baseline mortality model as outlined in Section~\ref{sec:cal.trad.mort.model}. More specifically, at the end of each iteration in the NR algorithm, we perform missing value imputation for the missing years in the period effects using an exponential weighted moving average technique.\footnote{We achieve this by using the \texttt{na\_ma} function of the \texttt{ImputeTS} package \citep{moritz2017imputets}.} This results in calibrated period effects that are defined over the complete calibration period $\mathcal{T}$ and that fulfill the identifiability constraints.

\begin{figure}[ht!]
\centering
\includegraphics[scale=0.65]{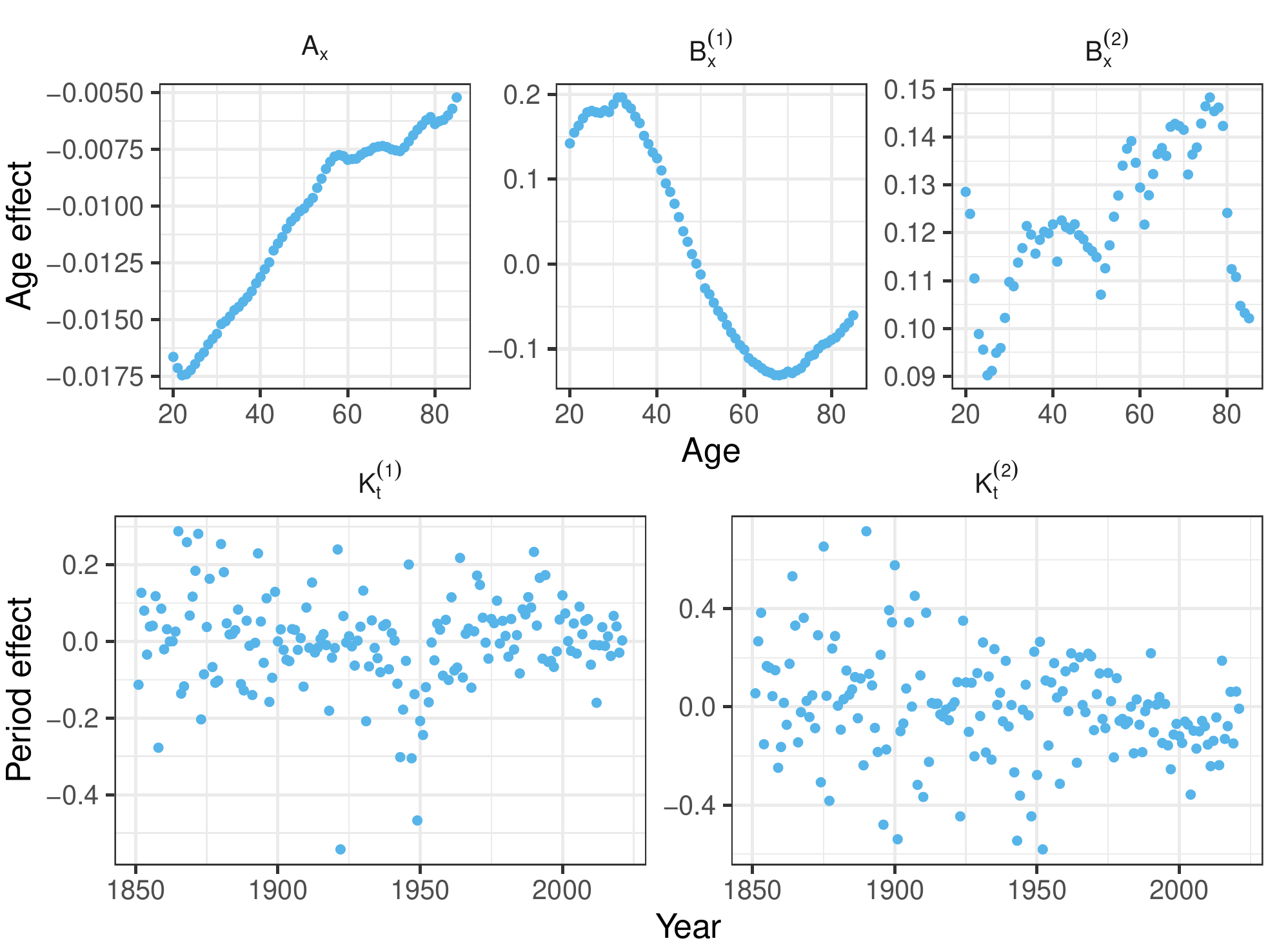}
\includegraphics[scale=0.65]{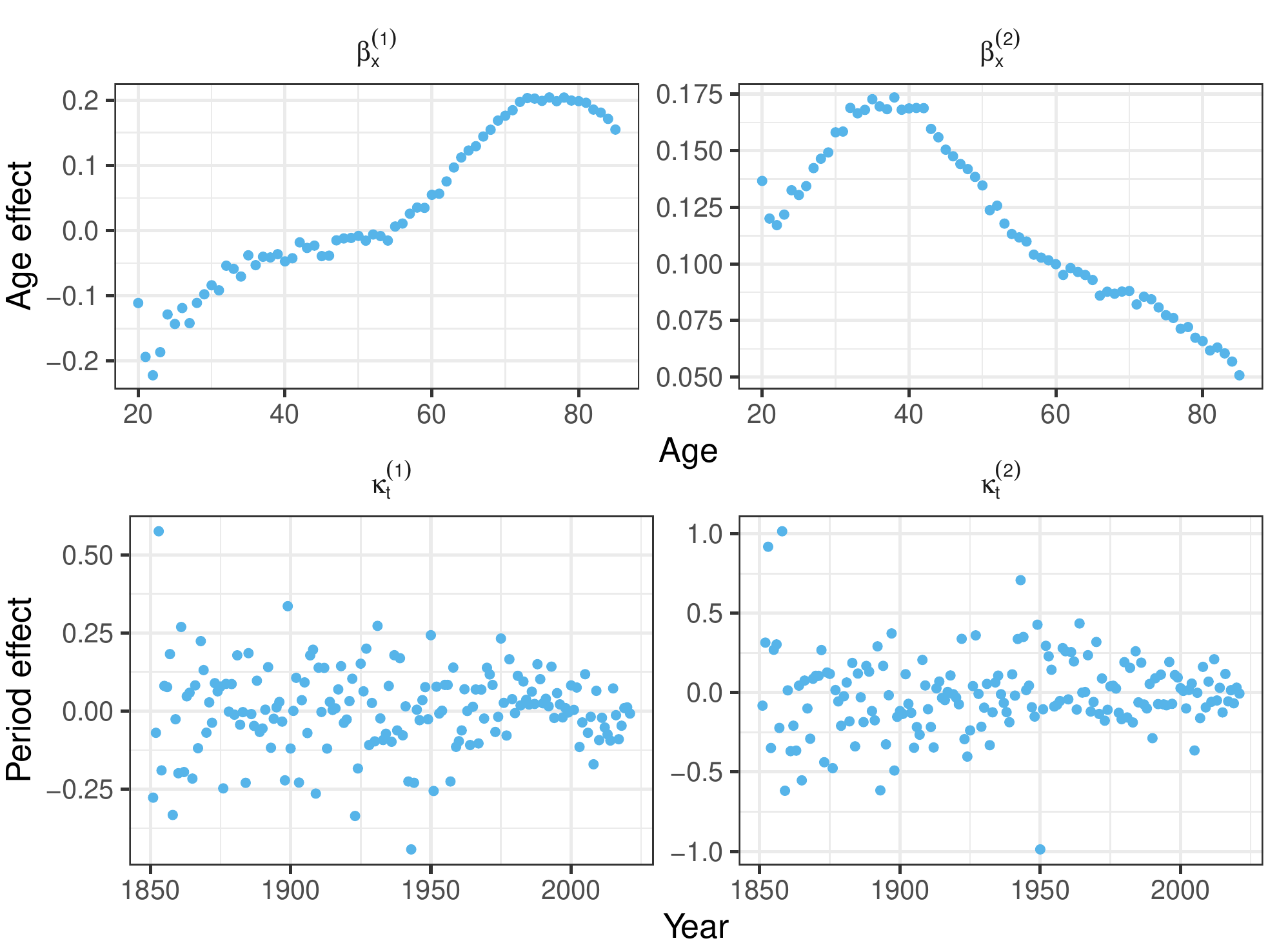}
\caption{The calibrated common (top) and country-specific (bottom) age and period effects in the baseline mortality improvement model. Outlier-free calibration period $\widetilde{\mathcal{T}} = \mathcal{T} \backslash \left\{\mathcal{T}^o\right\}$ combined with missing value imputation, age range 20-85, male data.\label{fig:originaldeviation}}
\end{figure}

Following Equation~\eqref{eq:recov.orig.mortmodel}, we transform the fitted parameters $\hat{L}_t^{(i)}$ for $i=1,2$ and $\lambda_t^{(j)}$ for $j=1,2$ to the parameters $\hat{K}_t^{(i)}$ for $i=1,2$ and $\hat{\kappa}_t^{(j)}$ for $j=1,2$ in the original baseline mortality improvement model outlined in Equation~\eqref{eq:cal.baseline.improvement}. Figure~\ref{fig:originaldeviation} shows the age and period effects in the common trend specification (top) and the specification of the Dutch deviation from the European trend (bottom). In line with our expectations, the calibrated period effects no longer exhibit any severe outliers. 

\begin{figure}[ht!]
\centering
\includegraphics[scale=0.6]{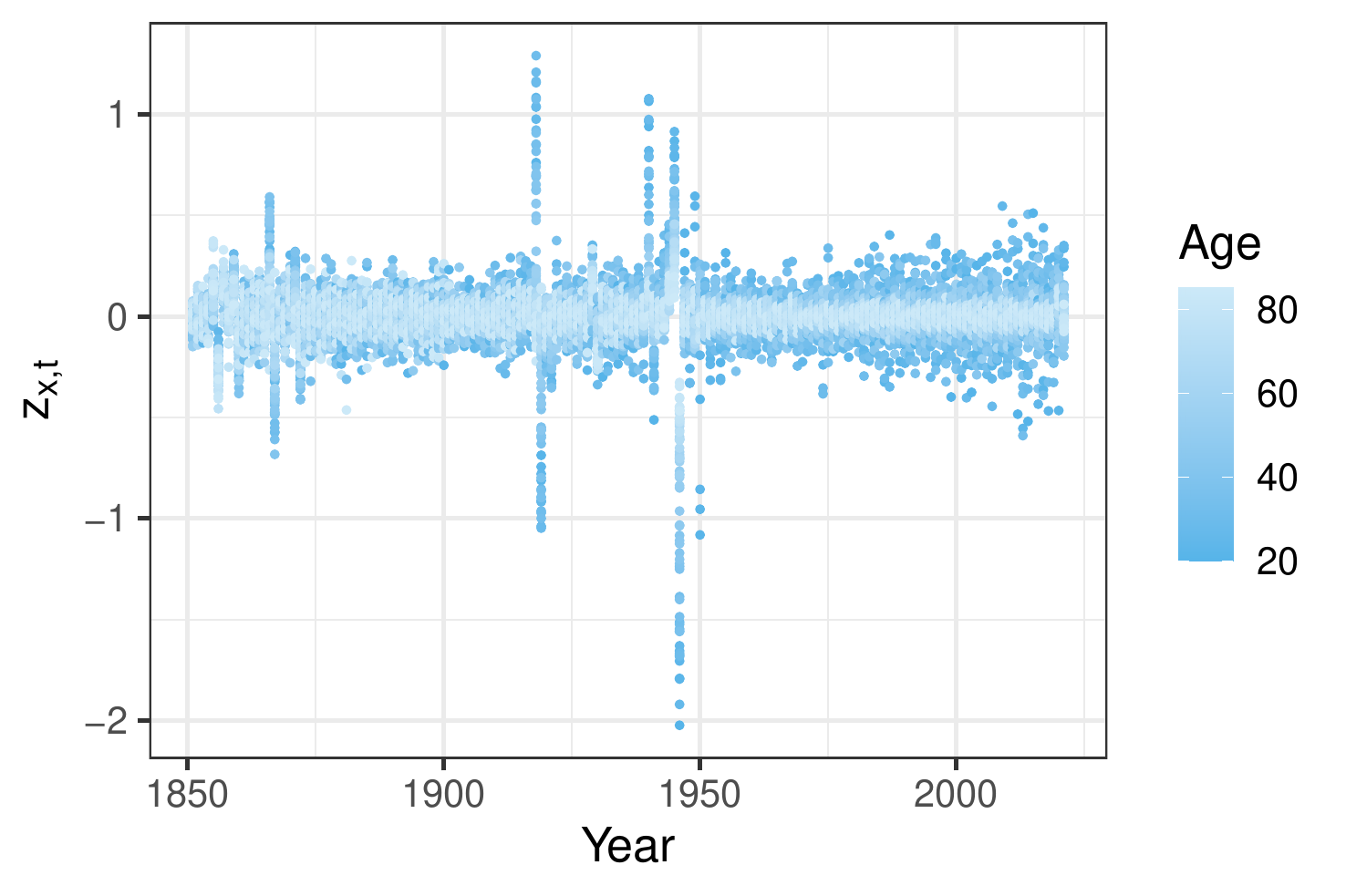}
\caption{The time series of realized residuals $z_{x,t}$ of the calibrated baseline mortality improvement model for all ages $x \in \mathcal{X}$. Calibration period 1851-2021, age range 20-85, male data. \label{fig:ztxperage}}
\end{figure}

We denote the calibrated age/period structure in the Dutch baseline mortality improvement model (see Equation~\eqref{eq:case:baseline.model}) as:
\begin{align*}
\hat{\eta}_{x,t}^{({\scriptscriptstyle \text{NL}})} 
:= \hat{A}_x + \hat{B}_x^{(1)} \hat{K}_t^{(1)} +  \hat{B}_x^{(2)} \hat{K}_t^{(2)} + \hat{\beta}_x^{(1)} \hat{\kappa}_t^{(1)} + \hat{B}_x^{(2)} \hat{\kappa}_t^{(2)}.
\end{align*}
Following Section~\ref{sec:regimeswitch}, we calculate the baseline model's residuals as:
\begin{align}\label{eq:ztx}
z_{x,t} := \log \hat{m}_{x,t}^{({\scriptscriptstyle \text{NL}})} - \log \hat {m}_{x,t-1}^{({\scriptscriptstyle \text{NL}})} - \hat{\eta}_{x,t}^{({\scriptscriptstyle \text{NL}})},
\end{align}
which are perturbed by periods of high volatility caused by mortality shocks. Figure~\ref{fig:ztxperage} shows the time series of observed $z_{x,t}$ for each age $x \in \mathcal{X}$ and $t \in \mathcal{T}$.

\subsection{Calibrating the regime-switching model} \label{sec:casestudyregimeswitch}

\subsubsection{Regime-switching model with age-specific effect}
We calibrate the regime-switching model separately on two distinct age groups: $\mathcal{X}^{(1)}$, covering ages 20-59, and $\mathcal{X}^{(2)}$, covering ages 60-85. This split enables us to distinguish the occurrence of different types of mortality shocks. By doing so, we can effectively distinguish the impact of events like the COVID-19 pandemic, which mainly affects older age groups, from pandemics and wars which primarily affect younger age groups. Therefore, we impose the following adjusted specifications for the baseline model's residuals:
\begin{align} \label{eq:modspecupdate}
z_{x,t} \sim \begin{cases}
\mathfrak{B}_x^{(1)} Y_t^{(1)} + \epsilon_{x,t}^{(1)} & \text{for } x \in \mathcal{X}^{(1)} \\
\mathfrak{B}_x^{(2)} Y_t^{(2)} + \epsilon_{x,t}^{(2)} & \text{for } x \in \mathcal{X}^{(2)},
\end{cases}
\end{align}
where $Y_t^{(1)}$ and $Y_t^{(2)}$ are both regime-switching and alternate between a low and high volatility state, characterized by two independent Markov chains $\rho_t^{(1)}$ and $\rho_t^{(2)}$, respectively. These regime-switching models and Markov chains have their own set of parameters and transition probabilities. The time series vector $\boldsymbol{z}_t := (z_{x,t})_{x \in \mathcal{X}^{(j)}}$ follows for each age group $\mathcal{X}^{(j)}$ an alternating multivariate normal distribution, as derived in Equation~\eqref{eq:cal:transdens} of Section~\ref{sec:cal.regime}. In the remainder of this section, we omit the superscript related to the age range under consideration for notational purposes.  

In accordance with the methodology outlined in Section~\ref{sec:cal.regime}, we adopt a flexible specification for the standard deviation of the residuals $\epsilon_{x,t}$. From Figure~\ref{fig:shockrates} we learn that the death rates in the years after 1970 exhibit lower volatility than those observed in previous years. Moreover, Figure~\ref{fig:ztxperage} reveals that the standard deviation of the error term $\epsilon_{x,t}$ decreases with age.\footnote{We ignore the outliers in Figure~\ref{fig:ztxperage} to assess the variability of $\epsilon_{x,t}$.} To accommodate these observations, we propose the following specification for $\sigma_e(x,t)$:
\begin{equation} \label{eq:volatility.adjustment}
\begin{aligned}
\sigma_e(x,t) &=  \sigma_{e_1}(x)  \mathbbm{1}_{\left\{t < 1970\right\}} + \sigma_{e_2}(x)  \mathbbm{1}_{\left\{t \geq 1970\right\}}, \\
\sigma_{e_j}(x) &= \sigma_{e_j} + \text{slope}_j \cdot (x - x_{\min}),
\end{aligned}
\end{equation}
for $j \in \{1,2\}$. Further, $x_{\min}$ refers to the minimum age in the age group under consideration and, e.g., $\mathbbm{1}_{\left\{t < 1970\right\}}$ is one for $t$ smaller than the year 1970 and zero elsewhere. This introduces extra flexibility regarding the variability of the residuals $\epsilon_{x,t}$ in the period before and after the year 1970. The parameter `$\text{slope}$' captures the linear decrease in the variability with age, before and after 1970. We use the same specification for $\sigma_e(x,t)$ for both age groups $\mathcal{X}^{(1)}$ and $\mathcal{X}^{(2)}$.

\subsubsection{Calibration results for the regime-switching model}
\begin{figure}[ht!]
\centering
\includegraphics[scale=0.65]{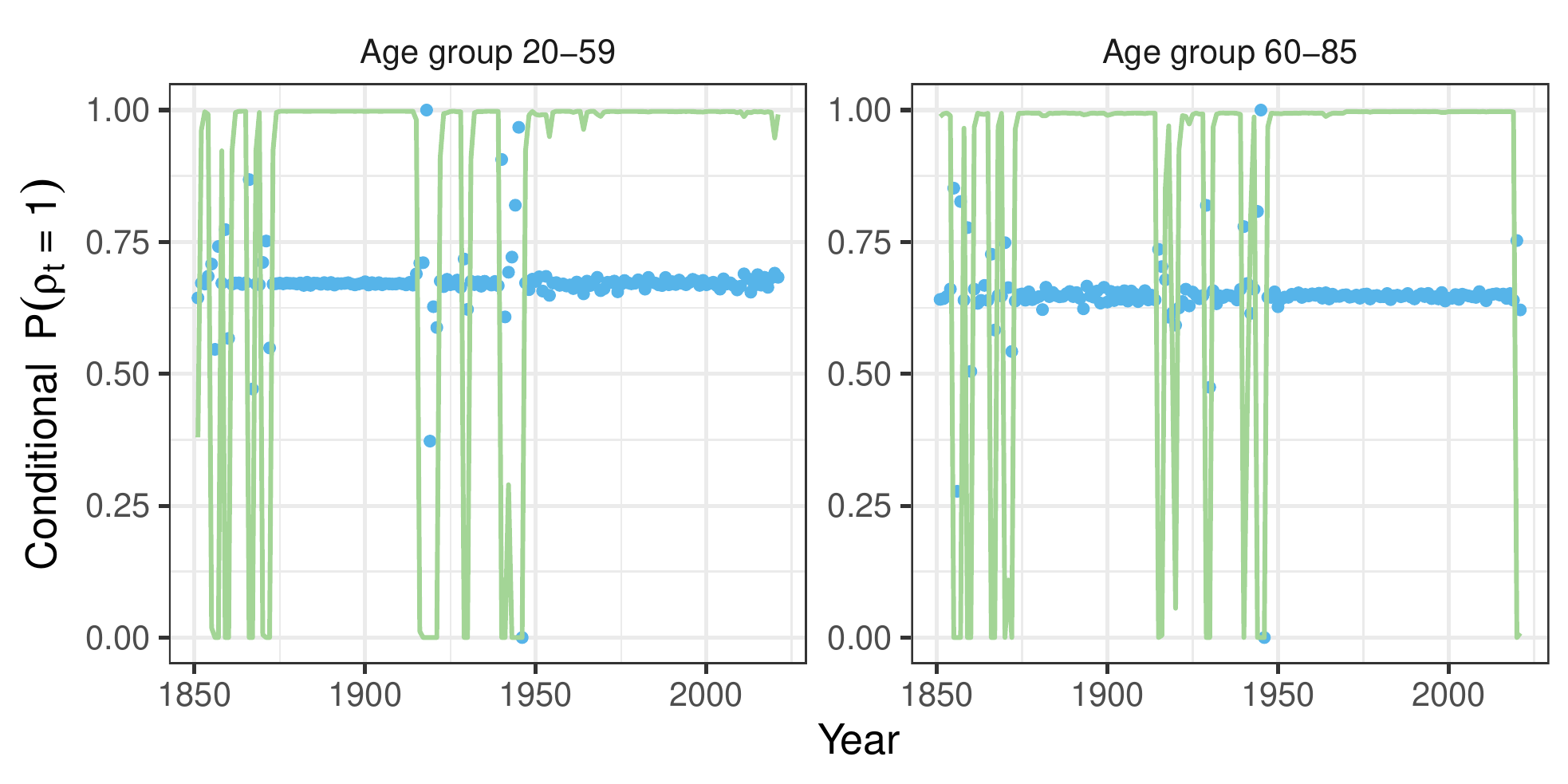}
\caption{The green line represents the conditional probability of being in the LVS (state 1) at year $t$, calculated using Equation~\eqref{eq:cal:regimprob}. These probabilities are calculated with the optimized parameter values in Table~\ref{tab:optTheta11} and Figure~\ref{fig:optC1}. The blue dots represent the age-averaged realized residuals $z_{x,t}$ versus year $t$. The left panel shows the results for the age group 20-59 and the right panel for the age group 60-85. Calibration period 1850-2021, male data. \label{fig:cprob1}}
\end{figure}

We provide all technical details about the calibration of the regime-switching model in Appendix~\ref{Appendix:calibrationRS}.  Table~\ref{tab:optTheta11} and Figure~\ref{fig:optC1} display the estimated parameters in the regime-switching model, which we calibrate separately for the age groups 20-59 and 60-85. We interpret the results for the age group 20-59. The estimated value for $p_{1,2}$ indicates that there is a $4.7\%$ probability of transitioning from the low to the high volatility state. The standard deviation of the shock random variable $Y_t^{(1)}$ in the high volatility regime, i.e.~$\sigma_H = 2.41$, is noticeably higher compared to that in the low volatility regime.\footnote{To compute the standard deviation in the low volatility state, we use Equation~\eqref{eq:volatility.adjustment}.} The calibrated age dependency $\hat{\mathfrak{B}}_x$ of mortality shocks for ages 20-59 exhibits a decreasing trend, resulting in less severe shocks for older individuals compared to younger ones. This can be attributed to the mortality shocks resulting from the two world wars where the majority of the victims during the world wars were young men. 
\begin{tabularx}{\textwidth}{p{0.38\columnwidth}p{0.57\columnwidth}}
   \centering
      \begin{tabular}{lrr}
      \hline
      Parameter  & \multicolumn{1}{c}{20-59} & \multicolumn{1}{c}{60-85} \\
      \hline
      $p_{1,2}$        & 0.04709  & 0.06656  \\
      $p_{2,1}$        & 0.34207  & 0.68966  \\
      $\sigma_{e_1}$   & 0.12750  & 0.05823  \\
      $\text{slope}_1$ & -0.00156 & 0.00069  \\
      $\sigma_{e_2}$   & 0.18173  & 0.03944  \\
      $\text{slope}_2$ & -0.00342 & -0.00048 \\
      $\mu_H$          & -0.01602 & 0.03155  \\
      $\sigma_H$       & 2.40648  & 0.89818  \\
      \hline 
      \end{tabular}  &
\includegraphics[width=0.975\linewidth,valign=m]{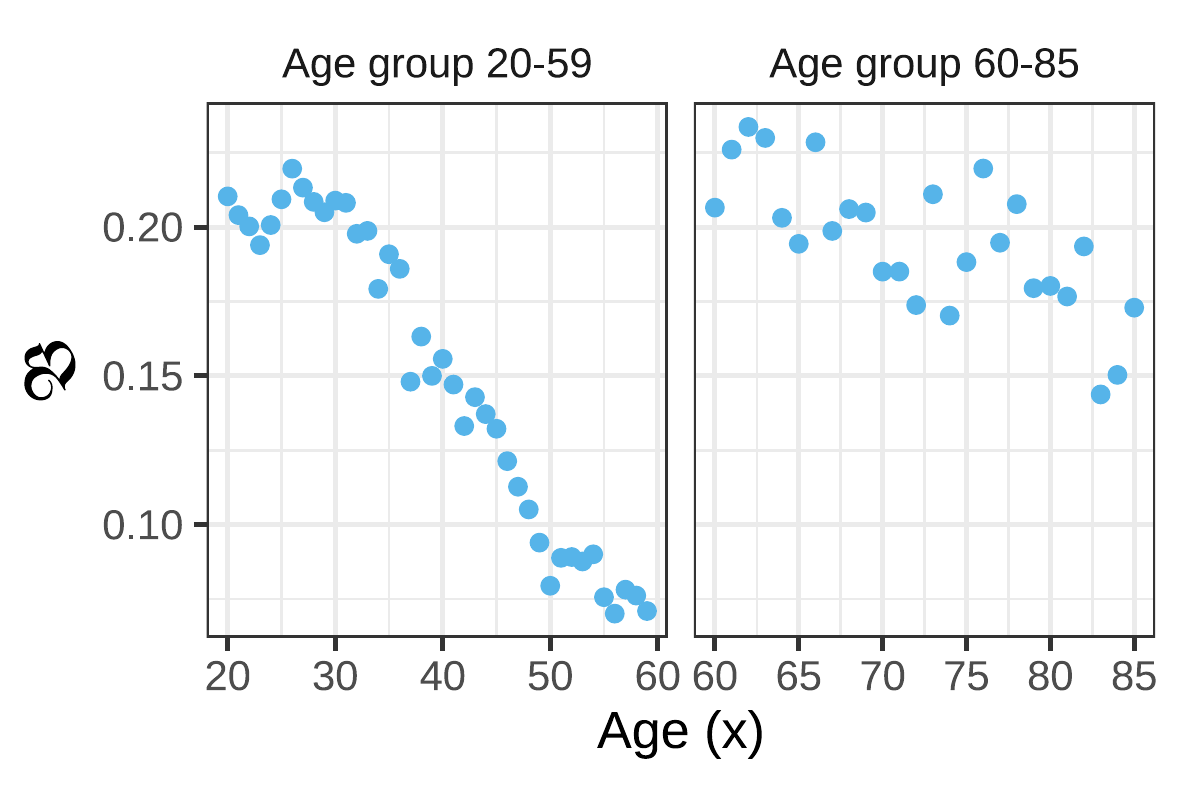}
\begin{tikzpicture}[overlay]
\node[inner sep=0pt,rotate=90] at (-8.4,0.68)
    {${}_{\boldsymbol{x}}$};
\end{tikzpicture} 
        \\ 
\captionof{table}{Calibrated parameters in the regime-switching model for the two age groups. Calibration period 1850-2021, male data.\label{tab:optTheta11}}
&     
    \captionof{figure}{Calibrated age-specific effect $\mathfrak{B}_x$ in the regime-switching model for age groups 20-59 and 60-85. Calibration period 1850-2021, male data.\label{fig:optC1}}
\end{tabularx}
The green line in Figure~\ref{fig:cprob1} shows the conditional probability of being in the low volatility state (LVS, state 1) in year $t$, as derived in Equation~\eqref{eq:cal:regimprob}. The figure reveals alternating periods of high and low volatility. The blue dots in this figure represent the age-averaged realized residuals $z_{x,t}$ in the calibrated baseline mortality improvement model. The outlying dots correspond to historical mortality shocks and are captured by the HVS. The results indicate, for instance, that during the COVID-19 pandemic period 2020-2021 the Markov chain occupies the HVS but only for the age group 60-85. This is in line with our understanding that the pandemic primarily affects older individuals.

\subsection{Time series models and resulting mortality projections}
We follow the projection strategy outlined in Section~\ref{sec:timedyn} to project mortality rates with our proposed mortality improvement model with \added{a} shock regime. 

\subsubsection{Projecting common and country-specific period effects using a multivariate time series model} \label{subsubsec1}
In accordance with Section~\ref{sec:spectimedyn} and inspired by the random behavior of the calibrated period effects illustrated in Figure~\ref{fig:originaldeviation}, we propose the following multivariate time series model for the four-dimensional vector of period effects $\boldsymbol{\mathcal{K}}_t = \left(K_t^{(1)}, K_t^{(2)}, \kappa_t^{(1)}, \kappa_t^{(2)}\right)$ for $t \in \{1851, 1852, \ldots, 2021\}$:
\begin{align} \label{eq:timedynformula}
\boldsymbol{\mathcal{K}}_t = \boldsymbol{c} + \boldsymbol{W}_t,
\end{align}
where $\boldsymbol{c} = [c_1, c_2, 0, 0]^T$ are the mean parameters or intercepts and $\boldsymbol{W}_t$ is a four-dimensional vector of white noise terms with mean $\boldsymbol{0}$ and covariance matrix $\boldsymbol{\Sigma}_w$.\footnote{We use zero intercepts in the time series models of the country-specific period effects to avoid divergence from the common improvement trend.} To estimate the time series models in Equation~\eqref{eq:timedynformula}, we use a weighted variant of maximum likelihood estimation, see Appendix~\ref{Appendix:weightedLL} for more details. This weighting principle is inspired by the work of \cite{mittnik2000conditional} and \cite{robben2022assessing}, and is motivated by the fact that earlier years' mortality dynamics are less representative compared to the dynamics observed in more recent years. This is in particular true when a lengthy calibration period is used as in our case study. The estimated mean parameter $\hat{\boldsymbol{c}}$ and covariance matrix $\hat{\boldsymbol{\Sigma}}_w$ equal:
\begin{align*}
\hat{\boldsymbol{c}} = \begin{bmatrix*}[r]
0.01170 \\ -0.06590 \\ 0 \\ 0
\end{bmatrix*}, \hspace{0.5cm}
\hat{\boldsymbol{\Sigma}}_w = \begin{bmatrix*}[r]
 0.00495 &  0.00012 &  0.00052 & -0.00120  \\
 0.00012 &  0.01491 &  0.00281 & -0.00320   \\
 0.00052 &  0.00281 &  0.00539 & -0.00321  \\
-0.00120 & -0.00320 & -0.00321 &  0.01564 
\end{bmatrix*}.
\end{align*}

\begin{figure}[ht!]
\centering
\includegraphics[scale=0.65]{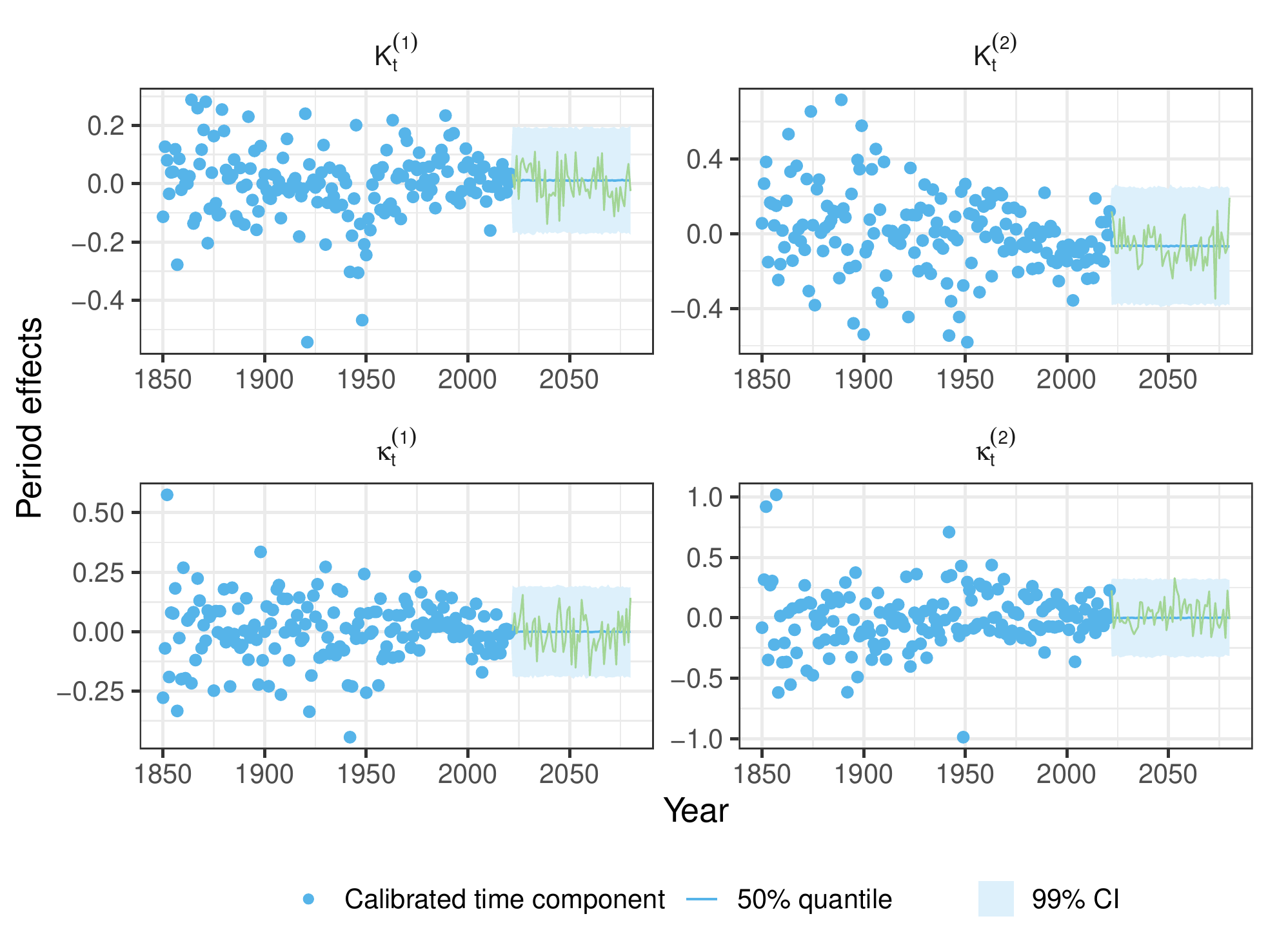}
\caption{The $0.5\%$, median and $99.5\%$ quantile, based on $10\ 000$ trajectories for the estimated time series models of the four-dimensional vector of period effects $\hat{\boldsymbol{\mathcal{K}}}_t$ (blue fan chart). The blue dots represent the calibrated period effects. The green line represents one trajectory for each period effect. Projection period 2022-2080. \label{fig:simKt}}
\end{figure}

We generate $10\ 000$ trajectories for the vector of period effects $\boldsymbol{\mathcal{K}}_t$ for $t \in \{1851,...,2021\}$ by taking random draws from a four-dimensional Gaussian distribution with the estimated $\hat{\boldsymbol{c}}$ and $\hat{\boldsymbol{\Sigma}}_w$ as parameters for the mean and covariance matrix, respectively. Figure~\ref{fig:simKt} shows the $0.5\%$, median and $99.5\%$ quantile, based on these $10\ 000$ trajectories for $K_t^{(1)}$, $K_t^{(2)}$, $\kappa_t^{(1)}$ and $\kappa_t^{(2)}$.

\subsubsection{Projecting the regime-switching model to generate age-specific mortality shocks} \label{subsubsec2}
\begin{figure}[ht!]
\centering
\includegraphics[scale=0.65]{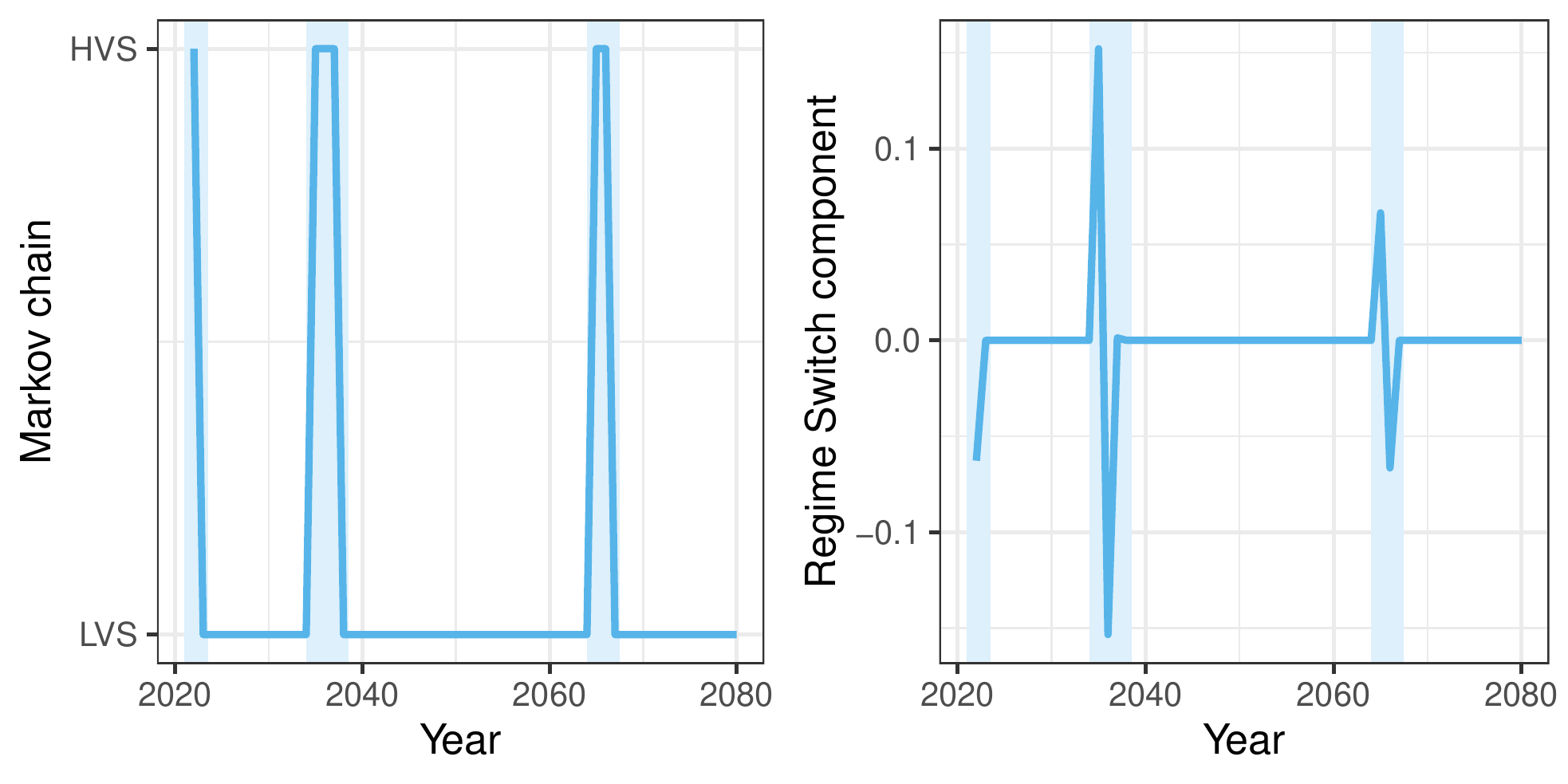}
\caption{In the left panel, we visualize one trajectory of the Markov chain $\rho_t^{(2)}$ for the age group 60-85 over the projection period 2022-2080. In the right panel, we show the corresponding generated trajectory for the regime-switching model with age-specific effect $\hat{\mathfrak{B}}_x Y_t^{(2)}$ averaged over the ages in the interval 60-85. The underlying Markov chain occupies the states pictured in the left panel. The light blue vertical bars refer to periods in the high volatility regime.\label{fig:simMarkRegime}}
\end{figure}

We now focus on generating the mortality shocks for the age groups $\mathcal{X}^{(1)}$, i.e.~ages 20-59, and $\mathcal{X}^{(2)}$, i.e.~ages 60-85. To better align the projections with real-world settings, we suggest two minor modifications to the projection strategy outlined in Section~\ref{sec:projmortratesubsub}. These modifications are explained in detail in Appendix~\ref{Appendix:C} and ensure that the generated shocks in a high volatility period do not have a long-lasting impact on future mortality rate projections. We then generate 10\ 000 trajectories for both the Markov chains and the regime-switching models, covering the period from 2022 to 2080. Figure~\ref{fig:simMarkRegime} displays one such trajectory. First, related to the COVID-19 pandemic, we assume that we leave the high volatility regime in the year 2023 for the age group 60-85.\footnote{Related to this, we also make sure that the mortality shock related to COVID-19 in the years 2020-2022 has been offset in 2023, see Appendix~\ref{Appendix:C}.} We further observe two future spells in the high volatility regime of three and two years respectively in which mortality shocks occur. 

\subsubsection{Mortality rate projections}
Following Section~\ref{sec:projmortratesubsub}, we forecast the force of mortality for an $x$-year old individual in year $t \in \mathcal{T}^{\text{{\tiny pred}}} := \{2022,...,2080\}$ recursively as
$$ \hat{\mu}_{x,t,\iota}^{({\scriptscriptstyle \text{NL}})} = \hat{\mu}_{x,t-1,\iota}^{({\scriptscriptstyle \text{NL}})} \cdot  \exp\left(\hat{A}_x + \displaystyle \sum_{i=1}^2 \hat{B}_x^{(i)} \hat{K}_{t,\iota}^{(i)} + \displaystyle \sum_{j=1}^2 \hat{\beta}_x^{(j)} \hat{\kappa}_{t,\iota}^{(j)} + \hat{\mathfrak{B}}_x^{(a)} \hat{Y}_{t,\iota}^{(a)} \right), $$
for age group $a \in \{[20,59], [60,85]\}$, for $t > 2021$ and referring to the $\iota$-th generated trajectory. As starting value of the recursion in the year $2021$ we use the observed crude central death rate, i.e.~$\hat{\mu}_{x,2021,\iota}^{({\scriptscriptstyle \text{NL}})} = d_{x,2021}^{({\scriptscriptstyle \text{NL}})}/E_{x,2021}^{({\scriptscriptstyle \text{NL}})}$ for all $\iota$. Using Equation~\eqref{eq:qxt.mort.rate}, we can calculate the corresponding mortality rates.

\begin{figure}[ht!]
\centering
\includegraphics[scale=0.7]{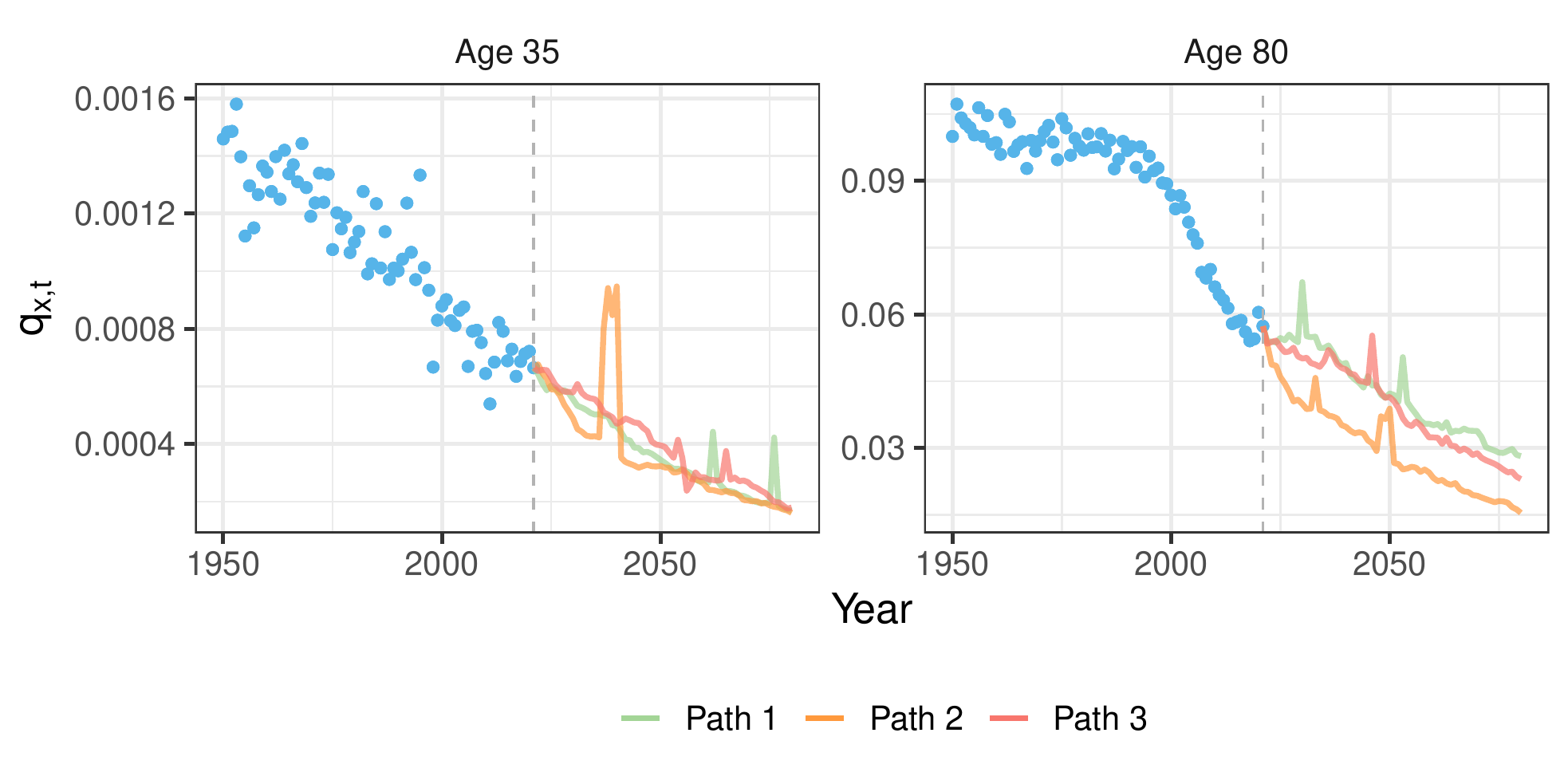}
\caption{Three generated trajectories for the mortality rates at ages 35 and 80. Calibration period 1850-2021 and projection period 2022-2080. We zoom in on the years starting from 1950 to better visualize the projections.\label{fig:3traj}}
\end{figure}

Figure~\ref{fig:3traj} illustrates three trajectories of the mortality rates at ages 35 (left panel) and 80 (right panel) over the projection period 2022-2080. The generated mortality decline is occasionally disrupted by some artificially generated mortality shocks. Multiple mortality shocks, sometimes with very high severity, occur in the left panel of Figure~\ref{fig:3traj}. This is mainly due to the integration of the two World Wars in the calibration procedure of the regime-switching model which mainly affects younger age groups. Section~\ref{sec:scenario.analysis} further elaborates on this.

\begin{figure}[ht!]
\centering
\includegraphics[scale=0.6]{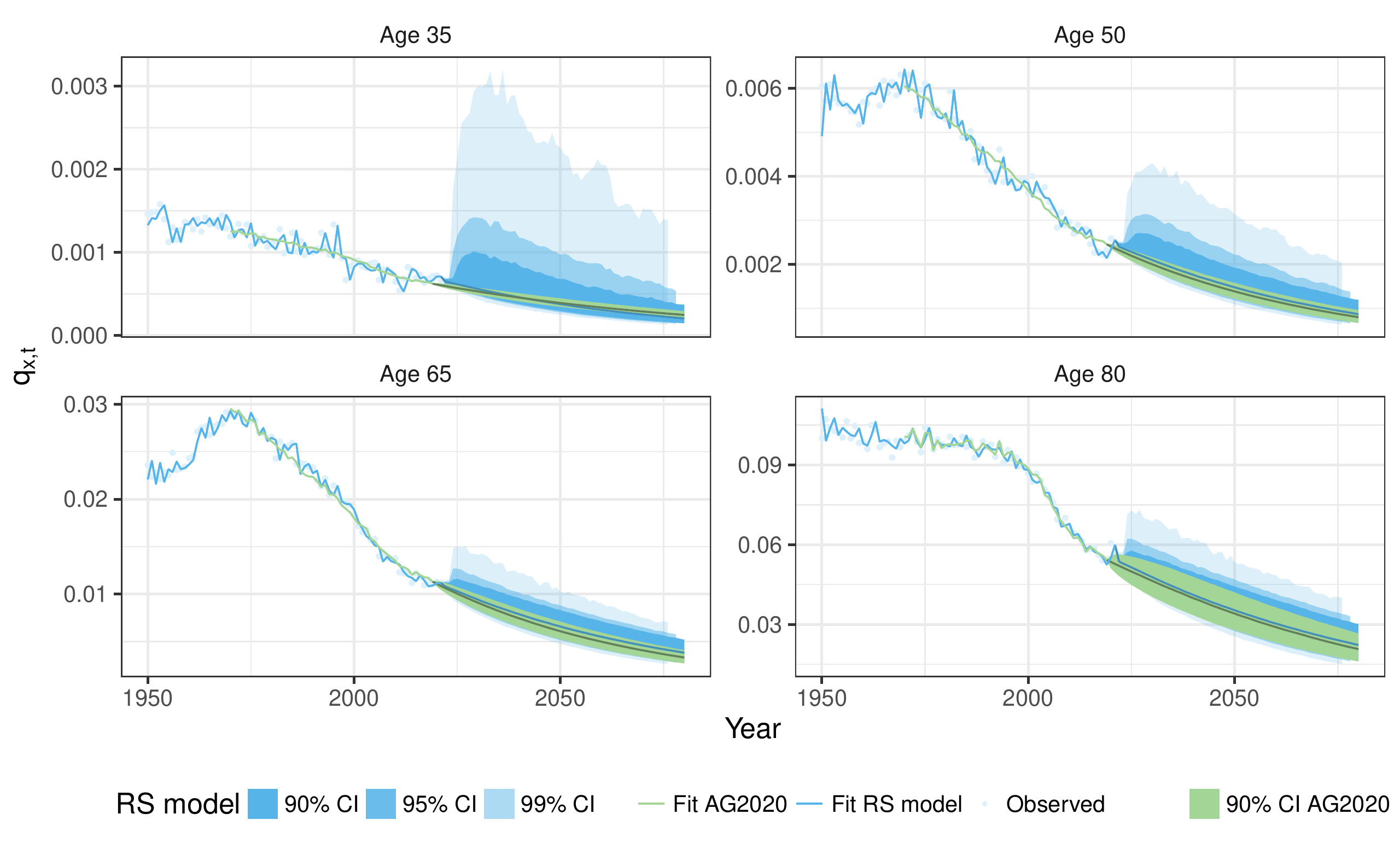}
\caption{The blue fan charts visualize the $90\%$, $95\%$, and $99\%$ prediction intervals based on $10\ 000$ generated trajectories of the mortality improvement model with shock regime. We use calibration period 1850-2021, age range 20-85, and projection period 2022-2080. The blue shaded dots are the observed mortality rates and the blue line shows the fit of the baseline mortality improvement model. Additionally, the green line shows the in-sample fit of the AG2020 model where we calibrated the Li-Lee model on data from 1970-2019. The green fan charts show the $90\%$, $95\%$, and $99\%$ prediction intervals based on the simulations resulting from this AG2020 model. The median quantiles of the simulations for both methods are shown in dark blue and dark green, respectively.\label{fig:simMortRegime}}
\end{figure}

Figure~\ref{fig:simMortRegime} shows the projected mortality rates for individuals aged 35, 50, 65, and 80. The plot displays the $90\%$, $95\%$, and $99\%$ prediction intervals based on $10\ 000$ generated trajectories. The blue dots indicate the observed mortality rates and the blue line represents the fit of the baseline mortality improvement model. The green line shows the in-sample fit of the mortality model constructed by the Dutch Actuarial Association in 2020, i.e.~the AG2020 model, a Li-Lee multi-population mortality model calibrated over the 1970-2019 period \citep{KAG2020}. The green fan charts display the $90\%$, $95\%$, and $99\%$ prediction intervals based on the simulations resulting from the AG2020 model. Our methodology results in wider fan charts than the AG2020 model due to the allowance for future mortality shocks. The median quantile of the simulations resulting from our proposed regime-switching mortality improvement model and the one resulting from the AG2020 model approximately coincide. As a result of the modifications made to the projection strategy described in Appendix~\ref{Appendix:C}, the bottom right panel of Figure~\ref{fig:simMortRegime} demonstrates that the COVID-19 shock has been offset by the year 2022 for individuals aged 80.

\subsection{Sensitivity analysis}\label{sec:scenario.analysis}
To reflect differences in risk appetite, we perform a sensitivity analysis with our proposed toolbox. The inclusion of World War I and World War II has a significant impact on the calibration of the regime-switching model. Figure~\ref{fig:ztxperage} shows that the largest outliers in the empirical distribution of the residuals $Z_{x,t}$ occur during the periods of the two world wars. This consequently leads to a large value for the calibrated volatility parameter $\hat{\sigma}_H$ in the high volatility regime, see Table~\ref{tab:optTheta11}. The wide confidence intervals observed in the projected mortality rates at ages 35 and 50 in Figure~\ref{fig:simMortRegime} confirm this issue. Furthermore, insurance contracts nowadays often include a war clause that excludes coverage during wars. Therefore, the risk modeler may want to exclude these war events when calibrating the regime-switching model. 

We explore a specific scenario that calibrates the improvement model on data from 1875, eliminating war events such as the Crimean and Franco-German wars that happened in the period 1850-1875, see Figure~\ref{fig:intro:overview}. Furthermore, we exclude World War I and II in the calibration of the shock regime by introducing weights in its log-likelihood function. Equation~\eqref{eq:cal:loglikregimeswitch} becomes:
\begin{equation} \label{eq:cal:loglikregimeswitch.weighted}
\begin{aligned} 
l(\Theta) &= \displaystyle \sum_{t\in\mathcal{T}} \nu_t \log f\left(\boldsymbol{z}_t \mid \boldsymbol{z}_{t -1}, ...,\boldsymbol{z}_{t_{\min}}, \Theta\right),
\end{aligned} 
\end{equation}
and we choose $\nu_t = 0$ for $t \in \{1914\text{-}1919,1940\text{-}1946\}$.\footnote{Note that we also exclude the years 1919 and 1946 since we work with a mortality improvement model and the years 1919 and 1946 are as such impacted by the world wars as well.} The calibration of the regime-switching model then occurs in the same way as explained in Section~\ref{sec:cal.regime} with details in Appendix~\ref{Appendix:regime.loglik}. 

Table~\ref{tab:optTheta11.1875} and Figure~\ref{fig:optC1.1875} display the calibrated parameters and age-specific effects $\hat{\mathfrak{B}}_x$ in the regime-switching model for the age ranges 20-59 and 60-85. Excluding war events leads to a significant decrease in $\sigma_H$, the volatility parameter in the high volatility regime. For the 20-59 age range, $\sigma_H$ drops from 2.41 to 0.62, and for the 60-85 age range, it drops from 0.90 to 0.39. The drop in $\sigma_H$ for the 20-59 age range is due to the exclusion of the war victims who died at a young age. Excluding all war events also reduces the probability of transitioning from the low to the high volatility regime, from 0.047 to 0.019 for the 20-59 age range, and from 0.67 to 0.052 for the 60-85 age range. The values of the calibrated volatility and slope parameters in the low volatility regime display only minor changes, which confirms the stability of the calibration strategy.

\begin{tabularx}{\textwidth}{p{0.38\columnwidth}p{0.57\columnwidth}}
   \centering
      \begin{tabular}{lrr}
      \hline
      Parameter  & \multicolumn{1}{c}{20-59} & \multicolumn{1}{c}{60-85} \\
      \hline  
      $p_{1,2}$        & 0.01873  & 0.05239  \\
      $p_{2,1}$        & 0.37218  & 0.78693  \\
      $\sigma_{e_1}$   & 0.12612  & 0.05394  \\
      $\text{slope}_1$ & -0.00171 & 0.00050  \\
      $\sigma_{e_2}$   & 0.18224  & 0.03919  \\
      $\text{slope}_2$ & -0.00344 & -0.00050 \\
      $\mu_H$          & 0.02684  & -0.01665 \\
      $\sigma_H$       & 0.61850  & 0.39284  \\
      \hline
      \end{tabular}  &
\includegraphics[width=0.975\linewidth,valign=m]{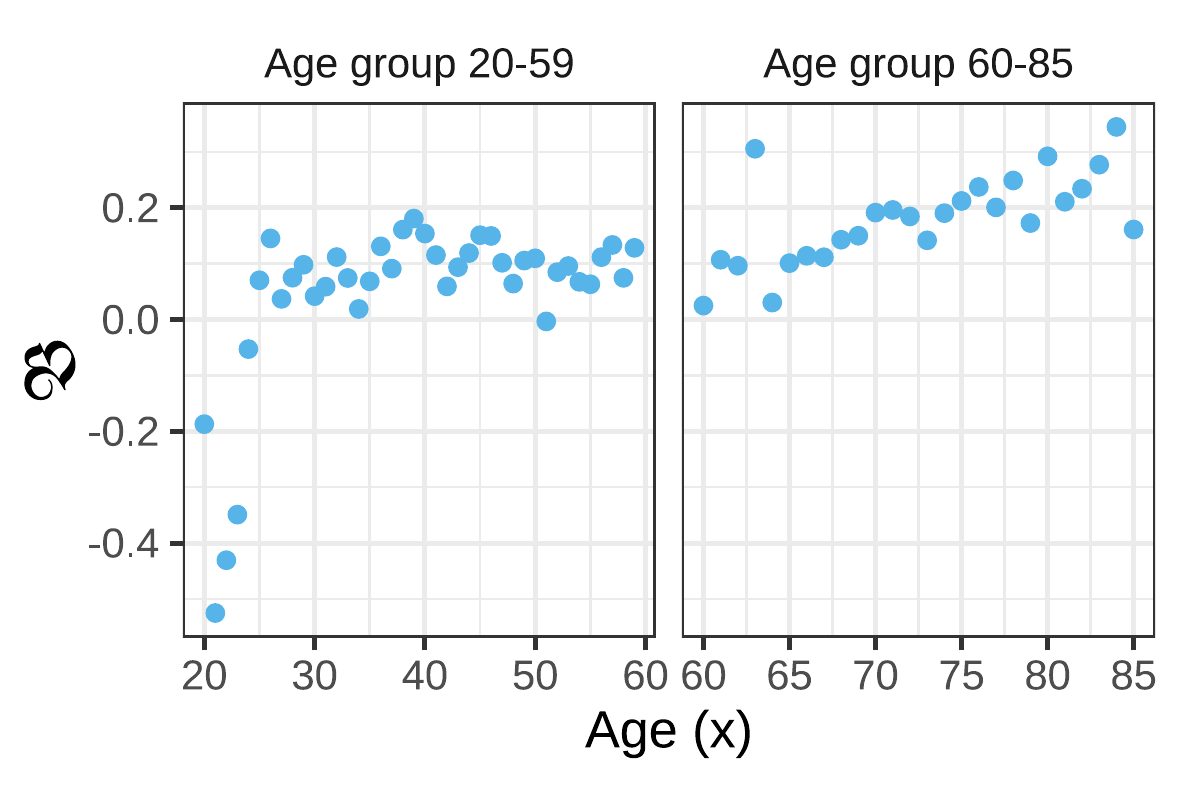} 
\begin{tikzpicture}[overlay]
\node[inner sep=0pt,rotate=90] at (-8.4,0.68)
    {${}_{\boldsymbol{x}}$};
\end{tikzpicture} 
        \\ 
\captionof{table}{Calibrated parameters in the regime-switching model for the two age groups. We exclude the world wars during the calibration. Calibration period 1875-2021, male data.\label{tab:optTheta11.1875}}
&     
    \captionof{figure}{Calibrated age-specific effect $\mathfrak{B}_x$ in the regime-switching model for the two age groups. We exclude the world wars during the calibration. Calibration period 1875-2021, male data.\label{fig:optC1.1875}}
\end{tabularx}

\begin{figure}[ht!]
\centering
\includegraphics[scale=0.62]{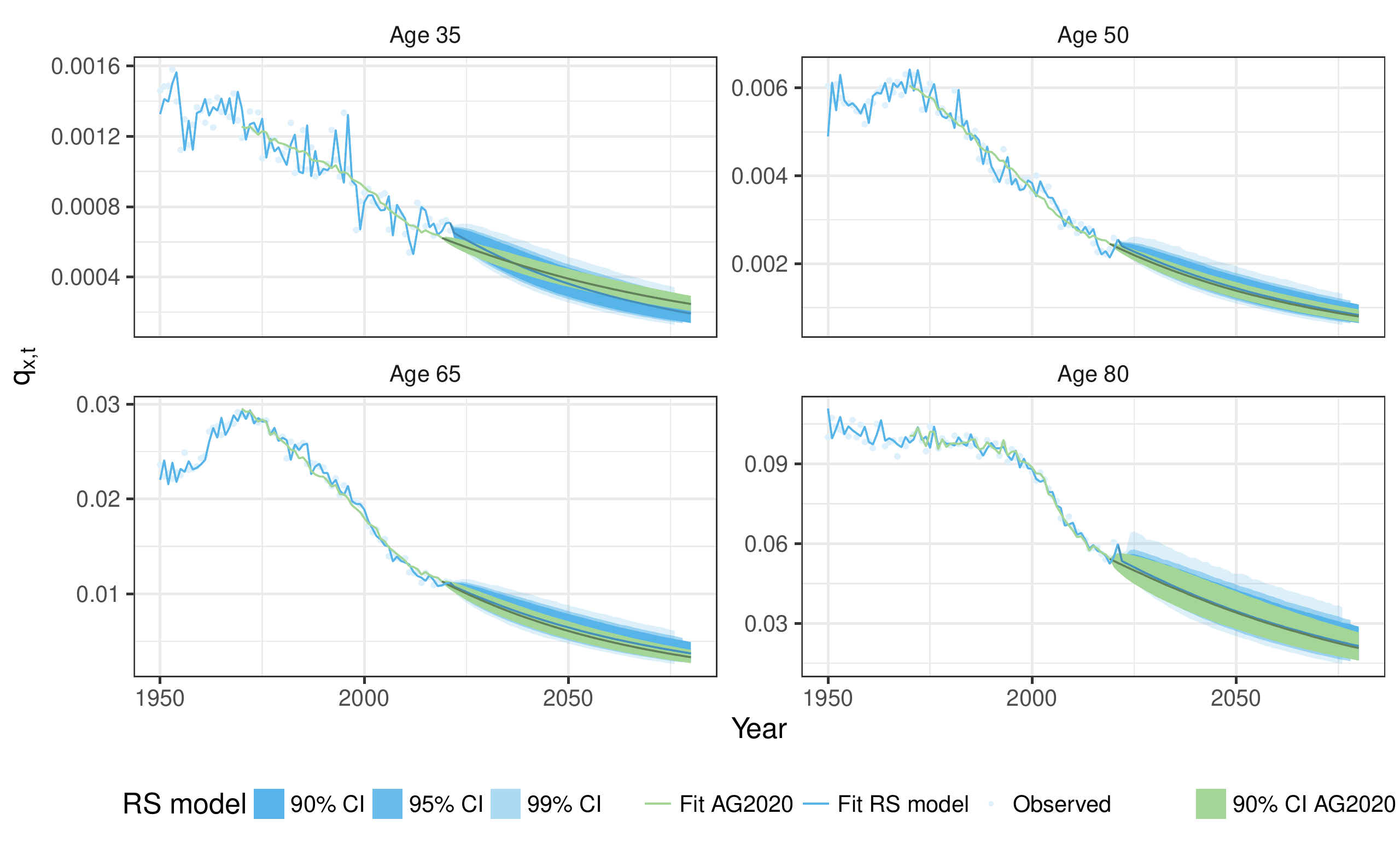}
\caption{The blue fan charts visualize the $90\%$, $95\%$, and $99\%$ prediction intervals based on $10\ 000$ generated trajectories constructed with the mortality improvement model with shock regime. We use calibration period 1875-2021, age range 20-85, and projection period 2022-2080. We further exclude the two world wars during the calibration of the regime-switching model.  The blue shaded dots are the observed mortality rates and the blue line shows the fit of the baseline mortality improvement model. Additionally, the green line shows the in-sample fit of the AG2020 model where we calibrated the Li-Lee model on data from 1970-2019. The green fan charts show the $90\%$, $95\%$, and $99\%$ prediction intervals based on the simulations resulting from this AG2020 model. The median quantiles of the simulations for both methods are shown in dark blue and dark green, respectively.\label{fig:simMortRegime.noww}}
\end{figure}

Figure~\ref{fig:simMortRegime.noww} displays the fan charts that represent the $90\%$, $95\%$, and $99\%$ prediction intervals based on the projected mortality rates spanning the years 2022-2080 for individuals aged 35, 50, 65, and 80. Notably, in contrast to Figure~\ref{fig:simMortRegime}, the variability in the generated mortality rates significantly reduces, but still mostly exceeds the variability from the AG2020 model.

\subsection{Solvency capital requirement} \label{sec:SCR}
We focus on the calculation of the solvency capital requirement (SCR) for the life underwriting risk module \added{in the Solvency II Directive (2009/138/EC)}, and examine the submodules of mortality, longevity, and catastrophe risk. We explore two approaches: the run-off VaR\added{, as discussed by \cite{risks7020058}, \cite{richards2014value} and \cite{levantesi2019application},} and the standard model outlined by \cite{ceiops2010qis}. \added{The standard model, unlike the VaR approach, does not require stochastic modeling of mortality rates. The run-off VaR relies on mortality rates generated from a stochastic model for the future years of the insurance coverage.}\footnote{\added{Alternatively, one could consider the one-year VaR which uses stochastically simulated mortality rates only for the first projection year. One approach is to recalibrate the mortality model based on the outcomes of the first-year simulations, and relies on the projected best-estimate mortality rates of these recalibrated models for the subsequent years \citep{richards2014value}. See, e.g., \cite{risks7020058} for more details and a comparison between the one-year VaR and the run-off VaR.}} \added{We hereby use the scenarios generated from our proposed mortality improvement model with shock regime.} \added{Appendix~\ref{Appendix:SolvencyCapitalRequirement} provides more details about the methodologies and their implementation.} \added{We put focus on the valuation of an immediate life annuity in Section~\ref{sec:SCR.ila} that is solely exposed to longevity risk and on the valuation of a term life insurance product in Section~\ref{sec:SCR.tli}, which is exposed to mortality and catastrophe risk.}

\subsubsection{SCR for an immediate life annuity} \label{sec:SCR.ila}
We consider the valuation of an immediate whole life annuity for an individual aged $x$ in the year 2021, with an annual payout at the end of each year of \euro{10\ 000} until the insured dies. An immediate life annuity is exposed to longevity risk \added{since} the insurer needs to pay out more annuities if the policyholder lives longer than expected. The standard model obtains the SCR associated with longevity risk as the change in Best Estimate Liabilities (BEL) under a longevity shock of $20\%$. In this shock scenario, all future mortality rates are multiplied by a factor of 0.8. The BEL in the year 2021 reflects the present value of future cash flows from the insurer to the policyholder, discounted to the year 2021, and is determined based on best-estimate projections of mortality rates and survival probabilities. To determine the run-off VaR, we use our proposed mortality improvement model \added{and generate} 10\ 000 simulations \added{of} the expected present value of the insurer's liabilities in the year 2021. Subsequently, we calculate the $99.5\%$ quantile of these simulated liabilities and subtract the BEL in the year 2021. Appendix~\ref{Appendix:ILA} further details the calculations of the SCR using both the run-off VaR and the standard model approach.  

\begin{figure}[ht!]
\centering
\includegraphics[scale=0.60]{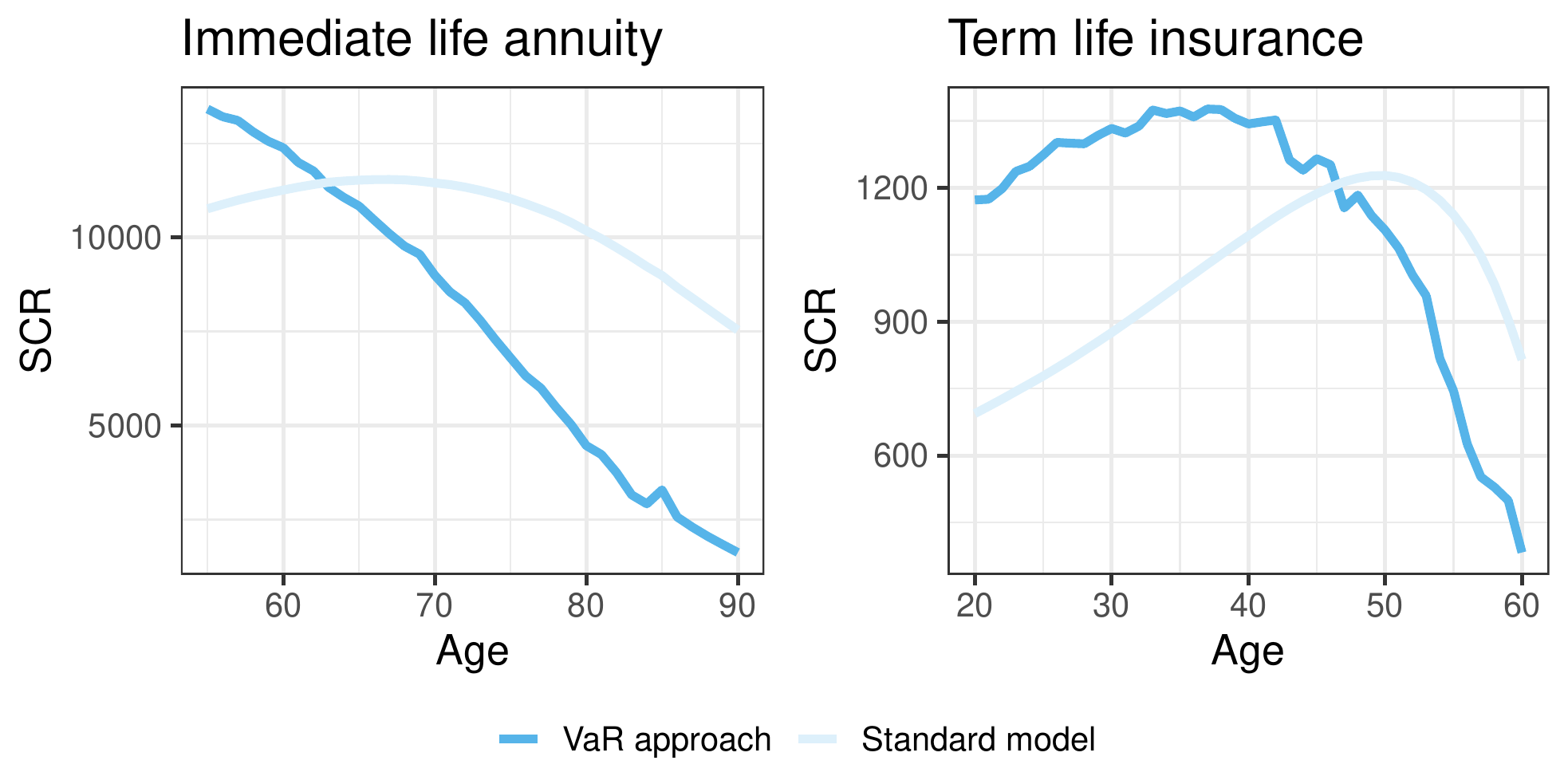}
\caption{In the left panel, the SCR for an immediate life annuity contract paying \euro{10 000} annually and starting in the year 2021 for a person aged 55-90. In the right panel, the SCR for a term life insurance contract with a death benefit of \euro{100 000} starting in the year 2021 for a person aged 20-60\added{, ending in the year} the policyholder reaches the age 65. The dark blue line shows the SCR calculated with the run-off VaR approach and the light blue line is the one calculated with the standard model.\label{fig:SCR.ILA.TLI}}
\end{figure}

The left panel of Figure~\ref{fig:SCR.ILA.TLI} presents the SCR for an immediate life annuity starting in 2021 for individuals aged between 55 and 90. The SCR obtained with the standard model first slightly increases and then decreases faster with age. Moreover, the discrepancy between the SCR derived through the VaR approach and \added{the SCR obtained with} the standard model is more \added{pronounced} for older ages. This phenomenon can be attributed to the fact that the impact of a 20$\%$ decrease in mortality rates on the Best Estimate Liabilities (BEL) \added{is larger for older age mortality rates compared to younger age mortality rates, see Equation~\eqref{eq:ilastm80}}. These findings are consistent with the results obtained in \cite{borger2010deterministic}. Therefore, a longevity portfolio predominantly comprising older individuals may be overly cautious regarding the amount of solvency capital when implementing the standard model. Instead of a fixed\added{, age-independent} longevity shock on \added{all} mortality rates, an age-dependent longevity shock would bring the results obtained with the standard model and the VaR approach closer to each other. \added{The findings from the longevity stress test in \cite{eiopa2018}, who derive age-dependent longevity shocks, support these results. However, EIOPA recommends retaining the age-independent longevity shock in the standard model, citing concerns about added complexity and implementation costs.} 

\subsubsection{Term life insurance}\label{sec:SCR.tli}
We examine a term life insurance policy with \added{a} terminal age \added{of} 65 that is issued to an individual aged $x$ in the year 2021. The insured death benefit equals \euro{150 000} and is payable at the end of the year of death. A term life insurance is exposed to mortality and catastrophe risk as the probability of paying out the death benefit \added{before reaching the term of the contract} increases when the mortality rates increase. The standard model obtains the SCR associated with mortality risk as the change in the BEL under \added{an upward} mortality shock of $15\%$ \citep{ceiops2010qis}. The SCR associated with catastrophe risk, under the standard model, is obtained as the change in BEL under an absolute increase in the mortality rates of the year 2021 by 0.0015. These two separate SCRs are aggregated according to the guidelines in \cite{ceiops2010qis}. Next, similarly to Section~\ref{sec:SCR.ila}, we obtain the SCR with the run-off VaR approach \added{by generating $10\ 000$ simulations for future mortality rates including the occurrence of shocks.} Appendix~\ref{Appendix:TLI} provides the relevant details.

The right panel of Figure~\ref{fig:SCR.ILA.TLI} shows the SCR for a term life insurance issued in the year 2021 to an individual aged between 20 and 60. For younger ages, i.e.~for term insurance policies with a longer term, the SCR obtained with the standard model is lower than the SCR using our proposed mortality model. Conversely, for ages above 45 years, we observe the opposite. \added{This stems from two factors: the increased impact of a $15\%$ mortality shock on the BEL for older age mortality rates in the standard model and the greater uncertainty in mortality rate predictions for younger ages compared to older ages in the VaR approach, see Figure~\ref{fig:simMortRegime.noww}.} In line with our findings in Section~\ref{sec:SCR.ila}, we emphasize the need for incorporating an age-dependent mortality and catastrophe shock in the standard model.

\section{\added{Discussion}}
In this paper, we have developed a framework for a stochastic multi-population mortality projection model that incorporates the occurrence and severity of \added{age-specific} mortality shocks into future mortality projections using a regime-switching model. We showed that the width of the mortality prediction intervals largely depends on which mortality shocks are considered in the calibration procedure of the regime-switching model. 


\added{From a technical point of view, we discussed the importance of} mortality improvement models as an alternative to the classic mortality models used in most academic literature. Overall, the proposed model can assist in modeling future mortality rates, which is crucial for actuaries, life insurers, and risk managers \added{in light of the need to} adhere to the capital requirements imposed by supervisory authorities. In terms of future research directions, a possible avenue is to extend the current two-state regime model to a three-state regime model. In this setting, regime 1 will represent slow mortality improvements, regime 2 fast mortality improvements, and regime 3 will capture mortality shocks. This extension incorporates a possible switch to slower mortality improvements, a behavior that has been observed in the last decade \citep{djeundje2022slowdown}. \added{We demonstrated the calibration of the regime-switching model under the assumption of normally distributed shocks. Future research can focus on investigating alternative distributional assumptions, their impact on the generated scenarios, and the development of a suitable calibration strategy for the regime-switching model.}

\added{From an economic point of view, our paper contributes to the literature on scenario thinking in risk management, which considers different hypothetical future scenarios and their corresponding implications. Hereto, we perform a quantitative impact assessment that investigates the impact of different scenarios related to the inclusion or exclusion of mortality shocks into future mortality rate projections. Using these different scenarios, (re)insurers can assess the corresponding impact on the calculation of, e.g., solvency capital requirements, and tailor the proposed framework to their risk management objectives. Scenario thinking is gaining attention in the actuarial literature. \cite{Gielen2014cancer} consider specific scenarios involving a 100$\%$, 50$\%$, or $20\%$ reduction in the one-year mortality numbers caused by cancer. They then evaluate the corresponding impact on capital requirements in comparison to traditional longevity models such as the Lee-Carter model. \cite{hanika2023covid} studies the combined effects of a financial and a mortality shock on a life insurer's risk during a COVID-19 stress test. He emphasizes the relevance of scenario thinking by evaluating practical risk mitigation strategies for mitigating the short-term impact of pandemic events, offering insights for future pandemic scenarios. \cite{bongiorno2022climate} exemplify how pension schemes can incorporate climate scenarios into their considerations, including aspects like funding strategies and long-term strategic asset allocation. }

{
\bibliography{References}}

\appendix 

\section{Identifiability constraints in a two-factor mortality improvement model} \label{secA:constraintsLC2}
\paragraph{Model specifications.} \cite{renshaw2003lee} introduced a two-factor version of the Lee-Carter model. We adopt a similar specification for the mortality improvement rates, as follows: 
\begin{align} \label{eqA:modelspec}
\log \mu_{x,t} - \log \mu_{x,t-1} = A_x + B_x^{(1)} K_t^{(1)} + B_x^{(2)} K_t^{(2)},
\end{align}
where $\mu_{x,t}$ denotes the force of mortality at age $x \in \mathcal{X}$ and year $t \in \mathcal{T}$. We denote the age-specific effects as $A_x,\: B_x^{(1)}$ and $B_x^{(2)}$ and the period-specific effects as $K_t^{(1)}$ and $K_t^{(2)}$. We introduce vector notations for these age and period effects as follows:
$$ \boldsymbol{B}^{(i)} = \left(B_x^{(i)}\right)_{x\in\mathcal{X}} \in \mathbb{R}^{|\mathcal{X}| \times 1}, \hspace{0.5cm} \boldsymbol{K}^{(i)} = \left(K_t^{(i)}\right)_{t\in\mathcal{T}} \in \mathbb{R}^{|\mathcal{T}|\times 1} \hspace{0.75cm} \text{for} \: i \in \{1,2\},$$
where $|\mathcal{X}|$ and $|\mathcal{T}|$ denote the number of ages and years considered in the age range $\mathcal{X}$ and calibration period $\mathcal{T}$, respectively.

\paragraph{Introducing the parameter constraints.} To make the mortality improvement model in Equation~\eqref{eqA:modelspec} fully identifiable, we need to impose parameter constraints. Straightforward extensions of the identifiability constraints in the Lee-Carter model are: 
\begin{equation} \label{eqA:const1}
||\boldsymbol{B}^{(i)}||_2^2 = \displaystyle \sum_{x\in \mathcal{X}} \left(B_x^{(i)}\right)^2 = 1, \hspace{0.5cm} \sum_{t \in \mathcal{T}} K_t^{(i)} = 0, \hspace{1cm} \text{for} \: i \in \{1,2\},
\end{equation}
where $||\cdot||_2$ denotes the Euclidean norm. We use constraints on the squared age effects $B_x^{(i)}$ for $i \in \{1,2\}$ to enhance numerical stability. However, these constraints alone are not sufficient to fully identify the model in Equation~\eqref{eqA:modelspec}. For instance, some transformations of the age and period effects can still lead to the same model fit:
\begin{equation}\label{eqA:transf.ex}
\begin{aligned} 
A_x &\mapsto A_x, \\
B_x^{(1)} &\mapsto B_x^{(1)} + B_x^{(2)}, \hspace{0.5cm} B_x^{(2)} \mapsto B_x^{(2)}, \\ 
K_t^{(1)} &\mapsto K_t^{(1)}, \hspace{1.6cm} K_t^{(2)} \mapsto K_t^{(2)} - K_t^{(1)}.
\end{aligned}
\end{equation}
We refer to such a transformation as an invariant transformation. \cite{hunt_blake_2020} note that for a Lee-Carter model with $N$ period effects, $N\cdot (N+1)$ parameter constraints are needed. We hence need an additional two constraints to fully identify the mortality improvement model. A convenient and useful set of constraints are the orthogonality constraints on the age and period effects \citep{hunt_blake_2020}, i.e.~we impose:
\begin{align} \label{eqA:const2}
\langle \boldsymbol{B}^{(1)},\boldsymbol{B}^{(2)}\rangle = \displaystyle \sum_{x\in\mathcal{X}} B_x^{(1)} B_x^{(2)} = 0, \hspace{0.5cm} \langle \boldsymbol{K}^{(1)},\boldsymbol{K}^{(2)}\rangle = \displaystyle \sum_{t\in\mathcal{T}} K_t^{(1)} K_t^{(2)} = 0,
\end{align}
where $\langle \cdot, \cdot \rangle$ refers to the Euclidean inner product.  This makes transformations as in Equation~\eqref{eqA:transf.ex} impossible. In the remainder of this appendix, we demonstrate how to convert a set of parameters $(A_x, B_x^{(1)}, B_x^{(2)}, K_t^{(1)}, K_t^{(2)})$ into a set of parameters $(\tilde{A}_x, \tilde{B}_x^{(1)}, \tilde{B}_x^{(2)}, \tilde{K}_t^{(1)}, \tilde{K}_t^{(2)})$ that satisfy the identifiability constraints defined in Equations~\eqref{eqA:const1} and \eqref{eqA:const2}, and result in equivalent model fits, see Equation~\eqref{eqA:modelspec}.

\paragraph{Linear transformations.} \cite{hunt_blake_2020} show that the only possible invariant transformations are of the form:
\begin{equation} \label{eq:invtransf}
\begin{aligned}
\tilde{A}_x &:= A_x - \boldsymbol{B}_x^T \boldsymbol{A}\boldsymbol{D}, \\
\tilde{\boldsymbol{B}}_x^T &= \left[\tilde{B}_x^{(1)}, \tilde{B}_x^{(2)}\right]^T := \boldsymbol{B}_x^T \boldsymbol{A} \\
\tilde{\boldsymbol{K}}_t &= \left[\tilde{K}_t^{(1)}, \tilde{K}_t^{(2)}\right] := \boldsymbol{A}^{-1} \boldsymbol{K}_t + \boldsymbol{D},
\end{aligned}
\end{equation}
where $\boldsymbol{B}_x = (B_x^{(1)}, B_x^{(2)})\in \mathbb{R}^{2\times 1}$, ${}^T$ denotes the transpose operator, $\boldsymbol{A} \in \mathbb{R}^{2\times2}$ is invertible and $\boldsymbol{D} \in \mathbb{R}^{2\times1}$. Note that the above transformations contain six free parameters, i.e.~four entries in $\boldsymbol{A}$ and two entries in $\boldsymbol{D}$. As a result, we need to impose six identifiability constraints. We denote:
\begin{align*}
\boldsymbol{A} = \begin{pmatrix}
\zeta_1 & \eta_1 \\ 
\zeta_2 & \eta_2
\end{pmatrix}, 
\hspace{0.5cm}
\boldsymbol{A}^{-1} = \dfrac{1}{\zeta_1\eta_2-\eta_1\zeta_2}\begin{pmatrix}
\eta_2 & -\eta_1 \\
-\zeta_2 & \zeta_1
\end{pmatrix}, \hspace{0.25cm} \text{and} \hspace{0.25cm}
\boldsymbol{D} =\begin{pmatrix}
\delta_1 \\ \delta_2
\end{pmatrix},
\end{align*}
with $\zeta_i, \eta_i, \delta_i \in \mathbb{R}$ for $i = 1,2$ such that  $\zeta_1 \eta_2 - \eta_1 \zeta_2 \neq 0$. The transformations from Equation~\eqref{eq:invtransf} then result in:
\begin{equation}
\begin{aligned} \label{eqA:transf}
\tilde{B}^{(1)}_x &:= \zeta_1 B_x^{(1)} + \zeta_2 B_x^{(2)}, \hspace{0.5cm} 
\tilde{K}^{(1)}_t := \dfrac{1}{\zeta_1\eta_2-\eta_1\zeta_2} \left(\eta_2 K_t^{(1)} - \eta_1 K_t^{(2)}\right) + \delta_1, \\
\tilde{B}^{(2)}_x &:= \eta_1 B_x^{(1)} + \eta_2 B_x^{(2)}, \hspace{0.5cm}
\tilde{K}^{(2)}_t := \dfrac{1}{\zeta_1\eta_2-\eta_1\zeta_2}\left(-\zeta_2 K_t^{(1)} + \zeta_1 K_t^{(2)}\right) + \delta_2.
\end{aligned}
\end{equation}
We then need to find the parameters $\zeta_i, \eta_i, \delta_i \in \mathbb{R}$ such that the constraints in Equations~\eqref{eqA:const1} and \eqref{eqA:const2} are fulfilled.

\paragraph{Location constraints.} Equation~\eqref{eqA:const1} specifies the location constraint that applies to $\tilde{K}_t^{(i)}$ for $i\in \{1,2\}$. Utilizing the transformation outlined in Equation~\eqref{eqA:transf} in conjunction with the location constraint yields the following result:
\begin{equation}\label{eqA:loc.param}
\begin{aligned} 
\delta_1 &= \dfrac{-1}{\zeta_1 \eta_2 - \eta_1 \zeta_2} \left( \dfrac{\eta_2}{|\mathcal{T}|}\displaystyle \sum_{t\in\mathcal{T}} K_t^{(1)} - \dfrac{\eta_1}{|\mathcal{T}|} \displaystyle \sum_{t\in\mathcal{T}} K_t^{(2)}\right)\\
\delta_2 &= \dfrac{-1}{\zeta_1 \eta_2 - \eta_1 \zeta_2} \left( -\dfrac{\zeta_2}{|\mathcal{T}|}\displaystyle \sum_{t\in\mathcal{T}} K_t^{(1)} + \dfrac{\zeta_1}{|\mathcal{T}|} \displaystyle \sum_{t\in\mathcal{T}} K_t^{(2)}\right).
\end{aligned}
\end{equation}
Next, we substitute the aforementioned expression into the transformed period effects from Equation~\eqref{eqA:transf} and obtain:
\begin{equation}\label{eqA:transfKt1Kt2}
\begin{aligned} 
\tilde{K}_t^{(1)} &= \dfrac{1}{\zeta_1 \eta_2 - \eta_1 \zeta_2} \left[\eta_2 \accentset{\circ}{K}_t^{(1)} - \eta_1 \accentset{\circ}{K}_t^{(2)}\right] \\
\tilde{K}_t^{(2)} &= \dfrac{1}{\zeta_1 \eta_2 - \eta_1 \zeta_2} \left[-\zeta_2 \accentset{\circ}{K}_t^{(1)} + \zeta_1 \accentset{\circ}{K}_t^{(2)}\right],
\end{aligned}
\end{equation}
where for $i \in \{1,2\}$:
\begin{equation*} 
\accentset{\circ}{K}_t^{(i)} = K_t^{(i)} - \frac{1}{|\mathcal{T}|} \sum_{t\in \mathcal{T}} K_t^{(i)}.
\end{equation*}
Note that, by construction, the transformed period effects in Equation~\eqref{eqA:transfKt1Kt2} satisfy the location constraints from Equation~\eqref{eqA:const1} for any $\eta_1$,  $\eta_2$, $\zeta_1$ and $\zeta_2$.

\paragraph{Orthogonal constraints.} We now put focus on the orthogonality constraints in Equation~\eqref{eqA:const2}. Define $\bar{\zeta}_2 = \zeta_2/\zeta_1$ and introduce an identical notation for $\bar{\eta}_2$. Using the notations in Equations~\eqref{eqA:transf} and~\eqref{eqA:transfKt1Kt2}, we obtain:
\begin{align}
\langle \tilde{\boldsymbol{B}}^{(1)}, \tilde{\boldsymbol{B}}^{(2)}\rangle &= \zeta_1 \eta_1 \left\{ ||\boldsymbol{B}^{(1)}||_2^2 + \bar{\zeta}_2 \bar{\eta}_2 ||\boldsymbol{B}^{(2)}||_2^2 + (\bar{\zeta}_2 + \bar{\eta}_2) \langle \boldsymbol{B}^{(1)}, \boldsymbol{B}^{(2)} \rangle \right\} = 0 \label{eqA:eq1}\\
\langle \tilde{\boldsymbol{K}}^{(1)}, \tilde{\boldsymbol{K}}^{(2)}\rangle &= \dfrac{\zeta_1 \eta_1}{(\zeta_1\eta_2-\eta_1\zeta_2)^2} \left\{ -\bar{\zeta }_2\bar{\eta}_2 ||\accentset{\circ}{\boldsymbol{K}}^{(1)}||_2^2 -  ||\accentset{\circ}{\boldsymbol{K}}^{(2)}||_2^2 + (\bar{\zeta}_2 + \bar{\eta}_2) \langle \accentset{\circ}{\boldsymbol{K}}^{(1)}, \accentset{\circ}{\boldsymbol{K}}^{(2)} \rangle \right\} = 0. \label{eqA:eq2}
\end{align}
We multiply Equation~\eqref{eqA:eq1} with $\langle \accentset{\circ}{\boldsymbol{K}}^{(1)}, \accentset{\circ}{\boldsymbol{K}}^{(2)} \rangle$ and Equation~\eqref{eqA:eq2} with $\langle \boldsymbol{B}^{(1)}, \boldsymbol{B}^{(2)} \rangle$ and then subtract both equations to obtain:
\begin{equation} \label{eqA:g1g2rel}
\begin{aligned}
& \overbrace{\big(  \langle \accentset{\circ}{\boldsymbol{K}}^{(1)}, \accentset{\circ}{\boldsymbol{K}}^{(2)} \rangle ||\boldsymbol{B}^{(1)}||_2^2 +  \langle \boldsymbol{B}^{(1)}, \boldsymbol{B}^{(2)} \rangle ||\accentset{\circ}{\boldsymbol{K}}^{(2)}||_2^2\big)}^{:=\: \xi_1} \: + \\ &\hspace{5cm}\bar{\zeta}_2 \bar{\eta}_2 \underbrace{\big( \langle \accentset{\circ}{\boldsymbol{K}}^{(1)}, \accentset{\circ}{\boldsymbol{K}}^{(2)} \rangle ||\boldsymbol{B}^{(2)}||_2^2 +  \langle \boldsymbol{B}^{(1)}, \boldsymbol{B}^{(2)} \rangle ||\accentset{\circ}{\boldsymbol{K}}^{(1)}||_2^2\big)}_{:= \: \xi_2}  = 0.
\end{aligned} 
\end{equation}
We proceed by substituting the above expression for $\bar{\zeta}_2$ into Equation~\eqref{eqA:eq2}, resulting in a quadratic equation in $\bar{\eta}_2$:
\begin{equation}\label{eqA:quadr}
\begin{aligned}
\overbrace{\left[\xi_2 \langle \accentset{\circ}{\boldsymbol{K}}^{(1)}, \accentset{\circ}{\boldsymbol{K}}^{(2)} \rangle \right]}^{:= \: a}\bar{\eta}_2^2 + \overbrace{\left[\xi_1||\accentset{\circ}{\boldsymbol{K}}^{(1)}||_2^2 - \xi_2 ||\accentset{\circ}{\boldsymbol{K}}^{(2)}||_2^2\right]}^{:= \: b}\bar{\eta}_2 + \overbrace{\left[-\xi_1 \langle \accentset{\circ}{\boldsymbol{K}}^{(1)}, \accentset{\circ}{\boldsymbol{K}}^{(2)} \rangle \right]}^{:= \: c} = 0.
\end{aligned}
\end{equation}
The corresponding discriminant equals:
\begin{align*}
D &:= b^2 - 4ac  \\
&= \left[\xi_1||\accentset{\circ}{\boldsymbol{K}}^{(1)}||_2^2 - \xi_2 ||\accentset{\circ}{\boldsymbol{K}}^{(2)}||_2^2\right]^2 + 4 \xi_1 \xi_2 \langle \accentset{\circ}{\boldsymbol{K}}^{(1)}, \accentset{\circ}{\boldsymbol{K}}^{(2)} \rangle^2.
\end{align*}
To prove that this discriminant is always positive, we consider two cases based on the sign of $\xi_1\xi_2$. In the case $\xi_1\xi_2 \geq 0$, it is evident that $D \geq 0$. On the other hand, if $\xi_1\xi_2 < 0$, we employ the Cauchy-Schwarz inequality and derive:
\begin{align*}
D &\geq \left[\xi_1||\accentset{\circ}{\boldsymbol{K}}^{(1)}||_2^2 - \xi_2 ||\accentset{\circ}{\boldsymbol{K}}^{(2)}||_2^2\right]^2 + 4 \xi_1 \xi_2 ||\accentset{\circ}{\boldsymbol{K}}^{(1)}||_2^2||\accentset{\circ}{\boldsymbol{K}}^{(2)}||_2^2 \\
&= \xi_1^2 ||\accentset{\circ}{\boldsymbol{K}}^{(1)}||_2^4 + \xi_2^2 ||\accentset{\circ}{\boldsymbol{K}}^{(2)}||_2^4 + 2 \xi_1\xi_2 ||\accentset{\circ}{\boldsymbol{K}}^{(1)}||_2^2 ||\accentset{\circ}{\boldsymbol{K}}^{(2)}||_2^2 \\
&= \left(\xi_1 ||\accentset{\circ}{\boldsymbol{K}}^{(1)}||_2^2 + \xi_2 ||\accentset{\circ}{\boldsymbol{K}}^{(2)}||_2^2 \right)^2 \\
&\geq 0.
\end{align*}
As a result, the quadratic equation in~\eqref{eqA:quadr} has two real solutions:
\begin{align*}
\bar{\eta}_2 \in \left\{\dfrac{-b - \sqrt{D}}{2a}, \dfrac{-b + \sqrt{D}}{2a}\right\}.
\end{align*}
By using the relation in Equation~\eqref{eqA:g1g2rel}, we deduce the expression for $\bar{\zeta}_2$:
\begin{align*}
\bar{\zeta}_2  = -\dfrac{\xi_1}{\xi_2 \bar{\eta}_2}.
\end{align*}
Next, we prove that the solution sets of $\bar{\zeta}_2$ and $\bar{\eta}_2$ coincide. For $\bar{\eta}_2 = (-b - \sqrt{D})/2a$, we obtain:
\begin{align*}
\bar{\zeta}_2 &= \dfrac{2a\xi_1}{\xi_2\left(b+\sqrt{D}\right)} \\
&= \dfrac{2 \xi_1 \xi_2 \langle \accentset{\circ}{\boldsymbol{K}}^{(1)}, \accentset{\circ}{\boldsymbol{K}}^{(2)} \rangle \left(b - \sqrt{D}\right)}{\xi_2 \left(b^2 - D^2\right)} \\
&= \dfrac{-2 \xi_1 \xi_2 \langle \accentset{\circ}{\boldsymbol{K}}^{(1)}, \accentset{\circ}{\boldsymbol{K}}^{(2)} \rangle \left(b - \sqrt{D}\right)}{4 \xi_1 \xi_2^2 \langle \accentset{\circ}{\boldsymbol{K}}^{(1)}, \accentset{\circ}{\boldsymbol{K}}^{(2)} \rangle^2} \\
&= \dfrac{-b + \sqrt{D}}{2\xi_2\langle \accentset{\circ}{\boldsymbol{K}}^{(1)}, \accentset{\circ}{\boldsymbol{K}}^{(2)} \rangle} \\
&= \dfrac{-b + \sqrt{D}}{2a}.
\end{align*}
Similarly, for $\bar{\eta}_2 = (-b + \sqrt{D})/2a$, we can prove that $\bar{\zeta}_2 = (-b - \sqrt{D})/2a$ and vice versa. 

\paragraph{Scaling constraints.} In the last step, we determine $\zeta_1$, $\eta_1$ such that the constraints on the age effects in Equation~\eqref{eqA:const1} are fulfilled. We obtain:
\begin{align*}
1 = \displaystyle \sum_{x\in \mathcal{X}} \left[\widetilde{B}_x^{(1)}\right]^2 = \sum_{x \in \mathcal{X}} \left(\zeta_1 B_x^{(1)} + \zeta_1 \bar{\zeta}_2 B_x^{(2)}\right)^2 = \zeta_1^2 \sum_{x \in \mathcal{X}} \left(B_x^{(1)} + \bar{\zeta}_2 B_x^{(2)}\right)^2.
\end{align*}
As a result and using a similar reasoning for the constrained on the squared $\widetilde{B}_x^{(2)}$, we deduce:
$$  \zeta_1 = \dfrac{\text{sign}\left\{\displaystyle \sum_{x \in \mathcal{X}} \left(B_x^{(1)} + \bar{\zeta}_2 B_x^{(2)}\right)\right\}}{\sqrt{\displaystyle \sum_{x \in \mathcal{X}} \left(B_x^{(1)} + \bar{\zeta}_2 B_x^{(2)}\right)^2}},
\hspace{0.5cm} \eta_1 = \dfrac{\text{sign}\left\{\displaystyle \sum_{x \in \mathcal{X}} \left(B_x^{(1)} + \bar{\eta}_2 B_x^{(2)}\right)\right\}}{\sqrt{\displaystyle \sum_{x \in \mathcal{X}} \left(B_x^{(1)} + \bar{\eta}_2 B_x^{(2)}\right)^2}},  $$
where we include the $\text{sign}(\cdot)$ term to have a unique value for $\zeta_1$ and $\zeta_2$. Note that the solution set of $\zeta_1$ and $\eta_1$ again coincide. Since, e.g., $\zeta_2 = \zeta_1 \bar{\zeta}_2$, we can easily obtain the (coinciding) solution sets of $\zeta_2$ and $\eta_2$. These results also define the (coinciding) solution set of $\delta_1$ and $\delta_2$ in Equation~\eqref{eqA:loc.param}.

In conclusion, we have derived two different sets of parameter transformations that lead to the same model fit and that satisfy the imposed location, scaling, and orthogonality constraints. However, because of the coinciding solution sets of $\eta_1$ and $\zeta_1$, $\eta_2$ and $\zeta_2$ and $\delta_1$ and $\delta_2$, we obtain that the parameters $\left\{\tilde{B}_x^{(1)},\tilde{K}_t^{(1)},\tilde{B}_x^{(2)},\tilde{K}_t^{(2)}\right\}$ in the first transformation coincides with the parameters $\left\{\tilde{B}_x^{(2)},\tilde{K}_t^{(2)},\tilde{B}_x^{(1)},\tilde{K}_t^{(1)}\right\}$ in the second transformation. This correspondence does not pose a particular issue as we can simply reassign the labels of the age and period effects. One viable approach to ensure uniqueness entails selecting the parameter solution such that the range of $\tilde{B}_x^{(1)}$ is larger than the one of $\tilde{B}_x^{(2)}$.

\section{Indirect estimation of the mortality improvement model} \label{appendix:indirect.estimation} 
We use the methodology proposed by \cite{hunt2021mortality} for indirect estimation of the baseline mortality improvement model. Hereto, we first transform the baseline mortality improvement model, as formulated in Equation~\eqref{eq:modelspec:baseline}, into a baseline mortality model for the logarithmic force of mortality. We iteratively calculate:
\begin{equation*}
\begin{aligned}
\log \mu_{x,t}^{(c)} &= \log \mu_{x,t-1}^{(c)} + A_x + \displaystyle \sum_{i=1}^m B_x^{(i)} K_t^{(i)} + \displaystyle \sum_{j=1}^l \beta_x^{(j,c)} \kappa_t^{(j,c)} \\
&= \log \mu_{x,t-2}^{(c)} + 2A_x + \displaystyle \sum_{i=1}^m B_x^{(i)} \left(K_{t-1}^{(i)} + K_t^{(i)}\right) + \displaystyle \sum_{j=1}^l \beta_x^{(j,c)} \left(\kappa_{t-1}^{(j,c)} + \kappa_t^{(j,c)}\right) \\
&= \ldots \\
&= \log \mu_{x,t_{\min}}^{(c)} + \left(t - t_{\min}\right) A_x + \displaystyle \sum_{i=1}^m B_x^{(i)} \left(\displaystyle \sum_{\tau = t_{\min}+1}^{t} K_{\tau}^{(i)}\right) + \displaystyle \sum_{j=1}^l \beta_x^{(j,c)} \left(\displaystyle \sum_{\tau = t_{\min}+1}^{t} \kappa_{\tau}^{(j,c)}\right),
\end{aligned}
\end{equation*}
for $x \in \mathcal{X}$ and $t \in \mathcal{T}$. We introduce the following notation:
\begin{align} \label{eqA:Ltlt.transf}
L_t^{(i)} = \displaystyle \sum_{\tau = t_{\min}+1}^{t} K_{\tau}^{(i)}, \hspace{0.5cm} \lambda_t^{(j,c)}  = \displaystyle \sum_{\tau = t_{\min}+1}^{t} \kappa_{\tau}^{(j,c)}.
\end{align}
We then obtain an equivalent specification for the baseline mortality improvement model in terms of the logarithmic force of mortality:
\begin{align} \label{eqA:tradmodel}
\log \mu_{x,t}^{(c)} = \log \mu_{x,t_{\min}}^{(c)} + \left(t - t_{\min}\right) A_x + \displaystyle \sum_{i=1}^m B_x^{(i)} L_t^{(i)} + \displaystyle \sum_{j=1}^l \beta_x^{(j,c)} \lambda_t^{(j,c)},
\end{align}
where $L_{t_{\min}}^{(i)} := 0$ for every $i \in \{1,...,m\}$ and $\lambda_{t_{\min}}^{(j,c)} := 0$ for every $j \in \{1,...,l\}$. 

Additionally, we need to impose identifiability constraints to make the mortality model specification in Equation~\eqref{eqA:tradmodel} fully identifiable.  Remark that the constraints formulated in Equations~\eqref{eq:constr1} and~\eqref{eq:constr2} serve as identifiability constraints in the baseline mortality improvement model. Consequently, we must translate these constraints in terms of the redefined model specification outlined in Equation~\eqref{eqA:tradmodel}. Using Equation~\eqref{eqA:Ltlt.transf}, we translate the constraints in Equation~\eqref{eq:constr1} into:
\begin{equation}\label{eq:constr1.1}
\begin{aligned} 
\displaystyle \sum_{x\:\in\: \mathcal{X}} \left(B_x^{(i)}\right)^2 &= 1, \hspace{0.5cm} L_{t_{\max}}^{(i)} = 0, \hspace{1cm} &\text{for} \: i \in \{1,\ldots,m\} \\
\displaystyle \sum_{x\:\in \:\mathcal{X}} &\left(\beta_x^{(j,c)}\right)^2 = 1,  &\text{for} \: j \in \{1,\ldots,l\}.
\end{aligned}
\end{equation}
The constraints in Equation~\eqref{eq:constr2} (for $m = 2$) translate into:
\begin{align}  \label{eq:constr2.1}
\displaystyle \sum_{x\:\in\:\mathcal{X}} B_x^{(1)} B_x^{(2)} = 0, \hspace{0.5cm}\displaystyle \sum_{t = t_{\min} + 1}^{t_{\max}} \left(L_t^{(1)} - L_{t-1}^{(1)}\right)\left(L_t^{(2)} - L_{t-1}^{(2)}\right) = 0,
\end{align}
with similar constraints on the country-specific parameters $\beta_x^{(j,c)}$ and $\lambda_t^{(j,c)}$ for $j\in\{1,2\}$. 

\section{Outlier detection using the Mahalanobis distance} \label{appendix:outlierdetection}
We perform outlier detection using the Mahalanobis distance and apply a robust Minimum Covariance Determinant (MCD) estimate of the location and scale. This can be done in a three-step procedure:
\begin{enumerate}
\item Compute the robust MCD estimate of location $\hat{\boldsymbol{\mu}}^{\text{rob}}$ and scale $\hat{\boldsymbol{\Sigma}}^{\text{rob}}$ for the $m$-dimensional time series of remainder components $ \mathcal{R} = \left\{\hat{\boldsymbol{R}}_t := \left(\hat{R}_t^{(1)}, ..., \hat{R}_t^{(m)} \right) \mid t \in \mathcal{T}\right\}$, i.e.~we use the \texttt{CovMcd} function from the \texttt{rrcov} package in \texttt{R}.
\item Compute the robust distances for each observation $\hat{\boldsymbol{R}}_t$:
\begin{align*}
d_t = \sqrt{\left(\hat{\boldsymbol{R}}_t - \hat{\boldsymbol{\mu}}^{\text{rob}}\right)^T \left(\hat{\boldsymbol{\Sigma}}^{\text{rob}}\right)^{-1} \left(\hat{\boldsymbol{R}}_t - \hat{\boldsymbol{\mu}}^{\text{rob}}\right)}.
\end{align*}
\item We identify year $t$ as an outlier whenever the robust distance $d_t$ exceeds the square root of the $99 \%$ quantile of the chi-square distribution with $m$ degrees of freedom, i.e.~$\sqrt{\scalebox{1.5}{$\chi$}_{m,0.99}^2}$. 
\end{enumerate}

\section{A weighted maximum likelihood estimation method for \\ projecting common and country-specific period effects} \label{Appendix:weightedLL}
We provide more details on the approach used for estimating the multivariate time series model for the joint vector of common and country-specific period effects, as formulated in Equation~\eqref{eq:timedynformula}. We implement a weighted variant of maximum likelihood estimation. Hereto, we minimize the following objective function:
\begin{align} \label{eq:objftimeseries}
- \displaystyle \sum_{t=1851}^{t_{\max}} \omega_t \log \left(\phi\left(\boldsymbol{\mathcal{K}}_t \mid \boldsymbol{c}, \boldsymbol{\Sigma}_w\right)\right),
\end{align}
where $\boldsymbol{\mathcal{K}}_t$ is the four-dimensional vector of period effects, $w_t = \gamma^{t_{\max} - t}$ are geometrically decaying weights and $\gamma$ is the rate of decay.\footnote{It should be noted that when the value of the parameter $\gamma$ is set equal to one, the resulting procedure is equivalent to traditional maximum likelihood estimation.} Furthermore, $\phi(\:\cdot\mid\boldsymbol{c},\boldsymbol{\Sigma}_w)$ denotes the four-dimensional Gaussian density function with mean $\boldsymbol{c}$ and covariance matrix $\boldsymbol{\Sigma}_w$.  The decay parameter $\gamma$ needs to be selected properly since it will have a major impact on the final projections. We follow an approach similar to \cite{mittnik2000conditional} and select the parameter $\gamma$ that maximizes the average of the log-likelihood values associated with the one-step ahead projections since the year 1900 onwards,\footnote{We start the one-step ahead projections at the year 1901, to have sufficient data available to calibrate the time series models, i.e.~at least 50 years.} i.e.~we maximize:
\begin{align}\label{eq:gamma.select}
\hat{\gamma} = \argmax \limits_{\gamma} \frac{1}{121} \sum_{t=1900}^{2020} \log \left(\hat{\phi}_{t+1 \mid t}\left(\boldsymbol{\mathcal{K}}_{t+1} \mid \gamma, \hat{\boldsymbol{c}}, \hat{\boldsymbol{\Sigma}}_w, \boldsymbol{\mathcal{K}}_{t}, \ldots, \boldsymbol{\mathcal{K}}_{1851}\right)\right).
\end{align}
Here, we calculate the one-step ahead projection by estimating the time series models up to the year $t$, followed by projecting the calibrated model to year $t+1$. Hence, the time series parameters $\hat{\boldsymbol{c}}$ and $\hat{\boldsymbol{\Sigma}}_w$ in Equation~\eqref{eq:gamma.select} are re-estimated for each value of $t$. The optimally selected value for $\gamma$ in the case study of Section~\ref{sec:case} equals $0.943$.

\section{Calibration and projection the regime-switching model}\label{Appendix:regime.loglik}
Our proposed calibration strategy builds upon prior work of \cite{HAINAUT2012236}. However, we modify his methodology to account for the age-specific mortality shock effect. Furthermore, to ensure that the Markov chain remains in the high volatility regime for at least two subsequent years, we introduce a memory state. This is necessary because a mortality shock in a given year $t$ leads to outlying values in the mortality improvement rates for both year $t$ and year $t+1$. In Section~\ref{Appendix:C}, we discuss projections from our regime-switching model. Specifically, we generate artificial mortality shocks in a manner that aims to reflect real-world scenarios.

\subsection{Calibration} \label{Appendix:calibrationRS}
\subsubsection{Setting up the log-likelihood}  \label{Appendix:setuplogl}
The log-likelihood of the regime-switching model equals:
\begin{equation} \label{eq:cal:loglikregimeswitcha}
\begin{aligned} 
l(\Theta) &= \log f\left(\boldsymbol{z}_{t_{\min}}, \boldsymbol{z}_{t_{\min+1}}, ..., \boldsymbol{z}_{t_{\max}}\mid \Theta \right) \\
&= \displaystyle \sum_{t\in\mathcal{T}} \log f\left(\boldsymbol{z}_t \mid \boldsymbol{z}_{t -1}, ...,\boldsymbol{z}_{t_{\min}}, \Theta\right),
\end{aligned} 
\end{equation}
with $\mathcal{T} := \left\{t_{\min}, t_{\min} + 1, \ldots, t_{\max}\right\}$ the calibration period, $\boldsymbol{z}_t$ the realized vector of residuals $(z_{x,t})_{x\in\mathcal{X}}$ from the calibrated baseline mortality improvement model and $f(\cdot)$ the density function of $\boldsymbol{z}_t$. \cite{Hamiltonfilter} proves that we can obtain the involved conditional probabilities recursively as:
\begin{equation} \label{eq:cal:recursion}
\begin{aligned}
f\left(\boldsymbol{z}_t \mid \boldsymbol{z}_{t -1}, ...,\boldsymbol{z}_{t_{\min}}, \Theta\right) &= \displaystyle \sum_{i \in \Omega} \sum_{j \in \Omega} \Big\{\mathbb{P}\left(\rho_{t-1} = i \mid \boldsymbol{z}_{t -1}, ...,\boldsymbol{z}_{t_{\min}}, \Theta\right)\cdot \mathbb{P}\left(\rho_t = j \mid \rho_{t-1} = i, \Theta\right) \cdot \\ & \hspace{4cm} f\left(\boldsymbol{z}_t \mid \rho_{t} = j, \Theta\right)\Big\},
\end{aligned}
\end{equation}
where:
\begin{itemize}
\item[\sbt] $\Omega$ is the state space of the Markov chain underlying the regime-switching model,
\item[\sbt] $\mathbb{P}\left(\rho_{t-1} = i \mid \boldsymbol{z}_{t -1}, ...,\boldsymbol{z}_{t_{\min}}, \Theta\right)$ is the probability of being in regime state $i$ at year $t-1$, conditionally on the residuals up to and including year $t-1$,
\item[\sbt] $\mathbb{P}\left(\rho_t = j \mid \rho_{t-1} = i, \Theta\right)$ is the transition probability from regime state $i$ to regime state $j$ in one time step,
\item[\sbt] $f\left(\boldsymbol{z}_t \mid \rho_{t} = j, \Theta\right)$ is the probability density of observing $\boldsymbol{z}_t$ conditionally on being at regime state $j$ at year $t$.
\end{itemize}
First, the conditional probability of being in regime state $i$ at year $t-1$ can be calculated recursively using the law of conditional probabilities and by applying the formula in Equation~\eqref{eq:cal:recursion} at time $t-1$:
\begin{equation} \label{eq:cal:regimprob}
\begin{aligned}
&\mathbb{P}\left(\rho_{t-1} = i \mid \boldsymbol{z}_{t -1}, ...,\boldsymbol{z}_{t_{\min}}, \Theta\right) = \dfrac{f\left(\rho_{t-1} = i, \boldsymbol{z}_{t-1} \mid \boldsymbol{z}_{t -2}, ...,\boldsymbol{z}_{t_{\min}}, \Theta\right)}{f\left(\boldsymbol{z}_{t-1} \mid \boldsymbol{z}_{t-2}, ...,\boldsymbol{z}_{t_{\min}}, \Theta\right)} \\
&\hspace{1cm}= \dfrac{\displaystyle \sum_{h \in \Omega}  \mathbb{P}\left(\rho_{t-2} = h \mid \boldsymbol{z}_{t-2}, ...,\boldsymbol{z}_{t_{\min}}, \Theta\right)\cdot \mathbb{P}\left(\rho_{t-1} = i \mid \rho_{t-2} = h, \Theta\right) \cdot f\left(\boldsymbol{z}_{t-1} \mid \rho_{t-1} = i, \Theta\right)}{f\left(\boldsymbol{z}_{t-1} \mid \boldsymbol{z}_{t-2}, ...,\boldsymbol{z}_{t_{\min}}, \Theta\right)}.
\end{aligned}
\end{equation}
Second, the transition probabilities $\mathbb{P}\left(\rho_t = j \mid \rho_{t-1} = i, \Theta\right)$ are time-invariant parameters in the Markov chain of the regime-switching model and we denote them as $p_{ij}$ with $i,j \in \Omega$. Sections~\ref{Appsec:memory} and~\ref{Appendix:startingvalues} further elaborate on the transition and stationary probabilities in this Markov chain. Third, we assume that $\boldsymbol{z}_t \mid \rho_{t} = j, \Theta$ follows a multivariate normal distribution with mean and variance depending on the occupied state $j \in \Omega$ of the Markov chain at time $t$, see Section~\ref{Appendix:cal.regime.two.state.markov.chain} for more details.

\subsubsection{Modification of the transition matrix for a two-state Markov chain with memory} \label{Appsec:memory}
Assume that the transition probability matrix of a two-state Markov chain $(\rho_t)_t$ with $t \in \mathcal{T}$ and with state space $\Omega = \{1,2\}$ is given by:
\begin{align*}
\boldsymbol{P} = \begin{pmatrix}
p_{11} & p_{12} \\
p_{21} & p_{22}
\end{pmatrix},
\end{align*}
where $p_{ij}$ for $i,j \in \{1,2\}$ is the probability of moving from state $i$ to state $j$ and where $p_{11} + p_{12} = 1$ and $p_{21} + p_{22} = 1$. Following Section~\ref{sec:regimeswitch}, we interpret state `1' as the low volatility state (LVS) and state `2' as the high volatility state (HVS). As discussed in Section~\ref{sec:cal.regime}, we want to ensure that once the Markov chain enters the HVS, i.e.~state 2, it stays there for at least two years. One possible way to account for this is by creating a  so-called memory state. Hereto we introduce the state ($X$,$1$), which refers to state 1 as the current state and the previous state can either be state 1 or 2, labeled as state $X$.\footnote{The first entry of the introduced vector notation for labeling the states denotes the state in the previous year, whereas the second entry denotes the current state.} State (1,2) represents state 2 as the current state with state 1 as the previous state and state (2,2) represents the current state 2 with previous state 2. So the state space now equals: 
$$ \Omega' = \{(X,1), (1,2), (2,2) \}.$$
Note that the state $(X,1)$ can also be decomposed in states $(1,1)$ and $(2,1)$. However, for our purposes this is not necessary as we only need to stay at least two periods in the HVS once entered. Based on this newly defined state space $\Omega'$, we introduce the following transition matrix:
\begin{align} \label{App:eq:transmat}
\boldsymbol{P}' = \begin{pmatrix}
p_{11} & p_{12} & 0 \\
0 & 0 & 1 \\
p_{21} & 0 & p_{22} 
\end{pmatrix}.
\end{align}
We can interpret it in the following way. The probability of staying in the LVS irrespective of its memory, i.e.~state $(X,1)$, equals $p_{11}$. The probability of going from state $(X,1)$ to $(1,2)$ equals $p_{12}$. Further, the probability of going from state $(X,1)$ to state $(2,2)$ equals $0$ because we can only move to state $(2,2)$ whenever the current state equals $2$. The second row in the matrix indicates that the probability of going from state $(1,2)$ to state $(2,2)$ equals $1$. This forces the Markov chain to be in the HVS for at least two periods. Row 3 of the transition matrix can be interpreted in a similar way. Note that we maintain the same transition probabilities as in the transition matrix $P$, and only add a memory to state $2$ of the Markov chain. 

\subsubsection{Starting values} \label{Appendix:startingvalues}
To launch the recursion in Equation~\eqref{eq:cal:recursion}, we need starting values for the probability of being in regime state $i$ at time $t_{\min}$ to calculate :
\begin{align}\label{eq:cal:startfztmin}
f\left(\boldsymbol{z}_{t_{\min}} \mid \Theta\right) = \displaystyle \sum_{i \in \Omega'} f\left(\boldsymbol{z}_{t_{\min}} \mid \rho_{t_{\min}} = i, \Theta\right) \cdot \mathbb{P}\left(\rho_{t_{\min}} = i\mid \Theta\right).
\end{align}
In this particular case, $|\Omega'| = 3$ equals three, i.e.~the dimension of the transition matrix $\boldsymbol{P}'$. We follow the approach of \cite{HAINAUT2012236} and \cite{maryhardyregime}, and take as starting values $\mathbb{P}(\rho_{t_{\min}} = i \mid \Theta)$ the stationary probabilities $\boldsymbol{\pi} = \left(\pi_{1}, \pi_2,\pi_3\right)^T$ of the Markov chain. The long-term, invariant probability of being in the low volatility state is $\pi_1$ and of being in the high volatility state is $\pi_2 + \pi_3$. To compute these stationary probabilities, we solve the system of equations:
$$ \boldsymbol{\pi} \boldsymbol{P}' = \boldsymbol{\pi},$$
which is equivalent to solving:
\begin{align*}
\begin{cases}
p_{11} \pi_1 + p_{21} \pi_3 \\
p_{12} \pi_1 = \pi_2 \\
\pi_2 + p_{22} \pi_3 = \pi_3,
\end{cases}
\end{align*}
subject to the constraint that $\pi_1 + \pi_2 + \pi_3 = 1$. Solving the above system of equations, leads to the following stationary distribution of the Markov chain:
\begin{align*}
\begin{cases}
\pi_1 = \dfrac{p_{21}}{p_{12} + p_{21} + p_{12}p_{21}} \vspace{0.25cm}\\
\pi_2 = \dfrac{p_{12}p_{21}}{p_{12} + p_{21} + p_{12}p_{21}} \vspace{0.25cm}\\
\pi_3 = \dfrac{p_{12}}{p_{12} + p_{21} + p_{12}p_{21}} .
\end{cases}
\end{align*}
To launch the recursion, we use:
\begin{align*}
f\left(\boldsymbol{z}_{t_{\min}} \mid \Theta\right) = \displaystyle \sum_{i \in \Omega'} \pi_i \cdot f\left(\boldsymbol{z}_{t_{\min}} \mid \rho_{t_{\min}} = i, \Theta\right).
\end{align*}

\subsubsection{Calibrating the regime-switching model using a two-state Markov chain with memory} \label{Appendix:cal.regime.two.state.markov.chain}
Following Equation~\eqref{eq:cal:transdens}, we impose a multivariate normal distribution to the time series vector of residuals $\boldsymbol{Z}_t$. We have:\footnote{We omit the superscript $(c)$ related to the country of interest $c$ for notational purposes.}
\begin{equation}\label{App:eq:cal:transdens}
\begin{aligned} 
\boldsymbol{Z}_t &\sim 
\begin{cases}
\mathcal{N}_{|\mathcal{X}|}\left(\boldsymbol{0},\: \sigma_e^2(x,t) \cdot \boldsymbol{I}_{|\mathcal{X}|}\right) \hspace{3cm} &\text{if} \: \rho_t = (X,1) \\
\mathcal{N}_{|\mathcal{X}|}\left(\boldsymbol{\mathfrak{B}}\mu_H, \: \boldsymbol{\mathfrak{B}} \boldsymbol{\mathfrak{B}}^T \sigma_H^2 + \sigma_e^2(x,t) \cdot \boldsymbol{I}_{|\mathcal{X}|}\right)  &\text{if} \: \rho_t \in \{(1,2), (2,2)\},
\end{cases}
\end{aligned}
\end{equation}
with $\boldsymbol{I}_{|\mathcal{X}|}$ the identity matrix of size $|\mathcal{X}|$ and $\mathcal{N}_{|\mathcal{X}|}(\cdot,\cdot)$ the $|\mathcal{X}|$-dimensional Gaussian distribution. In other words, in the HVS, irrespective of the previous state, we impose the multivariate normal distribution with the extra volatility term $\sigma_H$ to capture the mortality shocks. Equation~\eqref{App:eq:cal:transdens} defines the densities $f(\boldsymbol{z}_t \mid \rho_t = j, \Theta)$ used in Equation~\eqref{eq:cal:recursion}. By combining this with the transition probabilities in $\boldsymbol{P}'$ from Equation~\eqref{App:eq:transmat}, we obtain the optimal parameter vector $\hat{\Theta}$ by optimizing the log-likelihood in Equation~\eqref{eq:cal:loglikregimeswitcha}.

\subsubsection{Implementation details of the calibration of the regime-switching model} \label{Appendix:optimizelogl}
For the case study in Section~\ref{sec:case}, we employ a two-step procedure to optimize the log-likelihood in Equation~\eqref{eq:cal:loglikregimeswitcha}. This two-step optimization procedure is beneficial for complex or large-scale optimization problems, as it facilitates the search for a global, optimal solution. Let us denote the following parameter vectors:
\begin{align*}
\boldsymbol{\Theta}_{1,a} &:= \left(p_{12}^{(a)}, p_{21}^{(a)}, \sigma_{e_1}^{(a)}, \text{slope}_1^{(a)}, \sigma_{e_2}^{(a)}, \text{slope}_2^{(a)}, \mu_H^{(a)}, \sigma_H^{(a)} \right) \in \mathbb{R}^8  \\
\boldsymbol{\Theta}_{2,a} &:= \boldsymbol{\mathfrak{B}}^{(a)} \in \mathbb{R}^{n_a},
\end{align*}
where $a$ refers to the age group under consideration: age group $\mathcal{X}_1$, i.e.~ages 20-59, or age group $\mathcal{X}_2$, i.e.~ages 60-85. Furthermore, $n_a$ refers to the number of ages in age group $a$ and $\boldsymbol{\mathfrak{B}}^{(a)}$ refers to the age-specific effects of mortality shocks $\boldsymbol{\mathfrak{B}}$ restricted to the ages 20-59 for age group $\mathcal{X}_1$ and restricted to the ages 60-85 for age group $\mathcal{X}_2$. The parameters $\sigma_{e_i}^a$ and slope$_i^a$, for $i \in \{1,2\}$, are used in the specification for $\sigma_e(x,t)$, see Equation~\eqref{eq:volatility.adjustment}.

In the first step, we optimize the log-likelihood with respect to $\boldsymbol \Theta_{1,a}$ using an initial estimate of $\boldsymbol{\Theta}_{2,a}$. In the second step, the log-likelihood is optimized with respect to $\boldsymbol \Theta_{2,a}$ using the optimized values of $\boldsymbol \Theta_{1,a}$ from the first step. Both steps are repeated until a relative precision of $10^{-5}$ is reached. We opt for the differential evolution method \citep{price2006differential} to optimize the log-likelihood which is a popular approach for solving complex global optimization problems. We use the `jDE' implementation by \cite{4016057} in \texttt{R}.

\subsection{Modifications to the projection strategy of the regime-switching model} \label{Appendix:C}
The purpose of the high volatility state in the Markov chain is to capture mortality shocks, which may occur over a single year or over several years, such as during world wars or the COVID-19 pandemic. Whenever a mortality shock takes place, we enter the high volatility regime and remain there for the duration of that shock. Empirical observations suggest that the age-averaged residuals $z_{x,t}$ tend to offset each other, i.e.~sum up to zero, during each high volatility period caused by a mortality shock. This is because the baseline model's residuals $z_{x,t}$ are calculated from a mortality improvement model. If this offsetting does not occur, the generated mortality shock would affect the projection of the mortality rates in all future years. As a result, the fan chart of the generated trajectories for the mortality rates can become unrealistically wide. As a first modification, we therefore generate the residuals $z_{x,t}$ in such a way that they offset each other during each high volatility period while taking into account the duration of that period.

Denote the generated Markov chain as $\rho_t$ with $t \in \mathcal{T}^{\text{{\tiny pred}}} = \{2022, 2023, \ldots, 2080\}$, where $\rho_t$ switches between the LVS and the HVS. We denote the generated Markov chain as
$$ (\rho_t)_{t\in \mathcal{T}^{\text{{\tiny pred}}}} = \left(L_{t_1}, H_{t_2}, L_{t_3}, H_{t_4}, \ldots\right) $$
where $t_1 = 2022$ and where $L_{t_i}$ is a collection of consecutive years representing a low volatility period that starts in year $t_i$ and ends in year $t_{i+1} -1$. Similarly $H_{t_j}$ represents a high volatility period that starts in year $t_j$ and ends at year $t_{j+1}-1$. Next, we generate vectors $\boldsymbol{z}_t$ from a multivariate normal distribution where the mean and covariance matrix depend on the occupied state in year $t$. Each high volatility period $H_{t_j}$ consists of at least two consecutive years because of the memory state, as introduced in Section~\ref{Appsec:memory}. Suppose $H_{t_j}$ consists of $n_j := t_{j+1} - t_j$ years. We first generate $n_j-1$ vectors $\boldsymbol{z}_t$, for $t = t_j, t_{j} +1, ..., t_{j+1} - 2$, from the multivariate normal distribution with extra volatility term $\sigma_H$ (see Equation~\eqref{App:eq:cal:transdens}). Second, we take the last $n_j$-th generated vector equal to:
\begin{align*}
\boldsymbol{z}_{t_{j+1}-1} = - \displaystyle \sum_{\tau = t_{j}}^{t_{j+1}-2} \boldsymbol{z}_{\tau}.
\end{align*}
In this way we secure that the sum of the generated values $z_{x,t}$ over each high volatility period and for each age $x$ equals zero. Therefore, the mortality shocks, possibly spread over multiple years, do not influence future projections of mortality rates but will only affect the baseline mortality trend in that specific high volatility period. Furthermore, \added{since the focus in this paper is on mortality shocks that lead to sudden increases in the mortality rates}, we rearrange the vectors $\boldsymbol{z}_t$ generated in each high volatility period $H_{t_j}$. \added{We achieve this by positioning the simulated vector $\boldsymbol{z}_t$ with the highest average, across the different ages, at the beginning of each high volatility period $H_{t_j}$.} 

As evidenced in Figure~\ref{fig:cprob1}, the probability of being in the low volatility regime during the year 2021 is nearly one for the age group 20-59, and nearly zero for the age group 60-85. This aligns with our prior expectations, as described in Section~\ref{sec:casestudyregimeswitch}, considering the impact of the COVID-19 pandemic in the years 2020-2021 on mainly older individuals. To ensure that our mortality projections are up-to-date with the current information, we make the plausible assumption that there will be no mortality shocks in the years 2022 and 2023. As a second modification, we force the Markov chain $\rho_t^{(1)}$ in state $1$ for the age group 20-59 in the years 2022-2023. For the age group 60-85, we keep the Markov chain $\rho_t^{(2)}$ in state $2$ in the year 2022 and in state $1$ in the year 2023. 

\section{Solvency Capital Requirement} \label{Appendix:SolvencyCapitalRequirement}
The Solvency II Directive (2009/138/EC) states that the solvency capital requirement (SCR) corresponds to the Value-at-Risk (VaR) of an insurance or reinsurance company's basic own funds, with a confidence level of $99.5\%$, over a one-year horizon.\footnote{Basic own funds are calculated as the excess of assets over liabilities plus subordinated liabilities \citep{ceiops2010qis}} We put focus on the SCR for the life underwriting risk module, and examine the submodules related to mortality, longevity, and catastrophe risk. However, since longevity and mortality risk are not adequately captured within a one-year horizon, we adopt the run-off VaR approach, which is considered to be a suitable method for evaluating these risks \citep{richards2014value, risks7020058, eiopa2018}. Using this approach, the SCR is calculated as:
\begin{align} \label{eq:VaRapproach}
\text{SCR}^{\text{VaR}} = \text{VaR}_{0.995}\left(\boldsymbol{L}_0\right) - \text{BEL}_0,
\end{align}
where $\text{VaR}_{0.995}$ is the Value at Risk at level $0.995$ and $\boldsymbol{L}_0 := (L_{0,\iota})_\iota$ is the set of simulated liabilities \added{valued} at time 0. \added{These liabilities represent the expected present value of future cash flows of a particular insurance product and are calculated based on different simulations for the future mortality rates, see Section~\ref{sec:projmortratesubsub}.} Furthermore, $\text{BEL}_0$ is the liability \added{valued} at time 0 calculated \added{with} best-estimate projections of future mortality rates\added{, i.e., we set the error terms in Equation~\eqref{eq:TDF:timeseries} and the regime-switching $Y_t$ in Equation~\eqref{eq:simmutsfs} to zero when projecting future mortality rates.}  Furthermore, we assume, for simplicity, a fixed annual interest rate of 0.02 in the calculations of the (best-estimate) life contingent liabilities. 

In the standard model,\footnote{The European Commission, in collaboration with the Committee of Insurance and Occupational Pension Supervisors (CEIOPS), has established the standard model. Its configuration and calibration have been implemented through a series of Quantitative Impact Studies (QIS).} \cite{ceiops2010qis} approximates the SCR as:
\begin{align} \label{eq:shockapproach}
\text{SCR}^{\text{stm}} = \left(\text{BEL}_0 \mid \text{shock}\right) - \text{BEL}_0.
\end{align}
As such, the standard model calculates the SCR as the impact of a particular shock on the BEL at time 0. For longevity risk, the shock represents a permanent $20\%$ reduction in all future mortality rates for each applicable age $x$ \added{and only impacts policies linked to longevity risk}. For mortality risk, the shock represents a permanent $15\%$ increase in all future mortality rates for policies linked to mortality risk. For catastrophe risk, the shock represents a 0.0015 absolute increase in mortality rates only in the upcoming year\added{, for policies linked to mortality risk}. 

\subsection{SCR for an immediate life annuity} \label{Appendix:ILA}
We consider the valuation of an immediate \added{whole} life annuity \added{issued to} an individual aged $x$ in the year 2021, with an annual payout at the end of each year of \euro{10\ 000} until the insured dies.\footnote{We assume the premium payments are paid upfront, as is typical for an immediate life annuity, and therefore solely focus on the liabilities of the insurer.} The BEL at time 0 for this policy, i.e.~in the year 2021, is computed as the sum of discounted expected payouts, given by:
\begin{align} \label{eq:BEL00}
\text{BEL}_0 = \text{\euro{10\ 000}} \cdot \displaystyle \sum_{k=1}^{120-x}  v^k \cdot  {}_{k}p_{x,2021}^{(be)},
\end{align}
where we assume that 120 is the maximum age attainable and $v = 1/(1+i)$ is the discounting factor with a fixed interest rate $i = 0.02$.  The superscript ${(be)}$ refers to the best-estimate projection of the probability that an $x$ year old person in 2021 survives for $k$ years and equals:\footnote{\added{We consider the cohort survival probabilities ${}_{k}p_{x,t}$, i.e., the $k$-year survival probability of an $x$-year old in year $t$ taking evolutions in survival probabilities into account. }}
\begin{align} \label{eq:multiyearsurvprob}
{}_{k}p_{x,2021}^{(be)} = \displaystyle \prod_{j=0}^{k-1} p_{x+j,2021+j}^{(be)} = \displaystyle \prod_{j=0}^{k-1} \left(1 - q_{x+j,2021+j}^{(be)}\right).
\end{align}
In the above calculations, we use the best-estimate projections of the mortality rates obtained with the baseline mortality improvement model, as explained in Section~\ref{sec:casestudy:baseline}. We close the projected mortality rates using the method of \cite{Kannisto} up to age 120. 

We \added{evaluate} the SCR for this policy with both the VaR approach from Equation~\eqref{eq:VaRapproach} based on our proposed mortality improvement model with shock regime and with the standard model in Equation~\eqref{eq:shockapproach}. 

\paragraph{Standard model.} An immediate life annuity is solely exposed to longevity risk, as the insurer is obliged to pay out more annuities if the policyholder lives longer than expected. \added{Hence, we do not take mortality and catastrophe risk into consideration when calculating the SCR.} We apply the formula from Equation~\eqref{eq:shockapproach} and calculate the BEL at time 0 under the longevity shock \added{prescribed in the standard model as}:
\begin{align} \label{eq:ilastm80}
\text{BEL}_0 \mid \text{longevity shock} = \text{\euro{10\ 000}} \displaystyle \sum_{k=1}^{120-x}  v^k \left( \displaystyle \prod_{j=0}^{k-1} \left(1 - 0.8 \cdot q_{x+j,2021+j}^{(be)}\right)\right),
\end{align}
where we decrease the mortality rates with a factor of 0.8. We subtract $\text{BEL}_0$ to obtain the SCR under the standard model.

\paragraph{Run-off VaR.} We use the proposed mortality improvement model with shock regime, as developed in the case study presented in Section~\ref{sec:case}. To limit the war impact, we work with the \added{sensitivity analysis} proposed in Section~\ref{sec:scenario.analysis}. We generate 10 000 trajectories for the mortality rates \added{$q_{x+t, 2021+t, \iota}$} at each age $x$ over a sufficiently long projection period $t \in \mathcal{T}^{\text{{\tiny pred}}}$. Subsequently, we compute 10 000 simulated liabilities $L_{0,\iota}$ \added{valued at time 0 (here: 2021),} using the resulting simulated survival probabilities:  
\begin{align*} 
L_{0,\iota} = \text{\euro{10\ 000}} \cdot \displaystyle \sum_{k=1}^{120-x}  v^k \cdot  {}_{k}p_{x,2021,\iota},
\end{align*}
where the subscript $\iota$ refers to the $\iota$-th generated trajectory of the survival probabilities. We then compute the 99.5$\%$ VaR of these generated liabilities $L_{0,\iota}$ for $\iota = 1,2,...,10\ 000$, \added{which corresponds to scenarios where mortality rates improve substantially, such that people live longer, and, consequently, derive more income from this annuity. Subsequently, we subtract the BEL$_0$ from the $99.5\%$ VaR, see Equation~\eqref{eq:VaRapproach}, to obtain the SCR under the run-off VaR approach}. 

\subsection{Term life insurance} \label{Appendix:TLI}
We \added{consider} a term life insurance policy with terminal age of 65 that is issued to an individual aged $x$ in the year 2021. The death benefit of this policy equals \euro{150 000} and is payable at the end of the year of death, \added{if the death occurs before age 65}. We calculate the BEL at time 0 as the sum of the discounted expected payouts:
\begin{align*}
\text{BEL}_0 = \text{\euro{150 000}} \cdot \displaystyle \sum_{k=1}^{65-x} v^{k} \cdot {}_{k-1}p_{x,2021}^{(be)} \cdot q_{x+k-1,2021+k-1}^{(be)},
\end{align*}
where the superscript $(be)$ refers to the best-estimate projection of the mortality rates and survival probabilities.

\paragraph{Standard model.} A term life insurance contract is exposed to mortality and catastrophe risk. \added{Hence, we do not take longevity risk into consideration when calculating the SCR.} Indeed, if the \added{mortality rate} in a particular year increases, \added{it is more likely that the} insurance undertaking \added{has to pay out the death benefit of} \euro{150 000} \added{within the term of the contract}. We \added{calculate} the SCR with the standard model by aggregating the SCR of the mortality and catastrophe risk according to the guidelines in \cite{ceiops2010qis}:
\begin{align*}
\text{SCR}^{\text{stm}} = \sqrt{\left(\text{SCR}_{\text{mort}}^{\text{stm}}\right)^2 + \left(\text{SCR}_{\text{cat}}^{\text{stm}}\right)^2},
\end{align*}
where $\text{SCR}^{\text{stm}}_{\text{mort}}$ and $\text{SCR}^{\text{stm}}_{\text{cat}}$ are the SCRs attributed to mortality and catastrophe risk respectively, calculated with the standard model \citep{ceiops2010qis}. For the $\text{SCR}^{\text{stm}}_{\text{mort}}$, we calculate the BEL at time 0 subject to the \added{prescribed} mortality shock, by multiplying all future best-estimate mortality rates with a factor 1.15:
$$\begin{aligned}
\text{BEL}_0 \mid \text{mort shock} = \:&\text{\euro{150 000}} \cdot \Bigg(v\cdot 1.15\cdot q_{x,2021}^{(be)} + \\ \displaystyle &\sum_{k=2}^{65-x} v^{k} \cdot \left( \displaystyle \prod_{j=0}^{k-2} \left(1 - 1.15 \cdot q_{x+j,2021+j}^{(be)}\right)\right) \cdot1.15 \cdot  q_{x+k-1,2021+k-1}^{(be)}\Bigg).
\end{aligned}$$
Next, for the $\text{SCR}^{\text{stm}}_{\text{cat}}$, we calculate the BEL at time 0 subject to the catastrophe shock by adding a constant of 0.0015 to the mortality rates in the year 2021:
$$\begin{aligned}
\text{BEL}_0 \mid \text{cat shock} = \: \text{\euro{150 000}} \cdot \Bigg(&v\cdot\tilde{q}_{x,2021} + \\ \displaystyle &\sum_{k=2}^{65-x} v^{k} \cdot \left( \displaystyle \prod_{j=0}^{k-2} \left(1 - \tilde{q}_{x+j,2021+j}\right)\right) \cdot \tilde{q}_{x+k-1,2021+k-1}\Bigg),
\end{aligned}$$
where $\tilde{q}_{x,2021} = q_{x,2021}^{(be)} + 0.0015$ and $\tilde{q}_{x+j,2021+j} = q_{x+j,2021+j}^{(be)}$ for all $j>0$.

\paragraph{Run-off VaR.} We simulate 10 000 trajectories for the mortality rates at each age $x$ using our proposed mortality model from Section~\ref{sec:scenario.analysis} over a sufficiently long projection period. Consequently, we generate 10\ 000 trajectories for the liabilities of the insurer as follows:
\begin{align*}
L_{0,\iota} = \text{\euro{150 000}} \cdot \displaystyle \sum_{k=1}^{65-x} v^{k} \cdot {}_{k-1}p_{x,2021,\iota} \cdot q_{x+k-1,2021+k-1,\iota},
\end{align*}
where the subscript $\iota$ refers to the $\iota$-th generated trajectory of the survival probabilities and mortality rates.
We then calculate the 99.5$\%$ quantile of the liabilities $L_{0,\iota}$ for $\iota = 1,2,...,10\ 000$. \added{This $99.5\%$ quantile corresponds to a situation wherein mortality rates significantly worsen or wherein mortality shocks occur frequently. In such scenarios, policyholders experience earlier death, resulting in elevated losses for the insurer. Subsequently, we subtract the BEL$_0$ from the $99.5\%$ VaR, see Equation~\eqref{eq:VaRapproach}, to obtain the SCR under the run-off VaR approach}. 
\end{document}